\documentclass[aps,pra,reprint,showpacs]{revtex4-1}
\usepackage[english]{babel}
\usepackage{amsmath,amssymb,bbm,graphicx,color,comment,txfonts,chngcntr}
\counterwithin*{paragraph}{subsection}
\usepackage[bookmarks=true,colorlinks,citecolor=blue,urlcolor=blue]{hyperref}

\newcommand{\C}{\mathbb{C}}
\newcommand{\id}{\mathbbm{1}}
\newcommand{\ch}{\cosh(\beta \omega/2)}
\newcommand{\sh}{\sinh(\beta \omega/2)}
\newcommand{\ct}{\coth(\beta \omega/2)}
\newcommand{\bo}{(\beta \omega/2)}
\newcommand{\oq}{(\omega,\mathbf{q})}

\newcommand{\sdyn}{S_{\mathrm{dyn}}}
\newcommand{\xx}{\mathbf{x}}

\newcommand{\abs}[1]{\left\lvert #1 \right\rvert}

\renewcommand{\Im}{\mathop{\mathrm{Im}}}
\DeclareMathOperator{\tr}{tr}
\DeclareMathOperator{\Det}{Det}

\begin{document}

\date{\today}

\title{Thermodynamic Equilibrium as a Symmetry of the Schwinger-Keldysh Action}

\author{L. M. Sieberer$^{1,2}$}
\author{A. Chiocchetta$^3$}
\author{A. Gambassi$^3$}
\author{U. C. T\"auber$^4$}
\author{S. Diehl$^5$}

\affiliation{$^1$Department of Condensed Matter Physics, Weizmann Institute of
  Science, Rehovot 7610001, Israel}

\affiliation{$^2$Institute for Theoretical Physics, University of Innsbruck,
  A-6020 Innsbruck, Austria}

\affiliation{$^3$SISSA --- International School for Advanced Studies and INFN,
  via Bonomea 265, I-34136 Trieste, Italy}

\affiliation{$^4$Department of Physics (MC 0435), Robeson Hall, 850 West Campus
  Drive, Virginia Tech, Blacksburg, Virginia 24061, USA}

\affiliation{$^5$Institute of Theoretical Physics, TU Dresden, D-01062 Dresden,
  Germany}

\begin{abstract}
  Extended quantum systems can be theoretically described in terms of the
  Schwinger-Keldysh functional integral formalism, whose action conveniently
  describes both dynamical and static properties. We show here that in thermal
  equilibrium, defined by the validity of fluctuation-dissipation relations, the
  action of a quantum system is invariant under a certain symmetry
  transformation and thus it is distinguished from generic systems. In turn, the
  fluctuation-dissipation relations can be derived as the Ward-Takahashi
  identities associated with this symmetry. Accordingly, the latter provides an
  efficient test for the onset of thermodynamic equilibrium and it makes
  checking the validity of fluctuation-dissipation relations unnecessary. In the
  classical limit, this symmetry reduces to the well-known one which
  characterizes equilibrium in the stochastic dynamics of classical systems
  coupled to thermal baths, described by Langevin equations.
\end{abstract}



\maketitle

\section{Introduction}
\label{sec:introduction}

In recent years, the question under which conditions and how a quantum many-body
system thermalizes has received ever-growing attention. This interest has been
primarily triggered by the increasing ability to prepare and manipulate such
systems, which might be either
\emph{isolated}~\cite{Polkovnikovrev,Yukalov11,Eisert2015} --- as it is
typically the case in experiments with cold atoms ~\cite{Lamacraft12,Bloch2008}
--- or in contact with an environment (\emph{open}), and therefore subject to
losses and driving.

After an abrupt perturbation, \emph{isolated systems} are generically expected
to thermalize in the sense that expectation values of local quantities at long
times can be determined on the basis of suitable statistical
ensembles~\cite{Polkovnikovrev,Eisert2015}.
However, this might not be the case because of the presence of an
extensive amount of conserved quantities induced by integrability
~\cite{Jaynes1957,Kinoshita2006,Rigol2007,Kollar2011,Caux2012} or because of a breaking
of ergodicity due to the occurrence of many-body localization
~\cite{Basko2006,PalHuse10,Serbyn2013,VoskAltman13}.
Although it is possible to define a variety of effective temperatures based on
the static~\cite{Rossini2009,Mitra2011} and dynamic properties
~\cite{PhysRevB.84.212404,Foini2012} under such circumstances, the lack of
thermal behavior is witnessed by the fact that these temperatures do not
necessarily take all the same thermodynamic value.

Examples of \emph{open systems} include exciton-polaritons in semiconductor
heterostructures~\cite{Carusotto2013,Byrnes2014}, arrays of
microcavities~\cite{Hartmann2008,Houck2012}, trapped ions~\cite{Blatt2012}, as
well as optomechanical setups~\cite{Physics.2.40}.
In general it is unclear, \textit{a priori}, by which physical mechanism an
effective temperature is possibly established in these systems and, in case,
what determines its value. Recent work, however, suggests possible mechanisms
where an effective temperature can occur as a consequence of the competition
between driven-dissipative and coherent
dynamics~\cite{Mitra2006,Diehl2008,Diehl2010a,DallaTorre2010,DallaTorre2012,DallaTorre2013,Chiocchetta2013,Sieberer2013,Tauber14}.
Irrespective of its cause, effective thermalization often affects only the
low-energy degrees of
freedom~\cite{Mitra2006,Diehl2008,DallaTorre2010,Diehl2010a,DallaTorre2012,DallaTorre2013,Chiocchetta2013,Sieberer2013,Tauber14,Wouters2006,Mitra2011,Mitra2012,Oztop2012,Maghrebi2015}.

All these examples show clearly that the presence of effective thermodynamic
equilibrium (which might be established only in a subsystem or within a specific
range of frequencies) in the steady state of a system is often by no means
obvious.
Hence, before addressing the question of whether the time evolution of a certain
system leads to thermalization or not, it is imperative to identify criteria
which allow a clear-cut detection of thermodynamic equilibrium conditions in the
stationary state.
In this regard, it is important to consider not only the \emph{static}
properties of the density matrix of the system, which describes its stationary
state, but also the \emph{dynamics} of fluctuations: being encoded, e.g., in
two-time correlation and response functions, it might or might not be compatible
with equilibrium. As a fundamental difference between static and dynamic
properties, the latter necessarily involves the generator of time evolution,
while the former does not.
  
In this work we consider the following operative definition of thermal
equilibrium: a system is in thermal equilibrium at a certain temperature $T$ if
expectation values of arbitrary products of operators, evaluated at different
times, are connected by quantum fluctuation-dissipation relations (FDRs)
involving the temperature $T$.
These FDRs were shown~\cite{Chou1985,Wang2002,Jakobs2010} to be equivalent to a
combination of the quantum-mechanical time-reversal
transformation~\cite{Messiah:II} and the Kubo-Martin-Schwinger (KMS)
condition~\cite{Kubo1957,Martin1959}.  Heuristically, the latter condition
expresses the fact that the Hamiltonian ruling the time evolution of a system is
the same as that one determining the density matrix of the canonical ensemble,
which characterizes the system when it is weakly coupled to a thermal bath. In
both the generalized FDRs and the KMS condition the temperature appears as a
parameter.

From the theoretical point of view, static and dynamical properties of
statistical systems (both classical and quantum) are often conveniently studied
in terms of dynamical functionals, which are used in order to generate
expectation values of physical observables in the form of functional integrals
over a suitable set of fields.
Then, it is natural to address the issue of the possible equilibrium character
of the stationary state by investigating the properties of the corresponding
dynamical functional.
In the case of classical statistical systems evolving under the effect of an
external stochastic noise of thermal origin, this issue has been discussed to a
certain level of detail in the past
~\cite{Janssen1976,Bausch1976,Janssen1979,Janssen1992,aron10:_symmet_langev,Aron2014},
and it was found that the dynamical functional acquires a specific symmetry in
thermodynamic equilibrium. As in the case of the FDRs and the KMS condition, the
(inverse) temperature $\beta = 1/T$ enters as a parameter in this symmetry
transformation.  Remarkably, classical FDRs can be derived as a consequence of
this symmetry.
For quantum systems, instead, we are not aware of any analogous derivation based
on the symmetries of the corresponding dynamical functional, which takes the
form of a Schwinger-Keldysh action (see, e.g.,
Refs.~\onlinecite{Schwinger1961,Bakshi1963,Bakshi1963a,Mahanthappa1962,Keldysh1965,Kamenev2011,Altland/Simons,Stoof1999}).

The aim of the present work is to fill in this gap by showing that also the
Schwinger-Keldysh dynamical functional of a quantum system in thermal
equilibrium is characterized by a specific symmetry, i.e., it is invariant under
a certain transformation $\mathcal{T}_{\beta}$. This symmetry may be considered
as the generalization of the classical one mentioned above, to which it reduces
in a suitable classical limit~\cite{Altland2010}. In addition,
$\mathcal{T}_{\beta}$ can be written as a composition of the quantum-mechanical
time reversal expressed within the Schwinger-Keldysh formalism --- reflecting a
property of the generator of dynamics --- and of the transformation which
implements the KMS conditions, associated with a property of the state in
question.
The existence of this symmetry was already noticed in
Ref.~\onlinecite{Altland2010} for mesoscopic quantum devices, where it was used
to derive fluctuation relations for particle transport across them.
However, to our knowledge, the connection between this symmetry and the presence
of equilibrium conditions has not yet been established.

The rest of the presentation is organized as follows: the key results of this
work are anticipated and summarized in Sec.~\ref{sec:key-results}; in
Sec.~\ref{sec:symm-transf}, we specify the symmetry transformation
$\mathcal{T}_{\beta}$, provide its various representations, and list a number of
properties which are then detailed in Sec.~\ref{sec:invar-keldysh-acti}. In
particular, we discuss the invariance of unitary time evolution in
Sec.~\ref{sec:invariance-action}, while in Sec.~\ref{sec:diss-contr-equil} we
consider possible dissipative terms which are invariant under
$\mathcal{T}_{\beta}$.
We discuss how the quantum symmetry reduces in the limit $\hbar \to 0$ to the
one known in classical stochastic systems in Sec.~\ref{sec:classical-limit}. As
we discuss in Sec.~\ref{sec:equiv-symm-kms}, the symmetry 
can be interpreted as a practical implementation of the KMS
condition on the Schwinger-Keldysh functional
integral. 
Finally, Sec.~\ref{sec:implications} presents applications of the equilibrium
symmetry: in Sec.~\ref{sec:fluct-diss-theor} we derive the FDR for two-point
functions while in Sec.~\ref{sec:non-equil-nature} we show that the steady
states of a quantum master equation explicitly violate the symmetry. The case of
a system driven out of equilibrium by a coupling with two baths at different
temperature and chemical potential is considered in Sec.~\ref{sec:two-baths};
Sec.~\ref{sec:further-applications} briefly touches upon a number of other
applications of the symmetry.

\section{Key results}
\label{sec:key-results}

\paragraph{The invariance under $\mathcal{T}_{\beta}$ of the Schwinger-Keldysh
  action is a sufficient and necessary condition for a system to be in thermal
  equilibrium.}

As mentioned in Sec.~\ref{sec:introduction}, we consider a system to be in
thermal equilibrium if all the FDRs are satisfied with the same temperature
$T = \beta^{-1}$ or, equivalently~\cite{Chou1985,Wang2002,Jakobs2010}, if the
KMS condition (combined with time reversal) is satisfied. In
Sec.~\ref{sec:equiv-symm-kms}, we show that these conditions imply the thermal
symmetry $\mathcal{T}_{\beta}$ of the Schwinger-Keldysh action corresponding to
the stationary state of the system. Conversely, the fluctuation-dissipation
relations can be derived as consequences of the symmetry, proving their
equivalence.

\paragraph{A different perspective: thermal equilibrium as a symmetry.}

A key conceptual step forward we take in this work is to provide a compact
formulation of thermal equilibrium conditions of a quantum system --- i.e., the
KMS condition (or, alternatively, of the equivalent hierarchy of FDRs) --- in
terms of a single symmetry $\mathcal{T}_{\beta}$, which can be considered as the
fundamental property of quantum systems in thermal equilibrium.  This
perspective is especially fruitful within the field-theoretical formalism, where
various tools have been developed to work out the consequences of the symmetries
of the action of a given system.  In this context, for example, the hierarchy of
generalized quantum FDRs can be derived straightforwardly as the Ward-Takahashi
identities associated with the thermal symmetry (see
Secs.~\ref{sec:equiv-symm-kms} and~\ref{sec:fluct-diss-theor}).  In addition,
the Schwinger-Keldysh formalism provides a convenient framework to take
advantage of very powerful and efficient renormalization-group techniques for
studying the possible emergence of collective behaviors and for monitoring how
the effective description of a statistical system depends on the length and time
scale at which it is analyzed. The possible scale dependence of the
restoration/violation of the equilibrium symmetry could shed light on the
mechanism underlying the thermalization of extended systems.

As we mentioned above, the idea of viewing thermal equilibrium as a symmetry is
certainly not new. However, while previous studies were primarily concerned with
classical statistical
physics~\cite{Janssen1976,Bausch1976,Janssen1979,Janssen1992,aron10:_symmet_langev,Aron2014},
here we generalize this idea to the quantum case.

\paragraph{Unification of the quantum and classical description of equilibrium systems.}

As pointed out in Ref.~\onlinecite{Altland2010}, the equilibrium symmetry
reduces, in the classical limit, to a known symmetry which characterizes thermal
equilibrium in open classical
systems~\cite{Janssen1976,Bausch1976,Janssen1979,Janssen1992,aron10:_symmet_langev,Aron2014}. In
Sec.~\ref{sec:classical-limit} we review the classical limit of the
Schwinger-Keldysh action for a system coupled to a thermal
bath~\cite{Kamenev2011,Altland/Simons} and we discuss in detail how the
classical equilibrium symmetry is recovered. The comparison with the classical
symmetry highlights some remarkable differences with the quantum case: in fact,
in classical systems, thermal equilibrium can be regarded as a consequence of
detailed balance, which, in turn, is related to the property of
microreversibility of the underlying microscopic dynamics. In fact, the
classical equilibrium symmetry is derived by requiring the dynamical functional
to satisfy these
properties~\cite{Janssen1979,Janssen1992,aron10:_symmet_langev}. For quantum
system, instead, an analogous satisfactory definition of detailed balance and
microreversibility is seemingly still missing, leaving open the important
question about the very nature of thermal equilibrium of quantum systems.

\paragraph{Efficient check for the presence of thermodynamic equilibrium
  conditions.}

The symmetry is of great practical value, as it reduces answering the question
about the possible presence of thermodynamic equilibrium to verifying a symmetry
of the Schwinger-Keldysh action instead of having to check explicitly the
validity of all FDRs.
In particular, we show in Sec.~\ref{sec:non-equil-nature} that the Markovian
quantum dynamics described by a Lindblad master
equation~\cite{Kossakowski1972,Lindblad1976} explicitly violates the
symmetry. This reflects the driven nature of the system: indeed, the Lindblad
equation may be viewed as resulting from the coarse graining of the evolution of
an underlying time-dependent system-bath Hamiltonian, with a time dependence
dictated by coherent external driving fields.

Moreover, in Sec.~\ref{sec:two-baths} we consider a bosonic mode coupled to two
baths at different temperatures and chemical potentials: in this case, the
resulting net fluxes of energy or particles drive the system out of equilibrium
with a consequent violation of the symmetry.

\paragraph{A new perspective on the construction of the Schwinger-Keldysh
  action.}

At the conceptual level, the existence of the symmetry provides a new
perspective on the construction of Schwinger-Keldysh functional integrals. In
particular, as customary in quantum field theories, one may consider the
symmetry as the fundamental principle: indeed, it is explicitly present for any
time-independent (time-translation invariant) Hamiltonian which generates the
dynamics of a system at the microscopic scale. Then, requiring the symmetry to
hold for the full effective Keldysh action at a different scale fixes the
admissible dissipative terms so as to satisfy FDRs between response and
correlation functions of arbitrary order; translating back into the operator
language, this provides a concrete hint why stationary density matrices of the
form $\rho \sim e^{-\beta H}$ are favored over arbitrary functions $\rho(H)$ for
the description of static correlation functions.

\section{Symmetry transformation}
\label{sec:symm-transf}

As we anticipated above, a convenient framework for the theoretical description
of the time evolution of interacting quantum many-body systems is provided by
the Schwinger-Keldysh functional integral
formalism~\cite{Kamenev2011,Altland/Simons}. It offers full flexibility in
describing both non-equilibrium dynamics and equilibrium as well as
non-equilibrium stationary states, which is out of reach of the
finite-temperature Matsubara technique~\cite{Lifshitz1980}. In addition, it is
amenable to the well-established toolbox of quantum field theory.
The simplest way to illustrate the basic ingredients of the Schwinger-Keldysh
formalism is to consider the functional integral representation of the so-called
Schwinger-Keldysh partition function $Z$. For a system with unitary dynamics
generated by the Hamiltonian $H$ and initialized in a state described by a
density matrix $\rho_0$, this function is given by
$Z = \tr \left( e^{-i H t} \rho_0 e^{i H t} \right)$.
(Note that, as it stands, $Z=1$; however, it is instructive to focus on its
structure independently of its actual value.)
In this expression, time evolution can be interpreted as occurring along a
closed path: starting in the state described by $\rho_0$, the exponential
$e^{-i H t}$ to the left of $\rho_0$ corresponds to a ``forward'' evolution up
to the time $t$, while the exponential $e^{i H t}$ to its right corresponds to
an evolution going ``backward'' in time. The trace $\tr \left( \dotsb \right)$
connects, at time $t$, the forward with the backward branch of the time path and
therefore it produces a closed-time-path integral.
Along each of these two branches, the temporal evolution can be represented in a
standard way as a functional integral of an exponential weight $e^{i S}$ over
suitably introduced (generally complex) integration variables, i.e., fields,
$\psi_{+}(t, \mathbf{x})$ and $\psi_{-}(t, \mathbf{x})$ on the forward and
backward branches, respectively.
These fields are associated with the two sets of coherent states introduced as
resolutions of the identity in-between two consecutive infinitesimal time
evolutions in the Trotter decomposition of the unitary temporal evolution along
the two branches~\cite{Kamenev2011,Altland/Simons}.
The resulting Schwinger-Keldysh action $S$ is a functional of
$\psi_\pm(t,\mathbf{x})$ and it is generally obtained as a temporal integral
along the close path in time of a Lagrangian density. (Explicit forms of $S$
will be discussed further below, but they are not relevant for the present
discussion.)
By introducing different (time-dependent) sources $J_\pm$ for the fields
$\psi_\pm$ on the two branches, the partition function $Z[J_+,J_-]$ is no longer
identically equal to 1 and its functional derivatives can be used in order to
generate various time-dependent correlation functions (see, e.g.,
Refs.~\onlinecite{Kamenev2011,Altland/Simons,Stoof1999}).

As we show further below in Sec.~\ref{sec:equiv-symm-kms}, a system is in
thermodynamic equilibrium at a temperature $T = 1/\beta$, if the corresponding
Schwinger-Keldysh action is invariant under a certain transformation
$\mathcal{T}_{\beta}$ which acts on the fields along the closed time path.
In order to specify the form of $\mathcal{T}_{\beta}$, we focus on the dynamics
of a single complex bosonic field, which is sufficiently simple but general
enough to illustrate conveniently all the basic ideas.
In this case, the transformation $\mathcal{T}_{\beta}$ turns out to be composed
of a complex conjugation~\footnote{In Ref.~\onlinecite{Altland2010}, the symmetry is
  stated in terms of the real phase variables of complex fields. Then, the
  complex conjugation in Eq.~\eqref{eq:0} should be replaced by a change of
  sign.}  of the field components $\psi_{\sigma}$ with $\sigma = \pm$, an
inversion of the sign of the time variable, and a translation of the time
variable into the complex plane by an amount $i \sigma \beta/2$, i.e.,
\begin{equation}
\label{eq:0}
 \begin{split}
      \mathcal{T}_{\beta} \psi_{\sigma}(t, \mathbf{x}) & = 
      \psi_{\sigma}^{*}(-t + i \sigma \beta/2, \mathbf{x}), \\ 
      \mathcal{T}_{\beta}
      \psi_{\sigma}^{*}(t, \mathbf{x}) & =  \psi_{\sigma}(-t + i \sigma \beta/2, \mathbf{x}).
    \end{split}
\end{equation}
For convenience and future reference we provide an alternative compact
representation of the action of $\mathcal{T}_{\beta}$ both in the time and real
space domain $(t,\mathbf{x})$ as well as in the frequency-momentum domain
$(\omega,\mathbf{q})$. The convention for the Fourier transforms of the fields,
conveniently collected into two spinors
$\Psi_{\sigma}(t,\mathbf{x}) = \left( \psi_{\sigma}(t,\mathbf{x}),
  \psi_{\sigma}^{*}(t,\mathbf{x}) \right)^T$, is the following:
\begin{equation}
  \Psi_\sigma(t,\mathbf{x}) = \int \frac{\mathrm{d}^d \mathbf{q}}{(2\pi)^d}
  \int_{-\infty}^{+\infty} \frac{\mathrm{d} \omega}{2\pi}\, e^{i(\mathbf{q}\cdot\mathbf{x} -  \omega t)}\Psi_{\sigma}\oq.
\end{equation}
In this relation, $d$ is the spatial dimensionality of the system, and the field
spinors in the frequency-momentum domain are defined as
$\Psi_{\sigma}\oq = \left( \psi_{\sigma}\oq, \psi^*_{\sigma}
  (-\omega,-\mathbf{q}) \right)^T$.
Accordingly, we can write the symmetry transformation $\mathcal{T}_{\beta}$ in
the form
\begin{equation}
  \label{eq:1}  
  \begin{split}
    \mathcal{T}_{\beta} \Psi_{\sigma}(t,\mathbf{x}) & = \Psi_{\sigma}^{*}(-t + i \sigma
    \beta/2,\mathbf{x}) = \sigma_x \Psi_{\sigma}(-t + i \sigma
    \beta/2,\mathbf{x}), \\[2mm]
    \mathcal{T}_{\beta} \Psi_{\sigma}\oq & = e^{-\sigma \beta \omega/2}
    \Psi_{\sigma}^{*}(\omega,-\mathbf{q}) = e^{-\sigma \beta \omega/2} \sigma_x
    \Psi_{\sigma}(-\omega,\mathbf{q}),
  \end{split}
\end{equation}
where we introduced the Pauli matrix $\sigma_x = \left(
  \begin{smallmatrix}
    0 & 1 \\ 1 & 0
  \end{smallmatrix}
\right)$.
The transformation in real time requires evaluating the fields for complex
values of the time argument, which in principle is not defined; however, the
complementary representation in Fourier space indicates how this can be done in
practice: in frequency space, the shift of time by an imaginary part
$i \sigma \beta/2$ amounts to a multiplication by a prefactor
$e^{-\sigma \beta \omega/2}$.

As usual within the Schwinger-Keldysh formalism, it is convenient to introduce
what are known as \emph{classical} and \emph{quantum} fields. These are defined as the
symmetric and antisymmetric combinations, respectively, of fields on the
forward and backward branches:
\begin{equation}
  \label{eq:9}  
  \phi_c = \frac{1}{\sqrt{2}} \left( \psi_{+} + \psi_{-} \right), \quad \phi_q 
  = \frac{1}{\sqrt{2}} \left( \psi_{+} - \psi_{-} \right).
\end{equation}
Combining these fields into spinors
$\Phi_{\nu}\oq = (\phi_\nu\oq, \phi^*_\nu(-\omega,-\mathbf{q}))^T$ --- where the
index $\nu = c,q$ distinguishes classical and quantum fields --- the
transformation $\mathcal{T}_{\beta}$ takes the following form, which we report
here for future reference:
\begin{widetext}
  \begin{equation}
  \label{eq:3}
  \begin{split}
    \mathcal{T}_{\beta} \Phi_c\oq & = \sigma_x \left( \ch
      \Phi_c(-\omega,\mathbf{q}) - \sh \Phi_q(-\omega,\mathbf{q}) \right), \\
    \mathcal{T}_{\beta} \Phi_q\oq & = \sigma_x \left( - \sh
      \Phi_c(-\omega,\mathbf{q}) + \ch \Phi_q(-\omega,\mathbf{q}) \right).
  \end{split}
\end{equation}
\end{widetext}
We anticipate and summarize here a number of properties of the equilibrium
transformation $\mathcal{T}_{\beta}$, which are going to be discussed in detail
in Secs.~\ref{sec:invar-keldysh-acti} and \ref{sec:equiv-symm-kms}:
\begin{enumerate}
\item The transformation is linear, discrete and involutive, i.e.,
  $\mathcal{T}_{\beta}^2 = \id$. The last property follows straightforwardly
  from Eqs.~\eqref{eq:0} or \eqref{eq:1}.  Concerning linearity, note in
  particular that the complex conjugation in Eq.~\eqref{eq:0} affects only the
  field variables, i.e.,
  $\mathcal{T}_{\beta} \lambda \psi_{\sigma}(t, \mathbf{x}) = \lambda
  \psi_{\sigma}^{*}(-t + i \sigma \beta/2, \mathbf{x})$
  for $\lambda \in \C$ (see Sec.~\ref{sec:time-reversal}).
\item $\mathcal{T}_{\beta}$ can be written as a composition
  $\mathcal{T}_{\beta} = \mathsf{T} \circ \mathcal{K}_{\beta}$ of a
  time-reversal transformation $\mathsf{T}$ and an additional transformation
  $\mathcal{K}_{\beta}$, which we will identify in Sec.~\ref{sec:kms-condition}
  as the implementation of the KMS condition within the Schwinger-Keldysh
  functional integral formalism.
\item $\mathcal{T}_{\beta}$ is not uniquely defined, due to a certain freedom in
  implementing the time-reversal transformation within the Schwinger-Keldysh
  functional integral formalism, as discussed in
  Sec.~\ref{sec:time-reversal}. However, without loss of generality, we stick to
  the definition provided by Eq.~\eqref{eq:0} and we comment on the alternative
  forms in Sec.~\ref{sec:time-reversal}.
\item The transformation $\mathcal{T}_{\beta}$ leaves the functional measure
  invariant, i.e., the absolute value of the Jacobian determinant associated
  with $\mathcal{T}_{\beta}$ is equal to one, as discussed in
  Sec.~\ref{sec:from-kms-symmetry} and shown in App.~\ref{sec:jacob-equil-transf}.
\item \label{item:1}
  The various forms of the transformation $\mathcal{T}_{\beta}$ presented
  above apply to the case of a system of bosons with vanishing chemical
  potential $\mu$. In the presence of $\mu\neq 0$, Eq.~\eqref{eq:0} becomes
  \begin{equation}
    \label{eq:4}
    \begin{split}
      \mathcal{T}_{\beta, \mu} \psi_{\sigma}(t, \mathbf{x}) & = e^{\sigma \beta
        \mu/2} \psi_{\sigma}^{*}(-t + i \sigma \beta/2, \mathbf{x}), \\
      \mathcal{T}_{\beta, \mu} \psi_{\sigma}^{*}(t, \mathbf{x}) & = e^{- \sigma \beta
        \mu/2} \psi_{\sigma}(-t + i \sigma \beta/2, \mathbf{x}),
    \end{split}
  \end{equation}
  with a consequent modification of Eq.~\eqref{eq:1}, which can be easily worked
  out.
  After a transformation to the basis of classical and quantum fields according
  to Eq.~\eqref{eq:9}, this modification amounts to shifting the frequency
  $\omega$ in the arguments of the hyperbolic functions in Eq.~\eqref{eq:3},
  i.e., to $\omega \to \omega - \mu$.
\item\label{item:2} In taking the Fourier transforms in Eqs.~\eqref{eq:1}
  and~\eqref{eq:3} one implicitly assumes that the initial state of the system
  was prepared at time $t = - \infty$, while its evolution extends to
  $t = \infty$. In the following we will work under this assumption, commenting
  briefly on the role of an initial condition imposed at a finite time in
  Sec.~\ref{sec:classical-limit}.
\end{enumerate}

\section{Invariance of the Schwinger-Keldysh action}
\label{sec:invar-keldysh-acti}

As we demonstrate further below in Sec.~\ref{sec:equiv-symm-kms}, a system is in
thermodynamic equilibrium if its Schwinger-Keldysh action $S$ is invariant under
the transformation $\mathcal{T}_{\beta}$, i.e.,
\begin{equation}
  \label{eq:5}
  S[\Psi] = \tilde{S}[\mathcal{T}_{\beta} \Psi],
\end{equation}
where, for convenience of notation,
$\Psi 
= \left( \psi_{+}, \psi_{+}^{*}, \psi_{-}, \psi_{-}^{*} \right)^T$
collects all the fields introduced in the previous section into a single
vector. The tilde in $\tilde{S}$ indicates that all the parameters in $S$ which
are related to external fields have to be replaced by their corresponding
time-reversed values (e.g., the signs of magnetic fields have to be inverted),
while in the absence of these fields the tilde may be dropped.

According to the construction of the Schwinger-Keldysh functional integral
outlined at the beginning of the previous section, the action corresponding to
the unitary dynamics of a closed system is completely determined by its
Hamiltonian $H$. The initial state $\rho_0$ of the dynamics enters the
functional integral as a boundary condition: if the system was prepared in the
state $\rho_0$ at the time $t = 0$, the matrix element
$\langle \psi_{+, 0} | \rho_0 | \psi_{-, 0} \rangle$, where
$\lvert \psi_{\pm, 0} \rangle$ are coherent states, determines the (complex)
weight of field configurations at the initial time with
$\psi_{\pm}(0, \mathbf{x}) = \psi_{\pm, 0}(\mathbf{x})$.
In Sec.~\ref{sec:invariance-action}, we demonstrate the invariance of the
Schwinger-Keldysh action associated with a time-independent Hamiltonian dynamics
under the transformation $\mathcal{T}_{\beta}$. In particular, this invariance
holds for \emph{for any value of $\beta.$} Interestingly enough, the
Schwinger-Keldysh action associated with a Hamiltonian of a simple
non-interacting system --- which can be diagonalized in terms of single-particle
states --- turns out to be invariant under an enhanced version of this
transformation, involving possibly different values of $\beta$ for each of the
single-particle states (see Sec.~\ref{sec:enhanc-symm-simple}).  A constraint on
the value of $\beta$, however, comes from the inclusion of the boundary
condition for the functional integral which specifies the initial state
$\rho_0$.
Here we are interested in the stationary state of the system, which is
generically reached a long time after its preparation in the state
$\rho_0$. Hence, we assume that this was done in the distant past, i.e., at
$t = - \infty$, and that the evolution of the system extends to $t = +\infty$
(cf.\ point~\ref{item:2} in Sec.~\ref{sec:symm-transf}). In the construction of
the Schwinger-Keldysh functional integral for a system in thermodynamic
equilibrium~\cite{Kamenev2011,Altland/Simons}, a convenient alternative approach
for specifying the appropriate boundary conditions corresponding to the initial
equilibrium state $\rho_0$ of the system, consists in adding infinitesimal
dissipative contributions to the
action. Usually~\cite{Kamenev2011,Altland/Simons}, the form of these
contributions is determined by the requirement that the Green's functions of the
system are thermal with a specific temperature $T = 1/\beta$, i.e., that they
obey a fluctuation-dissipation relation; once these terms are included, any
reference to $\rho_0$ may be omitted. We demonstrate in
Sec.~\ref{sec:regularization}, that these dissipative contributions are
invariant under $\mathcal{T}_{\beta}$ with exactly the same $\beta$. Hence, the
thermal symmetry provides a different perspective on the construction of the
Schwinger-Keldysh functional integral for a system in thermal equilibrium: while
the unitary contributions are fixed by the Hamiltonian of the system, the
requirement of invariance under the symmetry transformation
$\mathcal{T}_{\beta}$ can be taken as the fundamental principle for specifying
the structure of the dissipative terms which can occur in the action if the
system is in thermodynamic equilibrium at temperature $T = 1/\beta$.  We
emphasize that only the simultaneous presence in the Schwinger-Keldysh action of
both the Hamiltonian and the dissipative contributions yields a well-defined
functional integral: the dissipative terms in the microscopic action are taken
to be infinitesimally small as for an isolated system, where they merely act as
a regularization which renders the functional integral finite and ensures that
the bare response and correlation functions satisfy a FDR; on the other hand, if
the isolated system is composed of a small subsystem of interest and a remainder
which can be considered as a bath, then finite dissipative contributions emerge
in the Schwinger-Keldysh action of the subsystem after the bath has been
integrated out. This scenario is considered in Secs.~\ref{sec:classical-limit}
and~\ref{sec:non-equil-nature}. Moreover, the system can act as its own bath: in
fact, one expects the effective action for the low-frequency and long-wavelength
dynamics of the system to contain dissipative contributions which are due to the
coupling to high-frequency fluctuations. In Sec.~\ref{sec:diss-contr-equil}, we
explicitly construct dissipative terms which comply with the thermal symmetry
$\mathcal{T}_{\beta}$. In particular, we find that the noise components
associated with these dissipative terms must necessarily have the form of the
equilibrium Bose-Einstein distribution function, as appropriate for the bosonic
fields which we are presently focussing on.

\subsection{Invariance of Hamiltonian dynamics}
\label{sec:invariance-action}

The Schwinger-Keldysh action $S$ associated with the dynamics generated by a
time-independent Hamiltonian $H$ can be written as the sum of a ``dynamical''
and a ``Hamiltonian'' part, $\sdyn$ and $S_{\mathcal{H}}$, respectively,
\begin{align}
  \label{eq:6}
  S & = \sdyn + S_{\mathcal{H}}, \\ \label{eq:7}
  \sdyn & = \frac{1}{2}\int_{t, \mathbf{x}} \left( \Psi_{+}^{\dagger} i \sigma_z\partial_t \Psi_{+} -
          \Psi_{-}^{\dagger} i\sigma_z\partial_t \Psi_{-} \right),
  \\ \label{eq:8} S_{\mathcal{H}} & = - \int_t \left( \mathcal{H}_{+} - \mathcal{H}_{-} \right),
\end{align}
where we used the shorthand
$\int_t \equiv \int_{-\infty}^{\infty} {\rm d} t, \int_{\mathbf{x}} \equiv \int {\rm d}^d
\mathbf{x}$, while $\sigma_z = \left(\begin{smallmatrix} 1 & 0 \\ 0 & -1
\end{smallmatrix}\right)
$
is the Pauli matrix. 
This structure of the Schwinger-Keldysh action results from the construction of
the functional integral outlined at the beginning of
Sec.~\ref{sec:symm-transf}. In particular, the Hamiltonians $\mathcal{H}_{\pm}$
are matrix elements of the Hamiltonian operator $H$ in the basis of coherent
states $| \psi_\pm \rangle$, i.e.,
$\mathcal{H}_{\sigma} = \langle \psi_{\sigma} | H | \psi_{\sigma}
\rangle/\langle \psi_{\sigma} | \psi_{\sigma} \rangle$,
where the amplitudes $\psi_\pm$ of the coherent states are just the integration
variables in the functional integral~\cite{Kamenev2011,Altland/Simons}.
Henceforth we focus on the case of a bosonic many-body system with contact
interaction, i.e., with Hamiltonians in Eq.~\eqref{eq:8} given by
\begin{equation}
  \label{eq:29}
  \mathcal{H}_{\sigma} = \int_{\mathbf{x}} \left( \frac{1}{2 m} \abs{\nabla \psi_{\sigma}}^2 + \tau
    \abs{\psi_{\sigma}}^2 + \lambda \abs{\psi_{\sigma}}^4 \right).
\end{equation}
Here $m$ is the mass of bosons, $\tau$ the chemical potential, 
and $\lambda$ parametrizes the strength of the $s$-wave two-body interaction.
We consider this case because it is sufficiently general for the purpose of
illustrating all basic concepts associated with the thermal symmetry and, in
addition, in the classical limit it allows a direct comparison with classical
stochastic models~\cite{Hohenberg1977,Folk2006}, where
$\phi_c = \left( \psi_{+} + \psi_{-} \right)/\sqrt{2}$ plays the role of a
bosonic order parameter field. This point is elaborated in
Sec.~\ref{sec:classical-limit}.
  
Below we show that the invariance of the Schwinger-Keldysh action $S$ under
$\mathcal{T}_{\beta}$ is intimately related to the structure of the action,
i.e., to the fact that it can be written as the sum of two terms containing,
separately, only fields on the forward and backward branches.

\subsubsection{Dynamical term}
\label{sec:dynamical-term}

To begin with, we show that the dynamical contribution $\sdyn$ to the
Schwinger-Keldysh action $S$ given in Eq.~\eqref{eq:7}, is invariant under
$\mathcal{T}_{\beta}$, i.e., that
$\sdyn[\mathcal{T}_{\beta} \Phi] = \sdyn[\Phi]$. To this end, it is convenient
to express the original fields $\{ \psi_\pm,\psi_\pm^*\}$ in the so-called
Keldysh basis, which is formed by the classical and quantum components
$\{ \phi_{c,q},\phi_{c,q}^*\}$ introduced in Eq.~\eqref{eq:9}. For the sake of
brevity, we arrange these fields into the vector
$\Phi = \left( \phi_c, \phi_c^{*}, \phi_q, \phi_q^{*} \right)^T$.
Rewriting $\sdyn$ in these terms and in frequency-momentum space, we obtain
($\int_{\omega, \mathbf{q}} \equiv \int {\rm d} \omega\, {\rm d}^d \mathbf{q}/(
2 \pi)^{d + 1}$)
\begin{multline}
  \label{eq:44}
  \sdyn[\mathcal{T}_{\beta} \Phi] = \int_{\omega, \mathbf{q}} \omega \left[
    \cosh^2\bo \Phi_q^{\dagger}\oq \sigma_z \Phi_c\oq \right. \\ -
  \sinh^2\bo \Phi_c^{\dagger}\oq \sigma_z \Phi_q\oq + \sh \ch \\
  \left. \times \left( \Phi_c^{\dagger}\oq \sigma_z \Phi_c^{\dagger}\oq -
      \Phi_q^{\dagger}\oq \Phi_q\oq \right) \right].
\end{multline}
The combination
$\Phi_{\nu}^{\dagger}\oq \sigma_z \Phi_{\nu}\oq = \phi_{\nu}^{*}\oq
\phi_{\nu}\oq - \phi_{\nu}(-\omega,-\mathbf{q})
\phi_{\nu}^{*}(-\omega,-\mathbf{q})$
with $\nu = c, q$ is an odd function of $\oq$, whereas $\omega \sinh\bo \ch$ is
even, and therefore the integral over the product of these terms vanishes. Then,
with some simple algebraic manipulation, the first two terms in
Eq.~\eqref{eq:44} are recognised to be nothing but $\sdyn[\Phi]$, from which the
invariance of $\sdyn$ follows straightforwardly.
Note that this property holds independently of the value of the parameter
$\beta$ in the transformation $\mathcal{T}_{\beta}$.

\subsubsection{Hamiltonian contribution}
\label{sec:coherent-vertex}

We consider now the transformation of the Hamiltonian
contribution $S_{\mathcal{H}}$ in Eq.~\eqref{eq:8} under $\mathcal{T}_{\beta}$. 
First, we argue that the strictly local terms (i.e., those which do not
involve spatial derivatives) in the Hamiltonian~\eqref{eq:29} are invariant
under $\mathcal{T}_{\beta}$; then, we extend the argument to the case of
quasilocal terms such as the kinetic energy contribution
$\propto |\nabla \psi_\pm|^2$ or non-local interactions.
Consider a contribution to $S_{\mathcal{H}}$ of the form
\begin{equation}
  \label{eq:10}
  \mathcal{V}[\Psi] = \int_{t,\mathbf{x}}
  \left( v_{+}(t,\mathbf{x}) - v_{-}(t,\mathbf{x}) \right),
\end{equation}
where
$v_{\sigma}(t,\mathbf{x}) = \left( \psi_{\sigma}^{*}(t,\mathbf{x})
  \psi_{\sigma}(t,\mathbf{x}) \right)^N$
is a generic local contribution to the Hamiltonian
$\mathcal{H}_{\sigma}$ and $N$ is an integer.
In particular, for $N = 1$ we obtain the term proportional to the chemical
potential in Eq.~\eqref{eq:29}, while for $N = 2$, $\mathcal{V}[\Psi] $ is just
the contact interaction. Since $v_{\sigma}(t, \mathbf{x})$ is real, under the
transformation $\mathcal{T}_{\beta}$ [see Eq.~\eqref{eq:4}] only its time
argument is shifted according to
$\mathcal{T}_{\beta} v_{\sigma}(t,\mathbf{x}) = v_{\sigma}(-t + i \sigma
\beta/2,\mathbf{x})$
and, taking the Fourier transform with respect to time of this relation, one
eventually finds
\begin{equation}
  \label{eq:11}
  \mathcal{T}_{\beta} v_{\sigma}(\omega,\mathbf{x}) = e^{-\sigma \beta \omega/2} v_{\sigma}(-\omega,\mathbf{x}).
\end{equation}
Accordingly, the vertex~\eqref{eq:10} is invariant under $\mathcal{T}_{\beta}$:
in fact, being local in time, its diagrammatic representation --- where the
fields $\psi_{\sigma}(t, \mathbf{x})$ and $\psi_{\sigma}^{*}(t,\mathbf{x})$ are
represented by ingoing and outgoing lines, respectively --- satisfies frequency
conservation for in- and outgoing lines, as can be seen by taking the Fourier
transform of each of the fields in $v_{\sigma}(t,\mathbf{x})$ individually. In
particular, the frequency $\omega$ in Eq.~\eqref{eq:11} corresponds to the
difference between the sums of the in- and outgoing frequencies and only the
$\omega = 0$ component contributes to Eq.~\eqref{eq:10}.
(As stated above, we assume that the time integrals in Eqs.~\eqref{eq:7},
\eqref{eq:8}, and therefore~\eqref{eq:11} extend over all possible real values,
i.e., we focus on the stationary state of the dynamics.)
This component, however, is invariant under $\mathcal{T}_{\beta}$ as follows
directly from Eq.~\eqref{eq:11}, and hence the same is true for the vertex, for
which $\mathcal{V}[\mathcal{T}_{\beta} \Psi] = \mathcal{V} [\Psi]$.

Clearly, the invariance of the vertex and of the dynamical term in
Eq.~\eqref{eq:7} relies on the fact that vertices, which are local in time, obey
frequency conservation. (Note that, as in Sec.~\ref{sec:dynamical-term}, this
invariance holds independently of the value of the parameter $\beta$ in
$\mathcal{T}_{\beta}$.)
Accordingly, one concludes that any contribution to the Hamiltonian, which is
local in time and does not explicitly depend on time, is invariant.
In particular, the proof of invariance presented here for the vertex in
Eq.~\eqref{eq:10} can be straightforwardly extended to expressions containing
spatial derivatives such as the kinetic energy $\propto |\nabla\psi_\pm|^2$ in
Eq.~\eqref{eq:29} and even to interactions which are not local in space, as long
as they are local in time, as anticipated above.
Note, however, that these considerations do not rule out the possible emergence
upon renormalization or coarse-graining of terms which are non-local in time, as
long as they are invariant under $\mathcal{T}_{\beta}$. This case is discussed
further below in Sec.~\ref{sec:diss-contr-equil}.

\subsubsection{Enhanced symmetry for non-interacting systems}
\label{sec:enhanc-symm-simple}

The equilibrium transformation $\mathcal{T}_{\beta}$ presented in
Sec.~\ref{sec:symm-transf} involves a single parameter $\beta$. While this form
is appropriate for the Gibbs ensemble describing the thermal equilibrium state
of the interacting many-body system with the Hamiltonian in Eq.~\eqref{eq:29},
an enhanced version of the symmetry is realized in non-interacting
systems. Since these systems can be diagonalized in terms of single-particle
states, they are trivially integrable. Statistically, integrable systems are
described by a generalized Gibbs
ensemble~\cite{Rigol2007,Iucci2009,Jaynes1957,Barthel2008,Goldstein2014,Pozsgay2014,Mierzejewski2014,Wouters2014,Essler2014},
constructed from the extensive number of conserved quantities (with possible
exceptions, see, e.g.,
Refs.~\onlinecite{Goldstein2014,Pozsgay2014,Mierzejewski2014,Wouters2014}). In
the case of non-interacting systems which we consider here (or, more generally,
for any system that can be mapped to a non-interacting one), these integrals of
motion are just the occupation numbers of single-particle states.
Below we provide an example, in which the Lagrange multipliers associated with
these conserved occupations enter as parameters in a generalization of the
equilibrium transformation Eq.~\eqref{eq:1}: more specifically, these
multipliers play the role of effective inverse temperatures of the individual
single-particle states. On the other hand, in non-integrable cases, the
eigenstates of the Hamiltonian are not single-particle states. Then one
generically expects the stationary state of the system to be in thermal
equilibrium at a temperature $T = 1/\beta$, which is determined by the initial
conditions of the dynamics of the system. Accordingly, the enhanced symmetry
that is present in the stationary state of the non-interacting integrable system
breaks down and the corresponding Schwinger-Keldysh action is invariant under a
single $\mathcal{T}_{\beta}$, only for that specific value of $\beta$. This
shows that the transformation $\mathcal{T}_{\beta}$ can be generalized in order
to account for the appearance of a generalized Gibbs ensemble in the trivial
case of a system that can be diagonalized in terms of single-particle
states. However, the question whether the generalized Gibbs ensemble emerging in
the stationary states of generic integrable systems is characterized by a
symmetry involving the Lagrangian multipliers associated with the respective
integrals of motion as parameters, is beyond the scope of the present work.

As an example, let us consider bosons on a $d$-dimensional lattice with
nearest-neighbour hopping and on-site interaction (i.e., the Bose-Hubbard
model~\cite{Fisher1989}), with Hamiltonian
\begin{equation}
  \label{eq:12}
  \begin{split}
    H & = H_{\mathrm{kin}} + H_{\mathrm{int}}, \\ H_{\mathrm{kin}} & = - t
    \sum_{\langle \mathbf{l}, \mathbf{l}' \rangle} a_{\mathbf{l}}^{\dagger}
    a_{\mathbf{l}'}, \\ H_{\mathrm{int}} & = \frac{U}{2} \sum_{\mathbf{l}}
    a_{\mathbf{l}}^{\dagger} a_{\mathbf{l}} \left( a_{\mathbf{l}}^{\dagger}
      a_{\mathbf{l}} - 1 \right),
  \end{split}
\end{equation}
where $a_{\mathbf{l}}$ is the annihilation operator for bosons on the lattice
site $\mathbf{l}$, $t$ is the hopping matrix element between site $\mathbf{l}$
and its nearest-neighbours $\mathbf{l}'$, while $U$ determines the strength
of on-site interactions.
We first consider the case $U = 0$, which is trivially integrable: the kinetic
energy contribution to the Hamiltonian is diagonal in momentum space and the
corresponding single-particle eigenstates are the Bloch states. These are
labelled by a quasi-momentum $\mathbf{q}$, and in terms of creation and
annihilation operators for particles in Bloch states, $a^{\dagger}_{\mathbf{q}}$
and $a_{\mathbf{q}}$ respectively, the kinetic energy can be written as
\begin{equation}
  \label{eq:14}
  H_{\mathrm{kin}} = \sum_{\mathbf{q}} \epsilon_{\mathbf{q}}
  a_{\mathbf{q}}^{\dagger} a_{\mathbf{q}}.
\end{equation}
Let us now consider a Schwinger-Keldysh functional integral description of the
stationary state of the system. Then, the kinetic energy in Eq.~\eqref{eq:14}
yields a contribution to the corresponding action which reads
\begin{equation}
  \label{eq:13}
  S_{\mathcal{H},\mathrm{kin}} = - \int_t \sum_{\mathbf{q}} \epsilon_{\mathbf{q}} \left(
    \psi_{\mathbf{q}, +}^{*} \psi_{\mathbf{q}, +} - \psi_{\mathbf{q}, -}^{*} \psi_{\mathbf{q}, -} \right),
\end{equation}
where $\psi_{\mathbf{q}, +}$ and $\psi_{\mathbf{q}, -}$ are the fields on the
forward and backward branches of the closed time path respectively, expressed in
the basis of Bloch states.  $S_{\mathcal{H}, \mathrm{kin}}$ is invariant under
the transformation of the fields
\begin{equation}
  \label{eq:15}
  \mathcal{T}_{\beta_{\mathbf{q}}} \Psi_{\mathbf{q}, \sigma}(\omega) =
  e^{-\sigma \beta_{\mathbf{q}} \omega/2} \Psi_{-\mathbf{q}, \sigma}^{*}(\omega),
\end{equation}
where, as in Eq.~\eqref{eq:1}, we arrange the fields in a spinor
$\Psi_{\mathbf{q}, \sigma}(\omega) = \left( \psi_{\mathbf{q}, \sigma}(\omega),
  \psi_{-\mathbf{q}, \sigma}^{*}(-\omega) \right)^T$.
The crucial point is that $\beta_{\mathbf{q}}$ can be chosen to depend on the
quasi-momentum $\mathbf{q}$, indicating that to each eigenstate of the system we
can assign an individual ``temperature'' $T_{\mathbf{q}} = 1/\beta_{\mathbf{q}}$
such that the corresponding mean occupation number
$n_{\mathbf{q}} = \langle a_{\mathbf{q}}^\dagger a_{\mathbf{q}}\rangle$ is
determined by a Bose distribution with precisely this
``temperature.''

Let us now consider the opposite limit in which the hopping amplitude $t$
vanishes while the interaction strength $U$ is finite. The interaction energy
$H_{\rm int}$ in Eq.~\eqref{eq:12} is diagonal in the basis of Wannier states
localized at specific lattice sites and the occupation numbers
$\hat n_{\mathbf{l}} = a_{\mathbf{l}}^\dagger a_{\mathbf{l}}$ of these sites are
conserved, rendering the system integrable. The generalized symmetry
transformation appropriate for this case can be obtained from Eq.~\eqref{eq:15}
by replacing the quasi-momentum $\mathbf{q}$ by the lattice site index
$\mathbf{l}$ and by introducing a set of ``local (inverse) temperatures''
$T_{\mathbf{l}}$ ($\beta_{\mathbf{l}}$) instead of $T_{\mathbf{q}}$
($\beta_{\mathbf{q}}$).

In the generic case, when both the hopping $t$ and the interaction $U$ are
non-zero, the system is not integrable. Then, neither the generalized
transformation Eq.~\eqref{eq:15} nor its variant with local ``temperatures'' are
symmetries of the corresponding Schwinger-Keldysh action, showing that this case
eventually admits only one single global temperature, which determines 
the statistical weight of individual many-body eigenstates of the system.

\subsection{Dissipative contributions in equilibrium}
\label{sec:diss-contr-equil}

The functional integral with the action $S$ in Eq.~\eqref{eq:6}, as it stands,
is not convergent but it can be made so by adding to $S$ an infinitesimally
small imaginary (i.e., dissipative)
contribution~\cite{Kamenev2011,Altland/Simons}.
Within a renormalization-group picture, this \emph{infinitesimal} dissipation
may be seen as the ``initial value,'' at a microscopic scale, of \emph{finite}
dissipative contributions, which are eventually obtained upon coarse graining
the original action $S$ and which result in, e.g., finite lifetimes of
excitations of the effective low-energy degrees of freedom.
The precise form of the corresponding effective low-energy action and, in
particular, of the dissipative contributions which appear therein, is strongly
constrained by the requirement of invariance under $\mathcal{T}_{\beta}$ of the
starting action at the microscopic scale: in fact, terms which violate this
symmetry will not be generated upon coarse-graining.
In the discussion below we identify those dissipative contributions to the
Schwinger-Keldysh action which are invariant under $\mathcal{T}_{\beta}$.  This
allows us to anticipate the structure of any low-energy effective action
possessing this a symmetry.
Note, however, that \emph{finite} dissipative terms may appear even at the
microscopic scale because of, e.g., the coupling of the system to an external
bath.
Below we consider two instances of this case: in
Sec.~\ref{sec:non-equil-nature} we show that $\mathcal{T}_{\beta}$ cannot be a symmetry 
of the action if the system is coupled to Markovian baths and driven --- a situation
described by a quantum master equation. 
Another specific example, in which the equilibrium symmetry is realized, 
is the particle number non-conserving coupling of the Schwinger-Keldysh action
Eq.~\eqref{eq:6} to an ohmic bath. This situation, which we discuss in
Sec.~\ref{sec:classical-limit}, is of particular interest, because its classical
limit renders what is known as the dynamical
model A~\cite{Hohenberg1977} with reversible mode couplings (termed model
$\mbox{A}^*$ in Ref.~\onlinecite{Folk2006}); this correspondence allows us to
establish a connection with the known equilibrium symmetry of the generating
functional associated with this classical stochastic dynamics.

Below we discuss dissipative terms of the action invariant under
$\mathcal{T}_{\beta}$, which involve first single particles (being quadratic in
the fields of the Schwinger-Keldysh action) in Sec.~\ref{sec:regularization},
and then their interactions in Sec.~\ref{sec:dissipative-vertices}.

\subsubsection{Single-particle sector}
\label{sec:regularization}

Dissipative contributions to the single-particle sector of the Schwinger-Keldysh
action which are invariant under $\mathcal{T}_{\beta}$ take the form
\begin{multline}
  \label{eq:17}
  S_d = i \int_{\omega, \mathbf{q}} h\oq \left( \phi_q^{*}\oq \phi_c\oq - \phi_q\oq
    \phi_c^{*}\oq \right. \\ \left. + 2 \ct \phi_q^{*}\oq \phi_q\oq \right),
\end{multline}
with an arbitrary real function $h\oq$ which transforms under time reversal as
$\tilde{h}\oq = h (\omega, -\mathbf{q})$.
When such dissipative terms are introduced in order to regularize the
Schwinger-Keldysh functional integral, a typical choice for $h\oq$ is
$h\oq = \epsilon$~\cite{Kamenev2011,Altland/Simons} with $\epsilon \to 0$. This
ensures that the Green's functions in the absence of interactions satisfy a
fluctuation-dissipation relation (we postpone the detailed discussion of such
relations to Sec.~\ref{sec:fluct-diss-theor}). The FDR for non-interacting
Green's functions, together with the invariance of interactions under the
transformation $\mathcal{T}_{\beta}$ shown in Sec.~\ref{sec:coherent-vertex},
guarantees that the FDR is satisfied to all orders in perturbation
theory~\cite{Jakobs2010}.

While there are no restrictions on the form of the function $h\oq$, the
hyperbolic cotangent $\ct$ appearing in the last term of $S_d$ is
\emph{uniquely} fixed by the requirement of invariance under
$\mathcal{T}_{\beta}$, as can be verified by following the line of argument
presented in Appendix~\ref{sec:invariance-s_d}. In particular, $S_d$ with a
certain value of $\beta$ in the argument of $\ct$ is invariant under
$\mathcal{T}_{\beta'}$ if and only if $\beta' = \beta$.
This shows that, remarkably, the appearance of the thermodynamic equilibrium
Bose distribution function $n(\omega) = 1/(e^{\beta \omega} - 1)$ at a
temperature $T = 1/\beta$ in $\ct = 2 n (\omega) + 1$, can be traced back to the
fact that $\mathcal{T}_{\beta}$ is a symmetry of the action.
Note that for simplicity we considered here only the case of vanishing chemical
potential, $\mu=0$. For finite $\mu$, the frequency $\omega$ in the argument of
the hyperbolic cotangent in Eq.~\eqref{eq:17} should be shifted according to
$\omega \to \omega - \mu$, as we discussed in point~\ref{item:1} in
Sec.~\ref{sec:symm-transf}.

\subsubsection{Dissipative vertices}
\label{sec:dissipative-vertices}

The dissipative contributions discussed in the previous section are quadratic in
the field operators and they naturally occur, e.g., when the system is coupled
to a thermal bath by means of an interaction which is linear in those fields.
However, this type of coupling necessarily breaks particle number conservation.
The number of particles is conserved if instead the system-bath interaction term
commutes with the total number of particles of the system,
$N = \int_{\mathbf{x}} n(\mathbf{x})$, where
$n(\mathbf{x}) = \psi^{\dagger}(\mathbf{x}) \psi(\mathbf{x})$ is the local
density.
In other words, to ensure particle number conservation, it is necessary that the
coupling terms are at least quadratic in the system operators.
Accordingly, dissipative \emph{vertices} appear in the Schwinger-Keldysh action
after integrating out the bath degrees of freedom. Then, the requirement of
invariance of these terms under $\mathcal{T}_{\beta}$ allows us to infer
\textit{a priori} their possible structure.
In particular, we find that a frequency-independent number-conserving quartic
vertex (i.e., the dissipative counterpart to the two-body interaction
$\propto \abs{\psi_{\sigma}}^4$ in the Hamiltonian~\eqref{eq:29}) is forbidden
by the thermal symmetry.

A generic quartic vertex, which conserves the number of particles and which is
local in time, can be parameterized as
\begin{multline}
  \label{eq:39}  
  S_d = - i \int_{\omega_1, \ldots ,\omega_4} \!\!\!\!\!\!\!\!\ \delta(\omega_1 - \omega_2 + \omega_3
  - \omega_4) \\ \times
  \begin{aligned}[t]
    \big[ & f_1(\omega_1,\omega_2,\omega_3,\omega_4) 
    \psi_{+}^*(\omega_1) \psi_{+}(\omega_2) \psi_{+}^*(\omega_3)
    \psi_{+}(\omega_4) \\
    + & f_2(\omega_1,\omega_2,\omega_3,\omega_4) 
  \psi_{-}^*(\omega_1) \psi_{-}(\omega_2) \psi_{-}^*(\omega_3)
  \psi_{-}(\omega_4)  \\
  + & f_3(\omega_1,\omega_2,\omega_3,\omega_4) 
    \psi_{+}^*(\omega_1) \psi_{+}(\omega_2) \psi_{-}^*(\omega_3)
    \psi_{-}(\omega_4)\big],
  \end{aligned}
\end{multline}
where $f_{1,2,3}$ are real functions; in order to simplify the notation we do
not indicate the (local) spatial dependence of the fields, which is understood
together with the corresponding integration in space.
Conservation of particle number is ensured by the $U(1)$ invariance
$\psi_\pm \mapsto e^{i\alpha_\pm} \psi_\pm$ on each contour separately, with
generic phases $\alpha_\pm$, while the overall $\delta$-function on the
frequencies guarantees locality in time. Restrictions on the functions
$f_{1,2,3}$ in the generic dissipative vertex in Eq.~\eqref{eq:39} follow from
the requirements of causality~\cite{Kamenev2011}, according to which $S_d$ to
the action must vanish for $\psi_{+} = \psi_{-}$, and invariance of the
dissipative vertex under the equilibrium transformation. These conditions are
studied in detail in Appendix~\ref{sec:invar-diss-vert}. In particular, we find
that they cannot be satisfied if $f_{1,2,3}$ are constant, i.e., do not depend
on the frequencies. One particular choice of these functions that is compatible
with the constraints is given by
\begin{equation}
  \label{eq:42}
  \begin{split}
    f_1(\omega_1, \omega_2, \omega_3, \omega_4) & = f_2(\omega_1, \omega_2,
    \omega_3, \omega_4) \\ & = \left( \omega_1 - \omega_2 \right) \coth \!
    \left( \beta \left(
        \omega_1 - \omega_2 \right)/2 \right), \\
    f_3(\omega_1, \omega_2, \omega_3, \omega_4) & = - 4 \left( \omega_1 -
      \omega_2 \right) \left( n(\omega_1 - \omega_2) + 1 \right),
  \end{split}
\end{equation}
with the Bose distribution function $n(\omega)$. It is interesting to note that,
in the basis of classical and quantum fields, this corresponds to a
generalization of Eq.~\eqref{eq:17} with $h(\omega, \mathbf{q}) = \omega$, in
which the fields are replaced by the respective densities defined as
$\rho_c = \left( \psi_{+}^{*} \psi_{+} + \psi_{-}^{*} \psi_{-} \right)/\sqrt{2}$
and
$\rho_q = \left( \psi_{+}^{*} \psi_{+} - \psi_{-}^{*} \psi_{-}
\right)/\sqrt{2}$.
Another notable property of this solution is that for $\omega_{1,2} \to 0$ we
have $f_{1, 2}(\omega_1, \omega_2, \omega_3, \omega_4) \to 2 T$ and
$f_3(\omega_1, \omega_2, \omega_3, \omega_4) \to - 4T$, i.e., these limits of
vanishing frequencies are finite. This implies that the form of $S_d$ with
$f_{1,2,3}$ given by Eq.~\eqref{eq:42} is to some extent universal: indeed, it
should be expected to give the leading dissipative contribution to the
Schwinger-Keldysh action of any number-conserving system in the low-frequency
limit. At higher frequencies, other less universal solutions might also be
important and one cannot make a general statement.

\subsection{Classical limit, detailed balance and microreversibility}
\label{sec:classical-limit}

A transformation analogous to $\mathcal{T}_{\beta}$ --- which becomes a symmetry
in equilibrium --- was previously derived for the stochastic evolution of
classical statistical systems in contact with an environment, within the
response functional formalism~\cite{MSR1973,Janssen1976,Bausch1976,Janssen1979,
  Janssen1992, DeDominicis1976,DeDominicis1978,Tauber2014}.
This formalism allows one to determine expectation values of relevant quantities
as a functional integral with a certain action known as response functional,
which can also be derived from a suitable classical limit of the
Schwinger-Keldysh action for quantum
systems~\cite{Kamenev2011,Altland/Simons}. In these classical systems, the
environment acts effectively as a source of stochastic noise over which the
expectation values are taken.

Here, we show that the classical limit of
$\mathcal{T}_{\beta}$~\cite{Altland2010} yields exactly the transformation which
becomes a symmetry when the classical system is at
equilibrium~\cite{aron10:_symmet_langev}.
In order to consider this limit within the Schwinger-Keldysh formalism, it is
convenient to express the Schwinger-Keldysh action in Eq.~\eqref{eq:6} in terms
of the classical and quantum fields $\phi_c$ and $\phi_q$, respectively, defined
in Eq.~\eqref{eq:9}, and to reinstate Planck's constant according
to~\cite{Kamenev2011,Altland/Simons}
\begin{equation}
  S \to S/\hbar, \quad \ct
  \to \coth(\beta \hbar \omega/2), \quad \phi_q \to \hbar \phi_q.
\end{equation}
Then, the action can be formally expanded in powers of $\hbar$ in order to take
the classical limit $\hbar \to 0$, and the classical part of the
Schwinger-Keldysh action is given by the contribution which remains for
$\hbar =0$.
Note that the limit $\hbar \to 0$ considered here is formally equivalent to
approaching criticality in equilibrium at finite temperature $T=\beta^{-1}$, for
which $\beta \Delta \to 0$, where $\Delta$ is the energy gap, which can be read
off from the retarded Green's function (see, e.g.,
Ref.~\onlinecite{Sieberer2014}). This equivalence conforms with the expectation
that quantum fluctuations generically play only a subdominant role in
determining the critical behavior of quantum systems at finite temperature.
In order to see the emergence of a stochastic dynamics driven by incoherent
(thermal) noise from a quantum coherent dynamics, we supplement the
Schwinger-Keldysh action in Eq.~\eqref{eq:6} (describing the latter) with
dissipative terms arising from its coupling to a bath. For simplicity, we assume
this bath to be characterized by an ohmic spectral density, while the system is
assumed to have the Hamiltonian in Eq.~\eqref{eq:29}.
Deferring to Sec.~\ref{sec:non-equil-nature} the discussion of the theoretical
description of such a system-bath coupling, we anticipate here that
the resulting contribution to the Schwinger-Keldysh action can be written as in
Eq.~\eqref{eq:19}, under the assumption that $\gamma(\omega) \nu(\omega)$ is
linear in the frequency, i.e., $\gamma(\omega) \nu(\omega) = 2 \kappa \omega$
and by choosing $L_{\sigma}(\omega)\to\psi_{\sigma}\oq$, with the thermal bath
acting independently on each momentum mode~\cite{Altland/Simons}.
Then, in the classical limit $\hbar \to 0$, we find
\begin{multline}
  \label{eq:20}
  S = \int_{t, \mathbf{x}} \Phi_q^{\dagger} \left\{ \left[ \left( \sigma_z + i
        \kappa \id \right) i \partial_t + \frac{ \nabla^2}{2 m} \right] \Phi_c +
    i 2 \kappa T \Phi_q \right\} \\ - \lambda \int_{t, \mathbf{x}} \left(
    \phi_c^{* 2} \phi_c \phi_q + \mathrm{c.c.}  \right).
\end{multline}
This action has the form of the response functional of the equilibrium dynamical
models considered in Ref.~\onlinecite{Hohenberg1977}: it includes both a linear
and a quadratic contribution in the quantum field $\phi_q$, but no higher-order
terms. After having transformed the quadratic term into a linear one via the
introduction of an auxiliary field (which is eventually interpreted as a
Gaussian additive noise), this quantum field can be integrated out and one is
left with an effective constraint on the dynamics of the classical field
$\phi_c$, which takes the form of a Langevin equation; here:
\begin{equation}
 \left( i - \kappa \right) \partial_t \phi_c = \left(-\frac{\nabla^2}{2 m}  + \lambda  |\phi_c|^2 \right) \phi_c + \eta,
\label{eq:stoch}
\end{equation}
where $\eta= \eta(t, \xx)$ is a (complex) Gaussian stochastic noise with zero
mean $\langle \eta (t, \xx)\rangle = 0$ and correlations
\begin{gather}
  \langle \eta(t, \xx) \eta^*(t', \xx')\rangle = \kappa T
  \delta(t-t')\delta^{(d)}(\xx-\xx'), \\ \langle \eta(t, \xx)\eta(t',
  \xx')\rangle = 0.
\end{gather}
Equation~\eqref{eq:stoch} describes the dynamics of the non-conserved
(complex scalar) field $\phi_c$ without additional conserved densities, 
which is known in the literature as \emph{model A}~\cite{Hohenberg1977}.
However, as can be seen from the complex prefactor $i - \kappa$ of the time
derivative on the left-hand side of Eq.~\eqref{eq:stoch}, the dynamics is not
purely relaxational as in model A, but it has additional coherent contributions,
also known as \emph{reversible mode couplings}~\cite{Tauber2014}.
The fact that the simultaneous appearance of dissipative and coherent dynamics
can be described by a complex prefactor of the time derivative is specific to
thermal equilibrium: in fact, dividing Eq.~\eqref{eq:stoch} by $i - \kappa$, the
reversible and irreversible parts of the resulting Langevin dynamics are not
independent of each other and in fact their coupling constants share a common
ratio~\cite{Janssen1979,Janssen1992,GardinerBook}. Under more general
non-equilibrium conditions, however, these reversible and irreversible
generators of the dynamics have different microscopic origins and no common
ratio generically exists.
In the present equilibrium context, however, the action Eq.~\eqref{eq:20}
corresponds to \emph{model A$^*$} in the notion of Ref.~\onlinecite{Folk2006},
and the form of the classical transformation appropriate for this case which
becomes a symmetry in equilibrium was given in Ref.~\onlinecite{Sieberer2014}.
This transformation emerges as the classical limit of $\mathcal{T}_{\beta}$
discussed in the previous sections~\cite{Altland2010}.
In fact, for $\beta = T^{-1}\to 0$ and neglecting the contribution of the
quantum fields in the transformation of the classical fields (i.e., at the
leading order in $\hbar$), Eq.~\eqref{eq:3} becomes
\begin{equation}
  \label{eq:21}
  \begin{split}
    \mathcal{T}_{\beta} \Phi_c(t,\mathbf{x}) & = \sigma_x \Phi_c(-t,\mathbf{x}), \\
    \mathcal{T}_{\beta} \Phi_q(t,\mathbf{x}) & = \sigma_x \left(
      \Phi_q(-t,\mathbf{x}) + \frac{i}{2 T} \partial_t \Phi_c(-t,\mathbf{x})
    \right),
  \end{split}
\end{equation}
after a transformation back to
the time and space domains.
Upon identifying the classical field $\Phi_c$ with the physical field and
$\Phi_q$ with the response field $\tilde{\Phi}$, according to
$\Phi_q = i\tilde{\Phi}$, Eq.~\eqref{eq:21} takes the form of the classical
symmetry introduced in Ref.~\onlinecite{aron10:_symmet_langev}. Note, however,
that the transformation \eqref{eq:21} is not the only form in which the
equilibrium symmetry in the classical context can be expressed.
In fact, the transformation of the response field $\tilde{\Phi}$ can also be
expressed~\cite{Janssen1979,Janssen1992} in terms of a functional derivative of
the equilibrium distribution rather than of the time derivative of the classical
field $\partial_t\Phi_c$ as in Eq.~\eqref{eq:21}.
The existence of these different but equivalent transformations might be related
to the freedom in the definition of the response field, which is introduced in
the theory as an auxiliary variable in order to enforce the dynamical constraint
represented by the Langevin
equation~\cite{Zinn-Justin,Janssen1979,Janssen1992,Tauber2014} such as
Eq.~\eqref{eq:stoch}. This implies~\cite{Zinn-Justin} that the related action
acquires the so-called Slavnov-Taylor symmetry. As far as we know, the
consequences of this symmetry have not been thoroughly investigated in the
classical case and its role for quantum dynamics surely represents an intriguing
issue for future studies.

We emphasize the fact that the derivation of the symmetry in the classical case
involves explicitly the equilibrium probability
density~\cite{Janssen1979,Janssen1992}.
Indeed, the response functional contains an additional contribution from the
probability distribution of the value of the fields at the initial time, after
which the dynamics is considered. This term generically breaks the
time-translational invariance of the theory~\cite{Janssen1979,Janssen1992},
unless the initial probability distribution is the equilibrium one. Accordingly,
when the classical equilibrium symmetry $\mathcal{T}_{\beta}$ is derived under
the assumption of time-translational invariance, its expression involves also
the equilibrium distribution.
In the quantum case discussed in the previous sections, instead,
time-translational symmetry was implicitly imposed by extending the time
integration in the action from $-\infty$ to $+\infty$, which is equivalent to
the explicit inclusion of the initial condition (in the form of an initial
density matrix) and makes the analysis simpler, though with a less transparent
interpretation from the physical standpoint.

Although in classical systems this equilibrium symmetry takes (at least) two
different but equivalent forms due to the arbitrariness in the definition of the
response functional mentioned above, it can always be traced back to the
condition of detailed balance~\cite{Janssen1979,Janssen1992,
  aron10:_symmet_langev}. Within this context, detailed balance is defined by
the requirement that the probability of observing a certain (stochastic)
realization of the dynamics of the system equals the probability of observing
the time-reversed realization, and therefore it encodes the notion of
\emph{microreversibility}.
This condition guarantees the existence and validity of fluctuation-dissipation
relations, which can be proved on the basis of this symmetry. In addition,
detailed balance constrains the form that the response functional can take as
well as the one of the equilibrium probability distribution for this stochastic
process.

The situation in the quantum case appears to be significantly less clear. In
fact, a precise and shared notion of \emph{quantum detailed balance} and
\emph{quantum microreversibility} is seemingly lacking. The first attempt to
introduce a principle of quantum detailed balance dates back to
Ref.~\onlinecite{Agarwal1973}, where it was derived from a condition of
microreversibility in the context of Markovian quantum dynamics described by a
Lindblad master equation. The mathematical properties of these conditions were
subsequently studied in detail (see, e.g.,
Refs.~\onlinecite{Alicki1976,Kossakowski1977,Frigerio1984,Majewski1984, Alicki2007})
and were shown to constrain the form of the Lindblad super-operator in order for
it to admit a Gibbs-like stationary density matrix. However, even when this
occurs, these operators are not able to reproduce the KMS condition and the
fluctuation-dissipation relations because of the underlying Markovian
approximation, as we discuss in Sec.~\ref{sec:non-equil-nature}.

The notion of microreversibility in quantum systems appears to have received
even less attention, as well as its connection with some sort of reversibility
expressed in terms of the probability of observing certain ``trajectories'' and
their time-reversed ones.
The definition proposed in Ref.~\onlinecite{Agarwal1973} (also discussed in
Ref.~\onlinecite{Carmichael1976}) appears to be a natural generalization of the
notion in the classical case, as it relates the correlation of two operators
evaluated at two different times with the correlation of the time-reversed ones.
However, to our knowledge, the relationship between this condition and
thermodynamic equilibrium has never been fully elucidated. Although addressing
these issues goes well beyond the scope of the present paper, they surely
represent an interesting subject for future investigations.

\section{Equivalence between the symmetry and the KMS condition}
\label{sec:equiv-symm-kms}

In this section we show that the invariance of the Schwinger-Keldysh action of a
certain system under $\mathcal{T}_{\beta}$ (as specified in
Sec.~\ref{sec:invar-keldysh-acti}) is equivalent to having multi-time
correlation functions of the relevant fields which satisfy the KMS
condition~\cite{Kubo1957,Martin1959}.
As the latter can be considered as the defining property of thermodynamic
equilibrium, this shows that the same applies to the invariance under the
equilibrium symmetry.

The KMS condition involves \emph{both} the Hamiltonian generator of dynamics
\emph{and} the thermal nature of the density matrix which describes the
stationary state of the system: heuristically this condition amounts to
requiring that the many-body Hamiltonian which determines the (canonical)
population of the various energy levels is the same as the one which rules the
dynamics of the system. The equivalence proved here allows us to think of the
problem from a different perspective: taking the invariance under
$\mathcal{T}_{\beta}$ as the fundamental property and observing that any
time-independent Hamiltonian respects it, we may require it to hold at any
scale, beyond the microscopic one governed by reversible Hamiltonian dynamics
alone.
In particular, upon coarse graining within a renormalization-group framework, only
irreversible dissipative terms which comply with the symmetry (such as those
discussed in Sec.~\ref{sec:diss-contr-equil}) can be generated in stationary
state and the hierarchy of correlation functions respect thermal
fluctuation-dissipation relations.
The validity of KMS conditions (and therefore of the symmetry
$\mathcal{T}_{\beta}$) hinges on the \emph{whole} system being prepared in a
canonical density matrix $\rho$.  Accordingly, if the system is described by a
microcanonical ensemble, the KMS condition holds only in a subsystem of it,
which is expected to be described by a canonical reduced density matrix.
Equivalently, this means that, in a microcanonical ensemble, only suitable local
observables satisfy this condition. In the case of quantum many-body systems
evolving from a pure state, an additional restriction on the class of
observables emerges due to the fact that, if thermalization occurs as
conjectured by the eigenstate thermalization hypothesis
(ETH)~\cite{Deutsch1991,Srednicki1994,Rigol2007}, the microcanonical ensemble is
appropriate only if the observable involves the creation and annihilation of a
small number of particles (low order correlation functions). This was shown to
be also the case for FDRs~\cite{Srednicki1999,Khatami2013}: however, as pointed
out above, the thermal symmetry implies the validity of FDRs involving an
arbitrary number of particles, which leads to the conclusion that it does not
apply to an isolated system thermalizing via the ETH. In other words, the
thermal symmetry implies that the whole density matrix takes the form of a Gibbs
ensemble, while in thermalization according to the ETH, only finite subsystems
are thermalized by the coupling to the remainder of the system, which acts as a
bath.
Thus we see how Hamiltonian dynamics favors thermal stationary states (with
density matrix $\rho$ proportional to $e^{-\beta H}$) over arbitrary functionals
$\rho = \rho(H)$. One explicit technical advantage of this perspective based on
symmetry is that it allows us to utilize the toolbox of quantum field theory
straightforwardly and to study the implications of $\mathcal{T}_{\beta}$ being a
symmetry; this is exemplified here by considering the associated Ward-Takahashi
identities and by showing the absence of this symmetry in dynamics described by
Markovian quantum master equations in Sec.~\ref{sec:non-equil-nature}.
We also note that the presence of this symmetry provides a criterion for
assessing the equilibrium nature of a certain dynamics by inspecting only the
dynamic action functional, instead of the whole hierarchy of
fluctuation-dissipation relations. In addition, this symmetry may be present in
the actions of open systems with both reversible and dissipative terms.

In the following, we consider a quantum system with unitary dynamics generated
by the (time-independent) Hamiltonian $H$, which is in thermal equilibrium at
temperature $T = \beta^{-1}$ and therefore has a density matrix
$\rho = e^{-\beta H}/\tr e^{-\beta H}$.  The KMS condition relies on the
observation that for an operator in the Heisenberg representation
$A(t) = e^{i H t} A e^{-i H t}$, one has
\begin{equation}
A(t) \rho = \rho A(t - i \beta)
\label{eq:KMS0}
\end{equation}
(for simplicity we do not include here a chemical potential, but at the end of
the discussion we indicate how to account for it).
This identity effectively corresponds, up to a translation of the time by an
imaginary amount, to exchanging the order of the density matrix and of the
operator $A$ and therefore, when Eq.~\eqref{eq:KMS0} is applied to a multi-time
correlation function, it inverts the time order of the involved times, which can be
subsequently restored by means of the quantum-mechanical time-reversal
operation.
Hence, the quantum-mechanical time reversal naturally appears as an element of the
equilibrium symmetry $\mathcal{T}_{\beta}$, while external fields have to be
transformed accordingly, as indicated in Eq.~\eqref{eq:5}.
The application of time reversal yields a representation of the KMS condition
which can be readily translated into the Schwinger-Keldysh formalism, as was
noted in Refs.~\onlinecite{Chou1985,Wang2002,Jakobs2010}. In particular, it
results in an infinite hierarchy of generalized multi-time quantum
fluctuations-dissipation relations which include the usual FDR for two-time
correlation and response functions of the bosonic fields as a special case (see
Ref.~\onlinecite{Jakobs2010} and Sec.~\ref{sec:fluct-diss-theor}).
One of the main points of this work is that these FDRs can also be regarded as
the Ward-Takahashi identities associated with the invariance of the
Schwinger-Keldysh action $S$ under the discrete symmetry~\footnote{Here we used
  the notion of ``Ward-Takahashi identity'' in the slightly generalized sense
  which encompasses the case of identities between correlation functions
  resulting from discrete symmetries (such as Eq.~\eqref{eq:1} in
  Sec.~\ref{sec:from-kms-symmetry}, which leads to, c.f., Eqs.~\eqref{eq:52} and
  \eqref{eq:53}) beyond the usual case of continuous
  symmetries~\cite{Zinn-Justin}.}  $\mathcal{T}_{\beta}$ and that, conversely,
the full hierarchy of FDRs implies the invariance of $S$ under
$\mathcal{T}_{\beta}$.

The argument outlined below, which shows the equivalence between the KMS
condition and the thermal symmetry, involves several steps: as a preliminary we
review in Secs.~\ref{sec:multi-time-corr} and~\ref{sec:time-reversal} how
time-ordered and anti-time-ordered correlation functions can be expressed using
the Schwinger-Keldysh technique and we specify how these correlation functions
transform under quantum mechanical time reversal. We apply these results to the
KMS condition in Sec.~\ref{sec:kms-condition}: first we discuss its
generalization to multi-time correlation functions and then we translate such a
generalization into the Schwinger-Keldysh formalism.
This part proceeds mainly along the lines of Ref.~\onlinecite{Jakobs2010}, with
some technical differences. Finally, we establish the equivalence between the
resulting hierarchy of FDRs and the thermal symmetry at the end of
Sec.~\ref{sec:kms-condition}.

\subsection{Multi-time correlation functions in the Schwinger-Keldysh formalism}
\label{sec:multi-time-corr}

\paragraph{Two-time correlation functions.}

Let us first consider a two-time correlation function
\begin{equation}
  \label{eq:16}
  \langle A(t_A) B(t_B) \rangle \equiv \tr \left( A(t_A) B(t_B) \rho \right)
\end{equation}
between two generic operators $A$ and $B$ (in the following we are particularly
interested in considering the case in which $A$ and $B$ are the field operators
$\psi(\mathbf{x})$ or $\psi^{\dagger}(\mathbf{x})$ at positions
$\mathbf{x} = \mathbf{x}_A$ and $\mathbf{x} = \mathbf{x}_B$) evaluated at
different times $t_A$ and $t_B$, respectively, in a quantum state described by
the density matrix $\rho$. We assume that the dynamics of the system is unitary
and generated by the Hamiltonian $H$. Then, the Heisenberg operator $A$ at time
$t_{A}$ is related to the Schr\"odinger operator at a certain initial time
$t_i< t_A$ via
\begin{equation}
  \label{eq:24}
  A(t_A) = e^{i H \left( t_A - t_i \right)} A e^{-i H \left( t_A - t_i \right)},
\end{equation}
with an analogous relation for $B$. 

The two-time correlation function can be represented within the
Schwinger-Keldysh formalism as (see Appendix~\ref{sec:repr-corr-funct})
\begin{equation}
  \label{eq:22}
  \begin{split}
    \langle A(t_A) B(t_B) \rangle 
   & = \langle A_{-}(t_A) B_{+}(t_B)  \rangle \\ & \equiv \int \mathcal{D}[\Psi]
   A_{-}(t_A) B_{+}(t_B) e^{i S[\Psi]},
  \end{split}
\end{equation}
irrespective of the relative order of the times $t_A$ and $t_B$. Here, the
functional integral is taken over the fields
$\Psi = \left( \psi_{+}, \psi_{+}^{*}, \psi_{-}, \psi_{-}^{*} \right)^T$, and
the exponential weight with which a specific field configuration contributes to
the integral is determined by the Schwinger-Keldysh action $S[\Psi]$.
In the following, by $O_{+/-}$ we indicate that a certain operator $O$ has been
evaluated in terms of the fields $\psi_\pm$ defined on the forward/backward
branch of the temporal contour associated with the Schwinger-Keldysh formalism
(see, e.g., Refs.~\onlinecite{Kamenev2011,Altland/Simons}).

\paragraph{Multi-time correlation functions.}

We define multi-time correlation functions in terms of time-ordered and
anti-time-ordered products of operators
\begin{equation}  
  \label{eq:26}
  \begin{split}
    A(t_{A,1}, \dotsc, t_{A,N}) & = a_1(t_{A,1}) a_2(t_{A,2}) \dotsm
    a_N(t_{A,N}), \\ B(t_{B,1}, \dotsc, t_{B,M}) & = b_M(t_{B,M})
    b_{M-1}(t_{B,M-1}) \dotsm b_1(t_{B,1}),
  \end{split}
\end{equation}
for $t_i < t_{A,1} < \dotsb < t_{A,N} < t_f$ and
$t_i < t_{B,1} < \dotsb < t_{B,M} < t_f$, where $t_f$ is an arbitrarily chosen
largest time. Here, $\{a_n,b_m\}_{n, m}$ are bosonic field operators. The
specific sequence of time arguments in $A$ and $B$ (increasing and decreasing
from left to right, respectively) leads to a time-ordering on the
Schwinger-Keldysh contour: indeed, as we show in
Appendix~\ref{sec:repr-corr-funct}, the multi-time correlation function can be
expressed as a Schwinger-Keldysh functional integral in the form
\begin{multline}
  \label{eq:27}
  \langle A(t_{A,1}, \dotsc, t_{A,N}) B(t_{B,1}, \dotsc, t_{B,M}) \rangle \\ =
  \langle B_{+}(t_{B,1}, \dotsc, t_{B,M}) A_{-}(t_{A,1}, \dotsc, t_{A,N}) \rangle.
\end{multline}

\paragraph{Anti-time-ordered correlation functions.}

Not only time-ordered correlation functions such as Eq.~\eqref{eq:27} can be
expressed in terms of functional integrals, but also correlation functions which
are anti-time-ordered and which, e.g., are obtained by exchanging the positions
of $A(t_{A,1}, \dotsc, t_{A,N})$ and $B(t_{B,1}, \dotsc, t_{B,M})$ on the
l.h.s.~of Eq.~\eqref{eq:27}. The construction of the corresponding functional
integral can be accomplished with a few straightforward modifications to the
procedure summarized in Appendix~\ref{sec:repr-corr-funct} (and presented, e.g.,
in Refs.~\onlinecite{ Kamenev2011,Altland/Simons}).
In a stationary state one has
$[\rho, H] = 0$ and  all the Heisenberg operators on the l.h.s.~of
Eq.~\eqref{eq:27} can be related to the Schr\"odinger operators at a later time
$t_f$. Then one finds
\begin{multline}
  \label{eq:31}
  \langle B(t_{B,1}, \dotsc, t_{B,M}) A(t_{A,1}, \dotsc, t_{A,N}) \rangle \\ =
  \langle A_{+}(t_{A,1}, \dotsc, t_{A,N}) B_{-}(t_{B,1}, \dotsc, t_{B,M}) \rangle_{S_b},
\end{multline}
where the action $S_b$ describes the backward evolution and it is related to the
action $S$ which enters the forward evolution in Eq.~\eqref{eq:27} simply by a
global change of sign $S_b = - S$.

\subsection{Quantum-mechanical time reversal}
\label{sec:time-reversal}

In this section we first recall some properties of the quantum-mechanical time
reversal operation $\mathsf{T}$~\cite{Messiah:II} and then discuss its
implementation within the Schwinger-Keldysh formalism. $\mathsf{T}$ is an
antiunitary operator, i.e., it is antilinear (such that
$\mathsf{T} \lambda |\psi\rangle = \lambda^{*} \mathsf{T}|\psi\rangle$ for
$\lambda \in \C$) and unitary ($\mathsf{T}^{\dagger} = \mathsf{T}^{-1}$). Scalar
products transform under antiunitary transformations into their complex
conjugates, i.e.,
$\langle \psi | A | \phi \rangle = \langle \tilde{\psi} | \tilde{A} |
\tilde{\phi} \rangle^{*}$,
where we denote by $| \tilde{\psi} \rangle = \mathsf{T} | \psi \rangle$ and
$\tilde{A} = \mathsf{T} A \mathsf{T}^{\dagger}$ the state and the Schr\"odinger
operator obtained from the state $| \psi \rangle$ and the operator $A$,
respectively, after time reversal.
Accordingly, expressing  the trace of
an operator in a certain basis $\{ |\psi_n\rangle \}_n$,  one finds
\begin{equation}
  \label{eq:35}
  \tr A = \sum_n \langle \psi_n | A | \psi_n \rangle = \sum_n \langle
  \tilde{\psi}_n | \tilde{A} | \tilde{\psi}_n \rangle^{*} = (\tr  \tilde{A})^{*}.
\end{equation}
In the last equality we used the fact that, due to the unitarity of
$\mathsf{T}$, also the time-reversed set $\{ |\tilde{\psi}_n\rangle \}_n$ forms
a basis.
For future convenience, we shall define the Heisenberg representation of
time-reversed operators such that it coincides with the Schr\"odinger
one at time $-t_f$, i.e., we set
\begin{equation}
  \label{eq:34}
  \tilde{A}(t_A) = e^{i \tilde{H} \left( t_A + t_f \right)} \tilde{A} e^{-i \tilde{H}
    \left( t_A + t_f \right)}.
\end{equation}
Note that this is distinct from the Heisenberg representation defined in
Eq.~\eqref{eq:24}, which coincides with the Schr\"odinger
one only at time $t_i$. 
In order to simplify the notation, we shall not distinguish these two different
Heisenberg representations, assuming implicitly that the latter and the former
are used, respectively, for operators and their time-reversed ones, such that
$A(t_i) = A$ while $\tilde{A}(-t_f) = \tilde{A}$.

Let us now study the effect of time reversal on the generic multi-time correlation
function in Eq.~\eqref{eq:27}. Due to translational invariance in time, the 
time arguments of the operators $A$ and $B$ can be shifted by $t_i - t_f$ without
affecting the correlation function. Then, by using Eqs.~\eqref{eq:35}
and~\eqref{eq:34}, one has
\begin{multline}
  \label{eq:36}
  \langle A(t_{A,1}, \dotsc, t_{A,N}) B(t_{B,1}, \dotsc, t_{B,M}) \rangle \\
  \begin{aligned}[t]
    & = \langle \tilde{A}(-t_{A,1},\dotsc,-t_{A,N}) \tilde{B}(-t_{B,1}, \dotsc,
    -t_{B,M}) \rangle_{\tilde{\rho}}^{*} \\ & = \langle
    \tilde{B}^{\dagger}(-t_{B,1}, \dotsc, -t_{B,M})
    \tilde{A}^{\dagger}(-t_{A,1},\dotsc,-t_{A,N}) \rangle_{\tilde{\rho}},
  \end{aligned}
\end{multline}
where the subscript in $\langle \dotsb \rangle_{\tilde{\rho}}$ indicates that
the expectation value is taken with respect to the time-reversed density
operator $\tilde{\rho} \equiv \mathsf{T}\rho\mathsf{T}^\dagger$, which is
time-independent. The expectation value on the r.h.s.~of Eq.~\eqref{eq:36} is
anti-time ordered and therefore it can be rewritten as a Schwinger-Keldysh
functional integral by using Eq.~\eqref{eq:31}. The l.h.s., instead, is
time-ordered and therefore it can be expressed as in Eq.~\eqref{eq:27}, such
that Eq.~\eqref{eq:36} becomes
\begin{multline}
  \label{eq:38}
  \langle B_{+}(t_{B,1}, \dotsc, t_{B,M}) A_{-}(t_{A,1}, \dotsc, t_{A,N})
  \rangle \\ = \langle \tilde{A}_{+}^{*}(-t_{A,1},\dotsc, -t_{A,N})
  \tilde{B}_{-}^{*}(-t_{B,1}, \dotsc, -t_{B,M}) \rangle_{\tilde{S}_b},
\end{multline}
where the subscript $b$ in $\tilde{S}_b$ indicates that the sign of the action
which describes the Hamiltonian evolution on the r.h.s.~of this relation has 
been reversed, as explained below Eq.~\eqref{eq:31}.
The time-reversed action $\tilde{S}$ differs from the action $S$ associated with
$H$ which enters (implicitly, cf.\ Eq.~\eqref{eq:22}) Eq.~\eqref{eq:27} because
in $\tilde{S}$ the time evolution is generated by $\tilde{H}$, the initial state
is the time-reversed density matrix $\tilde{\rho}$, and the integration over
time extends from $-t_f$ to $-t_i$. This latter difference becomes
inconsequential as $t_i \to -\infty$ and $t_f \to \infty$.

Let us now consider the case in which $A$ and $B$ are products of the
bosonic field operators $\psi$ and $\psi^{\dagger}$, such that $A_\pm$ and
$B_\pm$ involve the corresponding products of $\psi_\pm$ and their complex
conjugates.
As there are no further restrictions on $A$ and $B$, the l.h.s.~of
Eq.~\eqref{eq:38} can be generically indicated as
$\langle \mathcal{O}[\Psi] \rangle$, where $\mathcal{O}[\Psi]$ is the product of
various fields on the Schwinger-Keldysh contour corresponding to
$B_+(\ldots)A_-(\ldots)$ which, according to the notation introduced in
Sec.~\ref{sec:symm-transf}, are collectively indicated by
$\Psi = \left( \psi_{+}, \psi_{+}^{*}, \psi_{-}, \psi_{-}^{*} \right)^T$.
With this shorthand notation, Eq.~\eqref{eq:38} can be cast in the form
\begin{equation}
  \label{eq:40}
  \langle \mathcal{O}[\Psi] \rangle = \langle \mathcal{O}[\mathsf{T} \Psi] \rangle_{\tilde{S}_b},
\end{equation}
where the transformation
\begin{equation}
  \label{eq:41}
  \mathsf{T} \Psi_{\sigma}(t, \mathbf{x}) = \Psi_{-\sigma}^{*}(-t, \mathbf{x}),
\end{equation}
implements the quantum-mechanical time reversal within the Schwinger-Keldysh
formalism.
(With a slight abuse of notation, the same symbol $\mathsf{T}$ is used to
indicate here both the quantum-mechanical time-reversal operator introduced
above and the transformation of fields on the Schwinger-Keldysh contour in
Eqs.~\eqref{eq:40} and~\eqref{eq:41}.)
In Eq.~\eqref{eq:41} we took into account that the bosonic field operators in
the Schr\"odinger picture and in the real-space representation are time-reversal
invariant, i.e.,
$\tilde{\psi}(\mathbf{x}) = \mathsf{T} \psi(\mathbf{x}) \mathsf{T}^{\dagger} =
\psi(\mathbf{x})$,
which allows us to drop the tilde on the transformed field on the r.h.s.~of
Eq.~\eqref{eq:41}.
However, we note that in the last line of Eq.~\eqref{eq:36} the Hermitean
adjoint operators of those on the l.h.s.~appear and this is the reason why both
the r.h.s.~of Eq.~\eqref{eq:38} and the transformation prescription
Eq.~\eqref{eq:41} involve complex conjugation of the fields. 
Note that the time-reversal transformation in Eq.~\eqref{eq:41} is actually
linear, i.e., the complex conjugation affects only the fields variables and not
possible complex prefactors.  This follows again from the last line of
Eq.~\eqref{eq:36}: the combination of two antilinear transformations (time
reversal and Hermitean conjugation) results into a linear one.

Let us mention that while Eq.~\eqref{eq:38} follows from the second line of
Eq.~\eqref{eq:36}, one could have equivalently taken its first line as the
starting point for deriving a Schwinger-Keldysh time-reversal
transformation. Then one would have been lead to a different implementation
$\mathsf{T}'$ of time reversal:
\begin{equation}
  \label{eq:23}  
  \mathsf{T}' \Psi_{\sigma}(t, \mathbf{x}) = \Psi_{\sigma}(-t, \mathbf{x}),
\end{equation}
with an additional overall complex conjugation of the correlation function. In
some sense, the transformation $\mathsf{T}'$ is closer to the common way of
representing the quantum-mechanical time reversal than $\mathsf{T}$ is, as it
amounts to a mapping $t \mapsto -t$ and $i \mapsto -i$~\cite{Messiah:II}. For
our purposes, however, $\mathsf{T}$ is of main interest, since it is part of the
equilibrium transformation as we describe below.

We finally note that an alternative implementation of the time-reversal
transformation is based on the observation that the forward and backward
branches of the closed-time path integral are actually equivalent if the
dynamics of a system is time-reversal invariant (TRI)~\cite{Kamenev2011}. More
specifically, this means that the Schwinger-Keldysh action is invariant --- up
to a global change of sign --- upon exchanging the corresponding fields, i.e.,
$S[\psi_{+}, \psi_{-}] = - S[\psi_{-}, \psi_{+}]$.
This transformation is partly recovered in Eqs.~\eqref{eq:40} and~\eqref{eq:41}:
in fact, $\mathsf{T}$ in Eq.~\eqref{eq:41} involves an exchange of the contour
indices, while the corresponding global change of sign in the action is
indicated on the r.h.s.~of Eq.~\eqref{eq:40} by the subscript $\tilde{S}_b$
(cf.\ the definition of $S_b$ below Eq.~\eqref{eq:31}). However, the time
reversal transformation $\mathsf{T}$ in Eq.~\eqref{eq:41} --- derived from the
quantum-mechanical time-reversal operation --- additionally involves both
complex conjugation and the time inversion $t \mapsto - t$.

\subsection{KMS condition and generalized fluctuation-dissipation relations}
\label{sec:kms-condition}

\begin{figure*}
  \centering
  \includegraphics[width=.7\linewidth]{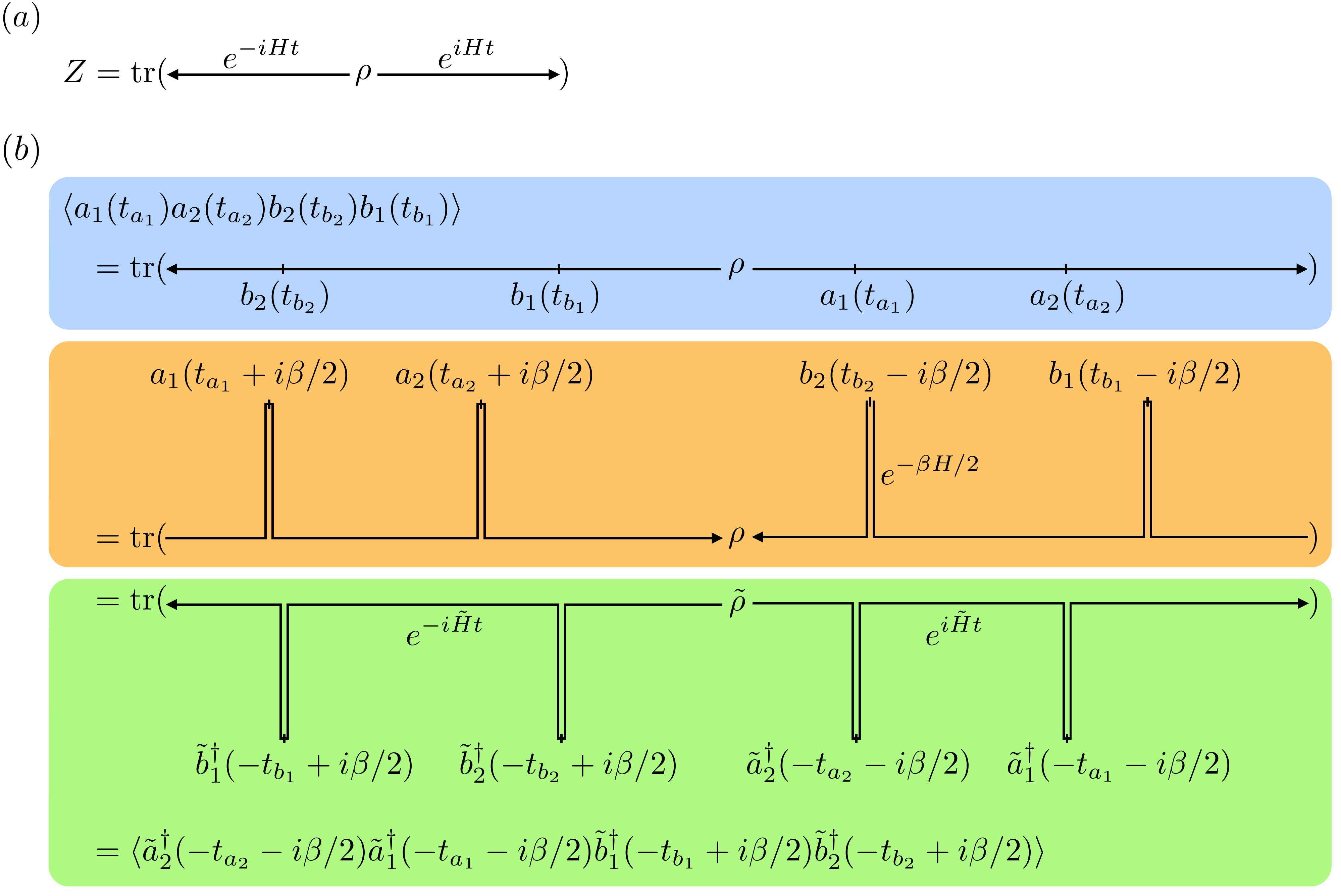}
  \caption{(Color online) $(a)$ Schematic representation of the
    Schwinger-Keldysh partition 
    function~\cite{Kamenev2011,Altland/Simons}. The time evolution of the
    density matrix $\rho(t) = e^{-i H t} \rho e^{i H t}$ can be represented by
    introducing two time lines to the left and right of $\rho$. These time lines
    correspond to the $+$ and $-$ parts of the Schwinger-Keldysh contour,
    respectively. $(b)$ Schematic representation of the KMS condition for a
    four-time correlation function
    $\langle a_1(t_{a,1}) a_2(t_{a,2}) b_2(t_{b,2}) b_1(t_{b,1}) \rangle$ with
    $t_{a,1} < t_{a,2}$ and $t_{b,1} < t_{b,2}$, where $a_{1,2}$ and
    $b_{1,2}$ are bosonic field operators.  %
    As illustrated by the first equality (light blue box), this correlation
    function is properly time-ordered and therefore it can be directly
    represented within the Schwinger-Keldysh formalism with the operators
    $a_{1,2}$ and $b_{1,2}$ evaluated along the $-$ and $+$ contours,
    respectively.  %
    The thermal density matrix $\rho = e^{-\beta H}/\tr e^{-\beta H}$ can be
    first split into the products of $e^{-\beta H/2}\times e^{-\beta H/2}$ and
    then these two factors can be moved in opposite directions along the two
    time lines, with the effect of
    adding $+i \beta/2$ and $-i \beta/2$ to the time arguments of $a_{1,2}$ and
    $b_{1,2}$, respectively.
    After these two factors have been moved to the end of the timelines, due to
    the cyclic property of the trace, they combine as represented by  
    the second equality (orange box), where the time lines now take detours into the
    complex plane and the overall time order is effectively reversed as
    indicated by the arrows, which converge towards $\rho$ instead of departing
    from it as in the case of sketch $(a)$ or of the first equality of sketch
    $(b)$. 
    The original time ordering
    can be then restored by means of the time-reversal operation $\mathsf{T}$,
    upon application of which operators are replaced by time-reversal
    transformed ones, $\tilde{\rho} = \mathsf{T} \rho \mathsf{T}^{\dagger}$
    etc., and the signs of time variables are reversed. In addition, due to the
    anti-unitarity of $\mathsf{T}$  
    one has to
    take the Hermitian adjoint of the expression inside the trace. 
    As a result, the
    order of operators is inverted and one obtains the third equality (green box)
    which is again properly time ordered. 
    This construction can be generalized to arbitrary correlation functions,
    leading to Eq.~\eqref{eq:52}.}
  \label{fig:KMS_proof_sketch}
\end{figure*}

The discussion of the previous sections about multi-time correlation functions
and the time-reversal transformation provides the basis for the formulation of
the KMS condition for multi-time correlation functions and of its representation
in terms of Schwinger-Keldysh functional integrals.
For the specific case of a four-time correlation function, the KMS condition is
pictorially illustrated in Fig.~\ref{fig:KMS_proof_sketch}. As described in the
caption, panel $(a)$ summarizes the convention as far as the forward/backward
contours are concerned. Panel $(b)$, instead, refers to the KMS condition which,
as anticipated after Eq.~\eqref{eq:KMS0}, involves a contour exchange of the
multi-time operators $A$ and $B$.
This exchange is the reason why the arrows in the second equality in
Fig.~\ref{fig:KMS_proof_sketch} $(b)$ are reversed, since both $A$ and $B$ turn
out to be anti-time ordered when moved to the opposite contour. The appropriate
time-ordering can be restored by means of the quantum-mechanical time reversal
transformation introduced in Sec.~\ref{sec:time-reversal}, as indicated by the
third equality in the figure.  This is a crucial step, because only time-ordered
correlation functions can be directly translated into the functional integral by
means of the usual Trotter decomposition, which makes the time-reversal
transformation indispensable in the construction.  However, this does not mean
that properties related to equilibrium conditions such as
fluctuation-dissipation relations are fulfilled only if the Hamiltonian is
invariant under the time-reversal transformation. Indeed, it turns out that
multi-time FDRs always involve both the Hamiltonian and its time-reversed
counterpart~\cite{Jakobs2010}, while as we show in
Sec.~\ref{sec:fluct-diss-theor} the FDR for two-time functions can be stated
without reference to the time-reversed Hamiltonian, even if the Hamiltonian is
not TRI.

The KMS condition for a two-time function reads~\footnote{We note that this
  condition is usually expressed in the form
  {$$\langle A(t_A) B(t_B) \rangle = \langle B(t_B) A(t_A + i \beta) \rangle.$$}
  However, an equilibrium state is also stationary and therefore both time
  arguments on the r.h.s.~can be translated by $-i \beta/2$ which leads
  immediately to Eq.~\eqref{eq:46}. Here we are assuming that the analytic
  continuation of real-time correlation functions into the complex plane is
  possible and unambiguous.}
\begin{equation}
  \label{eq:46}
  \langle A(t_A) B(t_B) \rangle = \langle B(t_B - i \beta/2) A(t_A + i \beta/2) \rangle.
\end{equation}
This relation can be proven by writing down explicitly the expectation value on
the l.h.s.~with $\rho = e^{-\beta H}/\tr e^{-\beta H}$ and by inserting the
definition of the Heisenberg operators reported in Eq.~\eqref{eq:24}.
The generalization of this procedure to the case of multi-time correlation
functions is straightforward and yields
\begin{multline}
  \label{eq:47}
  \langle A(t_{A,1}, \dotsc, t_{A,N}) B(t_{B,1}, \dotsc, t_{B,M}) \rangle \\ =
  \langle B(t_{B,1} - i \beta/2, \dotsc, t_{B,M} - i \beta/2) \\ \times
  A(t_{A,1} + i \beta/2, \dotsc, t_{A,N} + i \beta/2) \rangle.
\end{multline}
The real parts of the time variables on the r.h.s.~of this equation are such
that the corresponding product of operators is anti-time-ordered (see
Fig.~\ref{fig:KMS_proof_sketch}). According to their definition in
Eq.~\eqref{eq:26}, $A$ and $B$ correspond to products of operators with,
respectively, decreasing and increasing time arguments from right to
left. Consequently, Eq.~\eqref{eq:47} can be expressed as a functional integral
by using Eqs.~\eqref{eq:27} and~\eqref{eq:31} on the l.h.s.~and r.h.s.,
respectively. The presence of an imaginary part in the time arguments of
Eq.~\eqref{eq:47} does not constitute a problem: in fact, the functional
integral along the vertical parts of the time path in
Fig.~\ref{fig:KMS_proof_sketch} can be constructed by the same method as the
horizontal parts, which is summarized in Sec.~\ref{sec:multi-time-corr}. Hence,
we find
\begin{multline}
  \label{eq:48}
  \langle B_{+}(t_{B,1}, \dotsc, t_{B,M}) A_{-}(t_{A,1}, \dotsc, t_{A,N})
  \rangle \\ = \langle A_{+}(t_{A,1} + i \beta/2, \dotsc, t_{A,N} + i \beta/2)
  \\ \times B_{-}(t_{B,1} - i \beta/2, \dotsc, t_{B,M} - i \beta/2)
  \rangle_{S_b}.
\end{multline}
As we did in Eq.~\eqref{eq:40} for the case of quantum-mechanical time-reversal,
we may rewrite this equation in the form
\begin{equation}
  \label{eq:43}
  \langle \mathcal{O}[\Psi] \rangle = \langle \mathcal{O}[\mathcal{K}_{\beta} \Psi] \rangle_{S_b},
\end{equation}
where we define
\begin{equation}
  \label{eq:50}
  \mathcal{K}_{\beta} \Psi_{\sigma}(t) = \Psi_{-\sigma}(t - i \sigma \beta/2).
\end{equation}
This transformation $\mathcal{K}_{\beta}$ can be combined with the quantum
mechanical time reversal $\mathsf{T}$ defined in Eq.~\eqref{eq:41} in order to
express the equilibrium transformation $\mathcal{T}_{\beta}$ as
$\mathcal{T}_{\beta} = \mathsf{T} \circ \mathcal{K}_{\beta}$.~\footnote{It is
  straightforward to verify that the transformation $\mathsf{T}$ is not modified
  in the presence of complex time arguments.}
%
By using Eq.~\eqref{eq:40} on the r.h.s.~of Eq.~\eqref{eq:43}, one concludes
that the KMS condition implies
\begin{equation}
  \label{eq:52}
  \langle \mathcal{O}[\Psi] \rangle = \langle \mathcal{O}[\mathcal{T}_{\beta} \Psi]
  \rangle_{\tilde{S}},
\end{equation}
which indeed provides a generalized FDR for the correlation function
$\langle \mathcal{O}[\Psi] \rangle$~\cite{Jakobs2010}. For various choices of
the observable $\mathcal{O}[\Psi]$ we obtain the full hierarchy of multi-time
FDRs, which contains as a special case the usual FDR for two-time
functions (see Sec.~\ref{sec:fluct-diss-theor}). Before demonstrating that this
hierarchy is actually equivalent to the
invariance of the Schwinger-Keldysh action as expressed by Eq.~\eqref{eq:5}, several
remarks are in order:
\begin{enumerate}
\item Although by means of the time-reversal transformation $\mathsf{T}$ we were
  able to restore the time ordering in Eq.~\eqref{eq:48}, Eq.~\eqref{eq:52}
  still involves the time-reversed action $\tilde{S}$ and not the original
  action $S$. However, in practice it will typically be clear how $\tilde{S}$
  can be obtained from $S$, e.g., by reversing the signs of external magnetic
  fields. In the absence of fields which break time-reversal invariance the
  tilde in Eq.~\eqref{eq:52} may be dropped, i.e., $\tilde{S} = S$.
\item Equation~\eqref{eq:52} provides a generalized FDR expressed in terms of
  products $\mathcal{O}[\Psi]$ of fields on the forward and backward branches,
  which we collected in the four-component vector
  $\Psi = \left( \psi_{+}, \psi_{+}^{*}, \psi_{-}, \psi_{-}^{*} \right)^T$. A
  more familiar formulation of FDRs is provided in terms of classical and
  quantum fields ($\Phi_c$ and $\Phi_q$, see Eq.~\eqref{eq:9}), which allow one
  to identify correlations functions (i.e., expectation values involving only
  classical fields) and response functions or susceptibilities (expectation
  values involving both classical and quantum fields). FDRs provide relations
  between correlation and response functions.
  In order to express the KMS condition in Eq.~\eqref{eq:52} in terms of the
  classical and quantum fields $\Phi_c$ and $\Phi_q$ (or, alternatively, of
  $\Phi 
  = \left( \phi_c, \phi_c^{*}, \phi_q, \phi_q^{*} \right)^T$),
  we note that they are linearly related to $\Psi_+$ and $\Psi_-$ and therefore
  a generic product ${\cal O}[\Phi]$ of such fields can be expressed as a linear
  combination (with real coefficients) of products ${\cal O}_i[\Psi]$. According
  to Eq.~\eqref{eq:52}, the expectation value of such a combination and
  therefore $\langle {\cal O}[\Phi] \rangle$ can be expressed as on its
  r.h.s.~in terms of the same linear combination of
  $\langle {\cal O}_i[\mathcal{T}_{\beta}\Psi]\rangle_{\tilde{S}}$; since the
  transformation $\mathcal{T}_{\beta}$ is linear, it immediately follows that
  this linear combination is nothing but
  $\langle {\cal O}[\mathcal{T}_{\beta}\Phi] \rangle_{\tilde{S}}$, where the
  explicit form of the transformation of the components of the
  field $\Phi$ under $\mathcal{T}_{\beta}$ is reported in Eq.~\eqref{eq:3}. The
  KMS condition then becomes
\begin{equation}
  \label{eq:53}
  \langle \mathcal{O}[\Phi] \rangle = \langle \mathcal{O}[\mathcal{T}_{\beta} \Phi]
  \rangle_{\tilde{S}}.
\end{equation}
In Sec.~\ref{sec:fluct-diss-theor}, on the basis of Eq.~\eqref{eq:53}, we derive
the typical form of the FDR, which involves the correlation function of two
classical fields and the susceptibility expressed as a correlation between one
quantum and one classical field.

\item In the grand canonical ensemble with density matrix
  $\rho = e^{-\beta \left( H - \mu N \right)}/\tr e^{-\beta \left( H - \mu N
    \right)}$,
  where $N = \int_{\mathbf{x}} \psi^{\dagger}(\mathbf{x}) \psi(\mathbf{x})$ is
  the particle number operator, the KMS condition Eq.~\eqref{eq:47} has to be
  generalized. In order to derive it, we split again the density matrix into a
  product
  $e^{- \beta \left( H - \mu N \right)/2} \times e^{-\beta \left( H - \mu N
    \right)/2}$
  (cf.~the caption of Fig.~\ref{fig:KMS_proof_sketch}). Then, moving one of the
  two factors through each of the blocks of operators $A$ and $B$ as in the
  second (orange) box in panel \textit{(b)} of Fig.~\ref{fig:KMS_proof_sketch}
  has not only the effect of adding $+ i \beta/2$ and $-i \beta/2$ to the time
  arguments of the field operators in $A$ and $B$ respectively, as in the case
  $\mu = 0$; taking into account the canonical commutation relations, additional
  factors appears due to the fact that
\begin{equation}
    \label{eq:33}
    \begin{split}
      e^{\sigma \beta \mu N/2} \psi(\mathbf{x}) e^{-\sigma \beta \mu N/2} & 
      = e^{-\sigma \beta \mu/2} \psi(\mathbf{x}), \\ 
        e^{\sigma \beta \mu N/2}
      \psi^{\dagger}(\mathbf{x}) e^{-\sigma \beta \mu N/2} & = e^{\sigma \beta \mu/2}
      \psi^{\dagger}(\mathbf{x}),
    \end{split}
  \end{equation}
  where $\sigma = +1$ and $-1$ for operators which are part of the time-ordered
  and anti-time-ordered blocks of operators $A$ and $B$, respectively.  The
  factors $e^{\pm \sigma \beta \mu/2}$ can be taken out of the expectation value
  in Eq.~\eqref{eq:47} (corresponding to the trace in
  Fig.~\ref{fig:KMS_proof_sketch} \textit{(b)}) and do not affect the
  restoration of time order by means of the time-reversal transformation, which
  is illustrated in the last (green) box in panel \textit{(b)} of
  Fig.~\ref{fig:KMS_proof_sketch}.  Therefore, on the r.h.s.~of
  Eq.~\eqref{eq:52} they would appear as prefactors, which are absorbed in the
  modified transformation given in Eq.~\eqref{eq:4}.
\end{enumerate}

\subsection{From the KMS condition to a symmetry of the Schwinger-Keldysh action}
\label{sec:from-kms-symmetry}

In the previous section we showed that the KMS condition within the
Schwinger-Keldysh functional integral formalism takes the form of
Eq.~\eqref{eq:52} (with $\mathcal{T}_{\beta}$ given by either Eq.~\eqref{eq:1}
or \eqref{eq:4}).  Here we argue further, that the latter relation is equivalent
to requiring the invariance of the Schwinger-Keldysh action under the
equilibrium symmetry $\mathcal{T}_{\beta}$.
To this end, we express the expectation values on the left and right hand sides
of Eq.~\eqref{eq:52} as the functional integrals
\begin{equation}
  \label{eq:65}
  \langle \mathcal{O}[\Psi] \rangle = \int
  \mathcal{D}[\Psi] \mathcal{O}[\Psi] e^{i S[\Psi]}
\end{equation}
and 
\begin{equation}
  \label{eq:2}
  \langle \mathcal{O}[\mathcal{T}_{\beta} \Psi] \rangle_{\tilde{S}} = \int
  \mathcal{D}[\Psi] \mathcal{O}[\mathcal{T}_{\beta} \Psi] e^{i \tilde{S}[\Psi]},
\end{equation}
respectively.
Performing a change of integration variables $\Psi \to \mathcal{T}_{\beta} \Psi$
in the last expression, the argument of $\mathcal{O}$ simplifies because
$\mathcal{T}_{\beta} \Psi \to \mathcal{T}_{\beta}^2 \Psi = \Psi$, since
$\mathcal{T}_{\beta}$ is involutive (see Sec.~\ref{sec:symm-transf}).
In addition, we show in Appendix~\ref{sec:jacob-equil-transf} that the absolute
value of the determinant of the Jacobian
$\mathcal{J} = \delta(\mathcal{T}_{\beta} \Psi)/\delta\Psi$ associated with
$\mathcal{T}_{\beta}$ equals one, i.e., $\abs{\Det \mathcal{J} } = 1$, and
therefore the integration measure is not affected by the change of
variable. Accordingly, one has
\begin{equation}
  \label{eq:66}
  \langle \mathcal{O}[\mathcal{T}_{\beta} \Psi] \rangle_{\tilde{S}}= \int \mathcal{D}[\Psi]
  \mathcal{O}[\Psi] e^{i \tilde{S}[\mathcal{T}_{\beta}\Psi]},
\end{equation}
and by comparing this expression to Eq.~\eqref{eq:65} a \emph{sufficient}
condition for their equality is indeed Eq.~\eqref{eq:5}, which expresses the
invariance of the Schwinger-Keldysh action under the equilibrium transformation.
Since the observable $\mathcal{O}[\Psi]$ in Eqs.~\eqref{eq:52},~\eqref{eq:65}, and~\eqref{eq:66} is arbitrary, 
the condition is also \emph{necessary}, which proves that Eqs.~\eqref{eq:5} and~\eqref{eq:52} (and,
consequently, the KMS condition) are equivalent.

\section{Examples}
\label{sec:implications}

In this section we discuss some concrete examples of how the invariance of a
certain Schwinger-Keldysh action under the equilibrium transformation
$\mathcal{T}_{\beta}$ can be used in practice.
First we show that Eq.~\eqref{eq:52} (or, equivalently, Eq.~\eqref{eq:53})
contains as a special case the quantum FDR which establishes a relationship
between the two-time correlation function of the field $\phi_c$ (see
Eq. \eqref{eq:9}) and its response to an external perturbation which couples
linearly to it.
This was also noted in Ref.~\onlinecite{Jakobs2010}; however, the conceptual
advance done here consists in realizing that the FDR can be regarded as a
Ward-Takahashi identity associated with the equilibrium symmetry.
In Sec.~\ref{sec:non-equil-nature}, instead, we elaborate on the non-equilibrium
nature of Markovian dynamics described by a quantum master equation (of the
Lindblad form), which is seen to violate explicitly the equilibrium symmetry.

The case of a system which is driven out of equilibrium by a coupling to
different baths is considered in Sec.~\ref{sec:two-baths}. In particular, we
discuss a single bosonic mode coupled to two baths at different temperatures and
chemical potentials.
Finally, in Sec.~\ref{sec:further-applications} we briefly list a number of
additional applications of the symmetry. Some of those have been put into
practice in the context of classical dynamical systems, and could be generalized
to the quantum case with the aid of the symmetry transformation discussed in the
present paper.

\subsection{Fluctuation-dissipation relation for two-time functions}
\label{sec:fluct-diss-theor}

Considering the invariance of the Schwinger-Keldysh action under
$\mathcal{T}_{\beta}$ in Eq.~\eqref{eq:3} as the defining property of
thermodynamic equilibrium, the generalized FDR in Eq.~\eqref{eq:53} (which, as
discussed above, is nothing but the Ward-Takahashi identity associated with the
symmetry), emerges as a \emph{consequence} of equilibrium conditions.
Then, from the generalized FDR in Eq.~\eqref{eq:53}, the FDR for two-time
functions~\cite{Kamenev2011,Altland/Simons} can indeed be derived as a special
case. The latter reads
\begin{equation}
  \label{eq:57}
  \begin{split}
    G^K\oq & = \left( G^R\oq - G^A\oq \right) \ct \\ & = i\, 2 \Im G^R\oq \ct,
  \end{split}
\end{equation}
where the Keldysh, retarded, and advanced Green's functions $G^K$, $G^R$, and
$G^A$, respectively, are related to expectation values of classical and quantum
fields via
\begin{equation}
  \label{eq:61}
  \begin{split}
    i G^K\oq \left( 2 \pi \right)^{d + 1} \delta(\omega - \omega')
    \delta^{(d)}(\mathbf{q} - \mathbf{q}') & = \langle \phi_c\oq
    \phi_c^{*}(\omega',\mathbf{q}') \rangle, \\ i G^R\oq \left( 2 \pi \right)^{d
      + 1} \delta(\omega - \omega') \delta^{(d)}(\mathbf{q} - \mathbf{q}') & = \langle
    \phi_c\oq \phi_q^{*}(\omega',\mathbf{q}') \rangle, \\ i G^A\oq \left( 2 \pi
    \right)^{d + 1} \delta(\omega - \omega') \delta^{(d)}(\mathbf{q} - \mathbf{q}') &
    = \langle \phi_q\oq \phi_c^{*}(\omega',\mathbf{q}') \rangle.
  \end{split}
\end{equation}
Here we are assuming translational invariance in both time and space, which is
reflected in the appearance of frequency- and momentum-conserving
$\delta$-functions in the previous expressions. 
The FDR valid in classical systems~\cite{aron10:_symmet_langev,Aron2014} can be
recovered from the quantum FDR Eq.~\eqref{eq:57} by taking the classical limit
as described in Sec.~\ref{sec:classical-limit}. Contrary to what one might
suspect at a first glance from the appearance of the time-reversed action
$\tilde{S}$ in the generalized FDR in Eq.~\eqref{eq:53}, the derivation of the
FDR for two-time functions we present below is valid \emph{irrespective} of
whether the action contains external fields which break TRI or not.

Let us consider the identity Eq.~\eqref{eq:53} for specific
choices of the functional $\mathcal{O}[\Phi]$. In particular, by taking
$\mathcal{O}$ to be equal to $\phi_q\phi_q^*$, the expectation
value on the l.h.s.~of Eq.~\eqref{eq:53} has to vanish due to causality~\cite{Kamenev2011} and
therefore
\begin{equation}
  \label{eq:58}
  0 = \langle \phi_q\oq \phi_q^{*}(\omega',\mathbf{q}')
  \rangle = \langle \mathcal{T}_{\beta} \phi_q(\omega,-\mathbf{q}) \mathcal{T}_{\beta} \phi_q^{*}(\omega',-\mathbf{q}')
  \rangle_{\tilde{S}}.
\end{equation}
Upon inserting the expression of the fields transformed according to
Eq.~\eqref{eq:3}, one readily finds the FDR
\begin{equation}
  \label{eq:59}
  G^K_{\tilde{S}}\oq = \left( G^R_{\tilde{S}}\oq
    - G^A_{\tilde{S}}\oq \right) \ct
\end{equation}
with the time-reversed action $\tilde{S}$.  By setting, instead, $\mathcal{O}$
equal to the product of two classical fields, the l.h.s.~of Eq.~\eqref{eq:53}
renders the Keldysh Green's function $G^K$ in Eq.~\eqref{eq:61}; by using the
explicit form of $\mathcal{T}_{\beta}$ in Eq.~\eqref{eq:3} and the FDR derived
above, the r.h.s.~of that equation coincides with $G^K_{\tilde{S}}$ and
therefore one concludes that
\begin{equation}
  \label{eq:60}
  G^K_{\tilde{S}}\oq = G^K(\omega,-\mathbf{q}),
\end{equation}
which expresses the transformation behavior of the 
Keldysh Green's function under time reversal of the Hamiltonian.
Finally, by replacing $\mathcal{O}[\Phi]$ in Eq.~\eqref{eq:53} with the product
$\phi_c(\omega, \mathbf{q}) \phi_q^{*}(\omega', \mathbf{q})$ of one classical
and one quantum field, the l.h.s.~renders by definition the retarded Green's
function $G^R$ while the r.h.s.~can be worked out as explained above. Taking
into account Eq.~\eqref{eq:59}, one can eliminate the Keldysh Green's function
$G^K$ appearing on the r.h.s.~in favor of the retarded and advanced Green's
functions $G^R$ and $G^A$, respectively, and eventually finds
\begin{equation}
  \label{eq:55}
  G^R_{\tilde{S}}\oq = G^R(\omega,-\mathbf{q}).
\end{equation}
This relation, together with its complex conjugate (which relates the advanced
Green's functions calculated from the original and time-reversed Hamiltonians,
respectively) and Eqs.~\eqref{eq:60} and~\eqref{eq:59}, yields the
FDR~\eqref{eq:57}.

\subsection{Non-equilibrium nature of steady states of quantum master equations}
\label{sec:non-equil-nature}

In classical statistical physics, the coupling of a system to a thermal
bath and the resulting relaxation to thermodynamic equilibrium are commonly
modelled in terms of Markovian stochastic processes, which can be
described, e.g., by suitable Langevin equations with Gaussian white
noise~\cite{Tauber2014}. 
The Markovian dynamics of a \emph{quantum} system, on the other hand, is
described by a quantum master equation in the Lindblad
form~\cite{Kossakowski1972,Lindblad1976} (or by an equivalent Schwinger-Keldysh
functional integral).
Under specific conditions on the structure of the Lindblad
operator~\cite{Gardiner/Zoller,Alicki2007}, the stationary state of this
dynamics is described by a thermal Gibbs distribution, such that all
\emph{static} properties (equal-time correlation functions) are
indistinguishable from those in thermodynamic equilibrium.
In spite of this fact, however, \emph{dynamical} signatures of thermodynamic equilibrium such
as the KMS condition (see Sec.~\ref{sec:kms-condition}) and the FDR (see
Sec.~\ref{sec:fluct-diss-theor}) are violated~\cite{Talkner1986,Ford1996}.
This violation can be traced back to the fact that the Markovian and rotating
wave approximations, which are done in deriving this quantum master equation
cause an \emph{explicit} breaking of the equilibrium symmetry as we show in this
section --- i.e., although the system is coupled to a bath in thermodynamic
equilibrium, the system itself does not reach equilibrium. Physically, this can
be understood by noting that the microscopic dynamics underlying an approximate
Markovian behavior in this case is indeed driven.
A typical example in the context of quantum optics is an atom with two relevant
energy levels separated by a level spacing $\omega_0$ and subject to an external
driving laser with frequency $\nu$ detuned from resonance by an amount
$\Delta = \nu -\omega_0\ll \omega_0$.
Transitions between the ground and the excited state are made possible only by
the driving laser, and the energy scale which controls the validity of the
Markov approximation in the dynamics of the two-level system is set by
$\omega_0$. The excited state is assumed to be unstable and it can undergo
spontaneous decay by emitting a photon to the radiation field, which acts as a
reservoir. This illustrates the combined driven and dissipative nature of such
quantum optical systems.

In order to investigate in more detail the effect of the Markovian approximation
typically done in the driven context on the validity of the equilibrium symmetry
$\mathcal{T}_{\beta}$, we consider a system with a certain action $S$, whose
degrees of freedom are linearly coupled to those of a thermal bath. By
integrating out the latter degrees of freedom, a dissipative contribution to the
original action is generated. In order to simplify the discussion, we assume
that the bath consists of non-interacting harmonic oscillators
$b_{\mu,\sigma}(t)$, labelled by an index $\mu$ (with $\sigma$ referring to the
branch of the Schwinger-Keldysh contour), with proper frequency $\omega_\mu$,
which are in thermodynamic equilibrium at a temperature $T = 1/\beta$. The
Schwinger-Keldysh action of the bath is then given by
\begin{multline}
\label{eq:bath-action}
  S_b = \sum_{\mu} \int_{t, t'} \left( b_{\mu,+}^{*}(t),
    b_{\mu,-}^{*}(t) \right) \\ \times
  \begin{pmatrix}
    G_{\mu}^{++}(t,t') &  G_{\mu}^{+-}(t,t') \\
    G_{\mu}^{-+}(t,t') & G_{\mu}^{--}(t,t')
  \end{pmatrix}^{-1}
  \begin{pmatrix}
    b_{\mu,+}(t') \\ b_{\mu,-}(t')
  \end{pmatrix},
\end{multline}
where the Green's functions
$i G_{\mu}^{\sigma \sigma'}(t, t') = \langle b_{\mu,\sigma}(t) b_{\mu,
  \sigma'}^{*}(t') \rangle$
for the oscillators of the bath are fixed by requiring it to be in equilibrium and therefore they
read~\cite{Kamenev2011,Altland/Simons}
\begin{equation}
  \label{eq:117}
  \begin{split}
    G_{\mu}^{+-}(t,t') & = -in(\omega_{\mu}) e^{-i\omega_{\mu} \left( t-t'
      \right)}, \\
    G_{\mu}^{-+}(t,t') & = -i \left( n(\omega_{\mu}) +1 \right)
    e^{-i\omega_{\mu}\left( t-t' \right)}, \\ G_{\mu}^{++}(t,t') & = \theta (t -
    t')
    G_{\mu}^{-+}(t,t') + \theta (t' - t) G_{\mu}^{+-}(t,t'),\\
    G_{\mu}^{--}(t,t') & = \theta (t' - t) G_{\mu}^{-+}(t,t') + \theta (t - t')
    G_{\mu}^{+-}(t,t').
  \end{split}
\end{equation}
Here $n(\omega) = 1/\left( e^{\beta \omega} - 1 \right)$ is the Bose
distribution function and $\theta(t)$ denotes the Heaviside step function, which
is defined as
\begin{equation}
  \label{eq:54}
  \theta(t) =
  \begin{cases}
    1, & t \geq 0, \\
    0, & t < 0.
  \end{cases}
\end{equation}
The coupling $S_{sb}$ between the system and the bath is assumed to be linear in
the bath variables and have a strength $\sqrt{\gamma_{\mu}}$,
\begin{multline}
  \label{eq:118}
  S_{sb} = \sum_{\mu} \sqrt{\gamma_{\mu}} \int_t \left( L_{+}^{*}(t)
    b_{\mu,+}(t) + L_{+}(t) b_{\mu,+}^{*}(t) \right. \\ \left. - L_{-}^{*}(t)
    b_{\mu,-}(t) - L_{-}(t) b_{\mu,-}^{*}(t) \right).
\end{multline}
Here, $L_{\pm}(t)$ are associated with the quantum jump or Lindblad operators,
which we assume to be quasilocal polynomials of the system's bosonic fields
$\{\psi_\pm, \psi^*_\pm\}$ resulting from normally ordered operators in a second
quantized description (e.g., the simplest choice would be
$L_{\pm}(t) = \psi_{\pm}(t, \mathbf{x})$).  In order to simplify the notation,
we do not indicate here the spatial dependence of the fields (both of the system
and of the bath), which is understood together with the corresponding
integration over space. We assume the harmonic oscillators which constitute the
bath to be spatially uncorrelated.  The Schwinger-Keldysh functional integral
with total action $S + S_b + S_{sb}$ involving both system and bath degrees of
freedom is quadratic in the latter and, therefore, the bath can be integrated
out. The resulting contribution is
\begin{multline}
  \label{eq:121}
  S' = - \int_{\omega_0 - \vartheta}^{\omega_0 + \vartheta} \mathrm{d} \omega \,
  \gamma(\omega) \nu(\omega)\int_{t, t'} \left( L_{+}^{*}(t), - L_{-}^{*}(t)
  \right) \\ \times
  \begin{pmatrix}
    G_{\omega}^{++}(t, t') & G_{\omega}^{+-}(t, t') \\
    G_{\omega}^{-+}(t, t') & G_{\omega}^{--}(t, t')
  \end{pmatrix}
  \begin{pmatrix}
    L_{+}(t')\\ -L_{-}(t')
  \end{pmatrix},
\end{multline}
which eventually sums to $S$.
In deriving this action, we made the additional assumption that the bath modes
$\{ \omega_\mu \}_\mu$ form a dense continuum with a spectral density
$\nu(\omega) = \sum_\mu \delta (\omega - \omega_\mu)$,
centered around a frequency $\omega_0$, with a bandwidth $\vartheta$ (see
further below for its interpretation).
Then, sums of the form $\sum_{\mu} \gamma_{\mu} \cdots$ can be approximated as
integrals over frequencies
$\int_{\omega_0 - \vartheta}^{\omega_0 + \vartheta} \mathrm{d} \omega \,
\gamma(\omega) \nu(\omega) \cdots$,
where $\gamma(\omega)$ describes the frequency distribution of the oscillator
strengths.
Inserting the explicit expressions~\eqref{eq:117} for the bath Green's functions into Eq.~\eqref{eq:121},
we obtain
\begin{widetext}
  \begin{multline}
  \label{eq:19}
  S' = - i \int_{\omega_0-\vartheta}^{\omega_0 + \vartheta} \frac{\mathrm{d}
    \omega}{2\pi} \gamma(\omega) \nu(\omega) \left(
    \vphantom{\int_{-\infty}^{\infty}} n(\omega) L_{+}^{*}(\omega) L_{-}(\omega)
    + \left( n(\omega) +1 \right) L_{-}^{*}(\omega) L_{+}(\omega) - \int_{-
      \infty}^{\infty} \frac{\mathrm{d} \omega'}{2 \pi} \left\{ \left[ \theta (
        \omega' - \omega) \left( n(\omega) + 1 \right) \right. \right. \right. \\
    \left. \left. \left. + \theta (- \omega' + \omega) n(\omega) \right]
        L_{+}^{*}(\omega') L_{+}(\omega') + \left[ \theta (-\omega' + \omega)
          \left( n(\omega) + 1 \right) + \theta (\omega' - \omega) n(\omega)
        \right] L_{-}^{*}(\omega') L_{-}(\omega') \right\}
      \vphantom{\int_{-\infty}^{\infty}} \right),
\end{multline}
\end{widetext}
where $L_{\sigma}(\omega)$ describe the quantum jump operators in the frequency
space, and
$\theta(\omega) = i \mathcal{P} \frac{1}{\omega} + \pi \delta(\omega)$ (where
$\mathcal{P} $ denotes the Cauchy principal value) is the
Fourier transform of $\theta(t)$ in Eq.~\eqref{eq:54}.
The terms involving the principal value contribute to the Hamiltonian part of
the total Schwinger-Keldysh action of the system $S+S'$. 
Assuming that the jump operators are quasilocal polynomials of the bosonic field
operators of the system, $L_{\sigma}(\omega)$ transform under
$\mathcal{T}_{\beta'}$ as the field operators, i.e.,
\begin{equation}
  \mathcal{T}_{\beta'} L_{\sigma}(\omega) = e^{-\sigma \beta' \omega/2}
  L_{\sigma}^{*}(\omega)\quad \mbox{and}\quad \mathcal{T}_{\beta'} L_{\sigma}^{*}(\omega) = e^{\sigma
    \beta' \omega/2} L_{\sigma}(\omega);
    \label{eq:19b}
\end{equation}
inserting these expressions in Eq.~\eqref{eq:19} one finds that the
contour-diagonal terms (i.e., those proportional to
$L^{*}_{\sigma}(\omega) L_{\sigma}(\omega)$ with $\sigma = \pm 1$, which
include, in particular, the above-mentioned principal value terms) are invariant
due to frequency conservation (cf.\ the discussion in
Sec.~\ref{sec:coherent-vertex}).
On the other hand, for the contribution which is off-diagonal in the
Schwinger-Keldysh contour one finds
\begin{multline}
  S_{\mathrm{off-diag}}'[\mathcal{T}_{\beta'} \Psi] = - i
  \int_{\omega_0-\vartheta}^{\omega_0 + \vartheta} \frac{\mathrm{d}
    \omega}{2\pi} \gamma(\omega) \nu(\omega) \\ \times \left[ n(\omega)
    e^{\beta' \omega} L_{+}(\omega) L_{-}^{*}(\omega) + \left( n(\omega) +1
    \right) e^{-\beta' \omega} L_{-}(\omega) L_{+}^{*}(\omega) \right].
\label{eq:19c}
\end{multline}
If the value of $\beta'$ matches the inverse temperature $\beta = 1/T$ of the
bath modes, encoded in $n(\omega)$, it is easy to see that these terms are
invariant under $\mathcal{T}_{\beta}$ because
$n(\omega) e^{\beta \omega} = n(\omega) + 1$. In summary, one concludes that
$S'[\mathcal{T}_{\beta} \Psi] = S'[\Psi]$ and being also the action $S$ of the
system in isolation invariant under $\mathcal{T}_{\beta}$, the same holds for
the total effective action $S+S'$ of the system in contact with the thermal
bath.

In order to understand the effect of the Markovian approximation in the driven
context on the invariance under $\mathcal{T}_{\beta}$, let us now consider
Eq.~\eqref{eq:19} after this approximation has been done. In particular, in
order for these approximations to be valid, we assume that one can choose a
``rotating'' frame in which the evolution of the system is slow compared to the
energy scales $\omega_0$ and $\vartheta$ which characterize the bath.
This is possible if the system is driven by an external classical field such as
a laser, so that the frequency of the drive bridges the gap between the natural
time scales of the system and those of the bath.
Then, all jump operators in Eq.~\eqref{eq:121} may be evaluated at the same time
$t$, since the integral kernel in Eq.~\eqref{eq:121} (i.e., the product of bath
  Green's functions $G_\omega^{\sigma\sigma'}$, density of states $\nu(\omega)$
  and oscillator strength distribution $\gamma(\omega)$, integrated over the
  bath bandwidth) differs from zero only within a correlation time
  $\tau_c \approx 1/\vartheta$, which is assumed to be much shorter than the
  timescale over which $L_{\sigma}(t)$ evolves in the rotating frame.
  Additionally we assume that the spectral density $\nu(\omega)$ of the states of
  the bath and the corresponding coupling strength $\sqrt{\gamma(\omega)}$ to
  the system do not vary appreciably within the relevant window
  $\omega_0-\vartheta < \omega < \omega_0+\vartheta$, such that one
  can set $\gamma(\omega) \nu(\omega) \approx \gamma(\omega_0) \nu(\omega_0)$.
As before, the terms in Eq.~\eqref{eq:19} which are diagonal in the contour
indices are invariant under $\mathcal{T}_{\beta}$ (see Eq.~\eqref{eq:19b}) also
after the Markovian approximation; accordingly, we focus on the off-diagonal
terms in Eq.~\eqref{eq:19c}, which become
\begin{multline}
  S_{\mathrm{off-diag}}' [\Psi] = - i \gamma(\omega_0) \nu(\omega_0) \\ \times \int_{-
    \infty}^{\infty} \frac{\mathrm{d} \omega}{2 \pi} \left[ \bar{n}
    L_{+}^{*}(\omega) L_{-}(\omega) + \left( \bar{n} + 1
    \right) L_{-}^{*}(\omega) L_{+}(\omega) \right],
    \label{eq:19d}
\end{multline}
where $\bar{n} = n(\omega_0)$ is the occupation number of the bath modes at
frequency $\omega_0$. This makes it clear that $\vartheta$ acts as a
  high-frequency cutoff, whose precise value, under Markovian conditions, does
  not affect the physics.
Applying the transformation $\mathcal{T}_{\beta}$ to the fields one has
\begin{multline}
  S_{\mathrm{off-diag}}'[\mathcal{T}_{\beta} \Psi] = - i \gamma(\omega_0) \nu(\omega_0)
  \\ \times \int_{-\infty}^{\infty} \frac{\mathrm{d} \omega}{2 \pi} \left[ \bar{n} e^{\beta \omega} L_{+}(\omega)
    L_{-}^{*}(\omega) + \left( \bar{n} + 1 \right) e^{-\beta \omega}
    L_{-}(\omega) L_{+}^{*}(\omega) \right].
\end{multline}
In order for $S_{\mathrm{off-diag}}'$ to be invariant under
$\mathcal{T}_{\beta}$, this expression should be equal to
$S_{\mathrm{off-diag}}' [\Psi]$ in Eq.~\eqref{eq:19d}, which requires
$\bar{n} e^{\beta \omega} = \bar{n} +1$ for all values of the frequencies
$\omega$ within the relevant region
$\omega_0-\vartheta < \omega < \omega_0+\vartheta$. Clearly, this is not
possible and therefore the equilibrium symmetry is explicitly broken by the
Markovian approximation in the driven context.

\subsection{System coupled to different baths}
\label{sec:two-baths}
A simple way to drive a system out of equilibrium is to bring it in contact with
baths at different temperatures and chemical potentials. In this case, a net
flux of energy and particles is established across the system, preventing it
from thermalizing and, consequently, causing a violation of the symmetry
$\mathcal{T}_{\beta}$. This scenario occurs, e.g., in the context of quantum
electronics in a quantum dot connected by tunnel electrodes to two leads,
between which a finite voltage difference is maintained. Here, for simplicity,
we consider a minimal bosonic counterpart of this system, constituted by a
single bosonic mode --- described by the fields $\{\psi_\pm,\psi^*_\pm\}$
coupled to two baths of non-interacting harmonic oscillators kept at different
temperatures and chemical potentials. We show explicitly that the
non-equilibrium nature of this setup is accompanied by a violation of the
equilibrium symmetry. The generalization of this argument to a multi-mode system
or to a larger number of baths is straightforward.

We consider two baths of non-interacting harmonic oscillators $b_{i,\nu}$, where
the label $i = 1,2$ denotes the bath to which the operator belongs, while $\nu$
denotes the corresponding mode with frequency $\omega_{i,\nu}$. Each bath is
assumed to be in thermodynamic equilibrium with different inverse temperatures
$\beta_1$, $\beta_2$ and different chemical potentials $\mu_1$, $\mu_2$. The
Schwinger-Keldysh action $S_{b,i}$ of each bath takes the same form as in
Eq.~\eqref{eq:bath-action}. However, in the present case, the distribution
functions $n_i(\omega)$ entering the bath Green's functions depend on the
chemical potentials $\mu_i$ as
$n_i(\omega) = 1/(e^{\beta_i \left( \omega-\mu_i \right)} - 1)$. As in the
previous section, the bath is assumed to be coupled linearly to the system
variables $L_{\pm}(t)$ which are quasilocal polynomials of the bosonic fields
$\{\psi_\pm,\psi^*_\pm\}$. The dynamics of the system and the baths is then
controlled by a functional integral with the total Schwinger-Keldysh action
$S+S_{b,1}+S_{b,2}+ S_{sb}$, where $S$ is to the action of the system, i.e., the
single bosonic mode, and $S_{sb}$ is the system-bath coupling. As in the
previous section, an effective dynamics for the system's variables can be
obtained by integrating out those of the bath.
This yields an effective action
\begin{widetext}
  \begin{multline}
  \label{eq:eff-action2}
  S' = - i \int_{-\infty}^{+\infty} \frac{\mathrm{d} \omega}{2\pi}
  \tilde{J}(\omega) \left( \vphantom{\int_{-\infty}^{\infty}} \tilde{n}(\omega)
    L_{+}^{*}(\omega) L_{-}(\omega) + \left( \tilde{n}(\omega) +1 \right)
    L_{-}^{*}(\omega) L_{+}(\omega) - \int_{- \infty}^{\infty} \frac{\mathrm{d}
      \omega'}{2 \pi} {\bigg\{ } \left[ \theta (
        \omega' - \omega) \left( \tilde{n}(\omega) + 1 \right) \right. \right. \\
   \left. \left. + \theta (- \omega' + \omega) \tilde{n}(\omega) \right]
      L_{+}^{*}(\omega') L_{+}(\omega') + \left[ \theta (-\omega' + \omega)
        \left( \tilde{n}(\omega) + 1 \right) + \theta (\omega' - \omega)
        \tilde{n}(\omega) \right] L_{-}^{*}(\omega') L_{-}(\omega') {\bigg\}}
    \vphantom{\int_{-\infty}^{\infty}} \right).
\end{multline}
\end{widetext}
The action $S'$ is formally similar to the one in Eq.~\eqref{eq:19}, with the
spectral density $\gamma(\omega) \nu(\omega)$ replaced by the sum of the
spectral densities of the baths $\tilde{J}(\omega) = J_1(\omega)+J_2(\omega)$,
where $J_i(\omega) = \gamma_i(\omega)\nu_i(\omega)$ (with $\nu_i(\omega)$ and
$\gamma_i(\omega)$ defined as in Sec.~\ref{sec:non-equil-nature}, see after
Eq. \eqref{eq:121}). Analogously, the distribution function $n(\omega)$ of the
single bath we considered in Sec.~\ref{sec:non-equil-nature} is replaced by the
average of the distribution functions of the two baths $i = 1,2$ weighted by the
relative spectral densities, i.e.,
\begin{equation}
\label{eq:average-population}
\tilde{n}(\omega) = \frac{J_1(\omega)}{J_1(\omega)+J_2(\omega)} n_1(\omega) +
\frac{J_2(\omega)}{J_1(\omega)+J_2(\omega)}n_2(\omega).
\end{equation}
Now, we consider how the effective action \eqref{eq:eff-action2} transforms
under the thermal symmetry $\mathcal{T}_{\beta}$. Since here we are explicitly
considering the presence of chemical potentials, we will use the generalization
of the symmetry $\mathcal{T}_{\beta,\mu}$ in Eq. \eqref{eq:4}. As discussed in
Sec.~\ref{sec:non-equil-nature}, since $L_\pm, L_\pm^*$ are quasilocal
polynomials of the bosonic fields of the system, they transform under
$\mathcal{T}_{\beta,\mu}$ as
\begin{equation}
\begin{split}
  \mathcal{T}_{\beta,\mu} L_{\sigma}(\omega) &= e^{-\sigma \beta( \omega-\mu)/2}L_{\sigma}^{*}(\omega), \\
   \mathcal{T}_{\beta,\mu} L_{\sigma}^{*}(\omega)& = e^{\sigma \beta( \omega-\mu)/2} L_{\sigma}(\omega).
\end{split}  
\end{equation}
Accordingly, the products $L_\sigma(\omega)L^*_\sigma(\omega)$ are invariant
under the symmetry and therefore the contour-diagonal part of $S'$, which
contains such terms, is invariant. On the other hand, the part
$S_{\mathrm{off-diag}}'$ of $S'$ which is off-diagonal in the Schwinger-Keldysh
contour (i.e., the first two terms on the r.h.s. of Eq. \eqref{eq:eff-action2})
is modified as:
\begin{multline}
  S_{\mathrm{off-diag}}'[\mathcal{T}_{\beta,\mu} \Psi] = - i
  \int_{-\infty}^{+\infty} \frac{\mathrm{d} \omega}{2\pi} \tilde{J}(\omega) \\
  \times \left[ \tilde{n}(\omega) e^{\beta(\omega-\mu)} L_{+}(\omega)
    L_{-}^{*}(\omega) \right. \\ \left. + \left( \tilde{n}(\omega) +1 \right)
    e^{-\beta(\omega-\mu)} L_{-}(\omega) L_{+}^{*}(\omega) \right].
\end{multline}
Comparing this expression with Eq. \eqref{eq:eff-action2}, one readily sees that
the invariance of this term under $\mathcal{T}_{\beta,\mu}$ requires
$\tilde{n}(\omega)e^{\beta(\omega-\mu)} = \tilde{n}(\omega) + 1$, and therefore
$S_{\mathrm{off-diag}}'$ is not invariant under $\mathcal{T}_{\beta,\mu}$,
unless the two baths have the same temperature and chemical potential, i.e.,
$\beta_1=\beta_2 = \beta$ and $\mu_1 = \mu_2 = \mu$.  In this case, one can
easily verify from Eq.~\eqref{eq:average-population} that the average
distribution function $\tilde{n}(\omega)$ is just the Bose-Einstein
distribution $\tilde{n}(\omega) = 1/(e^{\beta \left( \omega-\mu \right)}-1)$
and, as a consequence,
$\tilde{n}(\omega)e^{\beta(\omega-\mu)} = \tilde{n}(\omega) + 1$.

In conclusion, when the system is driven out of equilibrium by a net flux of
energy or particles, induced by a difference between the temperatures or the
chemical potentials of the baths, the total action of the system is no longer
invariant under $\mathcal{T}_{\beta,\mu}$, as
$S'[\mathcal{T}_{\beta,\mu} \Psi] \neq S'[\Psi]$.

\subsection{Further applications}
\label{sec:further-applications}
Among the various possible applications of the symmetry $\mathcal{T}_{\beta}$,
we mention here:

\paragraph{Symmetry-preserving approximations.}

Properties of interacting many-body systems can usually be obtained only by
resorting to certain approximations. Then, while FDRs for correlation and
response functions can be established exactly in the absence of interactions, in
an approximate inclusion of the latter one has to make sure that the FDRs are
not broken. In other words, the approximation should conserve the thermal
symmetry. This requirement for classical statistical systems has been
implemented in the mode-coupling theory of the glass transition in
Ref.~\onlinecite{Andreanov2006}.

The field-theoretic formalism provides the natural framework for studying the
behavior of systems at long wavelengths and low energies by employing
renormalization-group methods. In any of these methods, an effective description
which is obtained by integrating out fast fluctuations must have the same
symmetries as those present at microscopic scales. For example, in the case of
the functional renormalization group (for reviews see
Refs.~\onlinecite{berges02:_nonper,pawlowski07:_aspec,delamotte08:_introd_nonper_renor_group,rosten12:_fundam,boettcher12:_ultrac_funct_renor_group}),
this is achieved by choosing an ansatz to approximate the scale-dependent
effective action which incorporates these symmetries. In this context, the
classical limit of the thermal symmetry discussed here has been used in
functional renormalization group studies of model A~\cite{Canet2007} and model
C~\cite{Mesterhazy2013}. The quantum thermal symmetry, instead, is analogously
preserved by the ansatz for the effective action chosen in
Ref.~\cite{Mesterhazy2015} (in the form of FDT), where the scale-dependent
crossover from quantum to classical dynamics is studied. Alternatively, one can
devise approximation schemes which are compatible with the equivalent KMS
conditions, as discussed in detail in Ref.~\onlinecite{Jakobs2010}. Note,
however, that in this work the KMS condition in the form of Eq.~\eqref{eq:52} is
imposed on the scale-dependent Green's functions (supplemented by the
corresponding condition on the vertex functions). On the other hand, the
symmetry constraint can directly be applied to the effective action itself,
which is the generating functional of vertex functions and contains information
on all correlation functions.

\paragraph{Fluctuation relations.}

Another concrete example of the usefulness of the thermal symmetry is provided
by the derivation of transient fluctuation relations~\cite{Campisi2011,
  Esposito2014} for time-dependent particle transport in
Ref.~\cite{Altland2010}. There, the symmetry is generalized in order to account
for the presence of a time-dependent counting field which probes the current
flowing through the system. This generalized symmetry yields a relation
analogous to Eq.~\eqref{eq:52} (however, formulated in terms of the generating
functional for correlation and response functions), from which, e.g., a
fluctuation relation for the probability distribution of work done on the system
and the transmitted charge can be derived.

\section{Conclusions}
\label{sec:conclusions}
 
We demonstrated here that the Schwinger-Keldysh action describing the dynamics
of a generic quantum many-body system acquires a certain symmetry
$\mathcal{T}_{\beta}$ if the evolution occurs in thermal equilibrium.  To a
certain extent, this symmetry was discussed in Ref.~\onlinecite{Altland2010} in
the specific context of fluctuation relations for particle transport. We traced
the origin of this symmetry back to the Kubo-Martin-Schwinger (KMS) condition
which establishes a relationship between multi-time correlation functions in
real and imaginary times of a system in canonical equilibrium at a certain
temperature. Fluctuation-dissipation relations are then derived as the
Ward-Takahashi identities associated with $\mathcal{T}_{\beta}$. Remarkably, in
the classical limit, this equilibrium symmetry reduces to the one known in
classical stochastic systems, where it was derived from the assumption of
detailed balance.
By comparing with this classical case, important questions on the nature of
equilibrium in quantum systems arise. In particular, while microreversibility
and detailed balance of the dynamics are deeply connected to the notion of
equilibrium in classical stochastic systems, an analogous relationship for
quantum systems does not clearly emerge and surely deserves further investigation.

The equilibrium symmetry $\mathcal{T}_{\beta}$ is expected to play a crucial
role in the study of thermalization in quantum systems, in particular when
combined with a renormalization-group analysis. In fact, on the one hand, it
provides a simple but powerful theoretical tool to assess whether a certain
system is able to reproduce thermal equilibrium. This can indeed be accomplished
by a direct inspection of the microscopic Schwinger-Keldysh action (or of the
effective one generated after integrating out some degrees of freedom, e.g.,
along a renormalization-group flow) which describes the dynamics of the system,
rather than checking, for instance, the validity of the fluctuation-dissipation
relations among various correlation functions.
On the other hand, the equilibrium symmetry might be useful also in order to
investigate or characterize possible departures from equilibrium and, in this
respect, it would be interesting to consider the case in which the system
evolves in a generalized Gibbs
ensemble~\cite{Rigol2007,Iucci2009,Jaynes1957,Barthel2008,Goldstein2014,Pozsgay2014,Mierzejewski2014,Wouters2014,Essler2014}.
Finally, while we focussed here on the case of bosons, the extension of our
analysis to different statistics, for instance fermionic and spin systems,
represents an interesting issue.

{\bf Acknowledgements.} --- We would like to thank A. Altland, J. Berges,
I. Carusotto, L. F. Cugliandolo, T. Gasenzer, J. M. Pawlowski, T. Prosen,
H. Spohn, C. Wetterich, and especially A. Rosch for useful
discussions. L. M. S. and S. D. gratefully acknowledge support by the START
Grant No. Y 581-N16. S. D. also thanks the German Research Foundation for
support via ZUK 64.

\appendix

\section{Invariance of quadratic dissipative contributions}
\label{sec:invariance-s_d}

In order to show explicitly the invariance of $S_d$ in Eq.~\eqref{eq:17} under
the transformation $\mathcal{T}_{\beta}$ in Eq.~\eqref{eq:3}, it is convenient
to assume that $h\oq$ as a function of the frequency $\omega$ has definite
parity and to consider separately the cases of odd and even functions,
$h_o(-\omega, \mathbf{q}) = - h_o(\omega, \mathbf{q})$ and
$h_e(-\omega, \mathbf{q}) = h_e(\omega, \mathbf{q})$, respectively.
The generic case follows straightforwardly by linear combination. 
We then consider how the actions
\begin{multline}
  \label{eq:18}  
  S_d = i \epsilon \int_{\omega, \mathbf{q}}
  \Phi_q^{\dagger}\oq
  \begin{Bmatrix}
    h_o\oq \\ h_e\oq \sigma_z
  \end{Bmatrix}
  \\ \times \left( \Phi_c\oq + \ct \Phi_q\oq \right)
\end{multline}
--- where $\Phi$ is introduced in Sec.~\ref{sec:dynamical-term}, see
Eq.~\eqref{eq:44} --- with a certain $\beta$ transform under a transformation
$\mathcal{T}_{\beta'}$ [see Eq.~\eqref{eq:3}] of the fields, with a generic
parameter $\beta'$. One finds
\begin{widetext}
  \begin{multline}
  S_d[\mathcal{T}_{\beta'} \Phi] = i \epsilon \int_{\omega, \mathbf{q}} \left( \sinh(\beta'
    \omega/2) \Phi_c^{\dagger}(-\omega,\mathbf{q}) + \cosh(\beta' \omega/2)
    \Phi_q^{\dagger}(-\omega,\mathbf{q}) \right) \sigma_x
  \begin{Bmatrix}
    h_o\oq \\ h_e \oq \sigma_z
  \end{Bmatrix}
  \sigma_x \\ \times \left[ \cosh(\beta' \omega/2) \Phi_c(-\omega,\mathbf{q}) -
    \sinh(\beta' \omega/2) \Phi_q(-\omega,\mathbf{q}) \ct
    \left( - \sinh(\beta' \omega/2) \Phi_c(-\omega,\mathbf{q}) + \cosh(\beta'
      \omega/2) \Phi_q(-\omega,\mathbf{q}) \right) \right].
\end{multline}
Note that the terms involving solely the classical field spinor $\Phi_c$ cancel
each other only if $\beta' = \beta$.
Otherwise, terms $\propto \Phi_c^\dag \Phi_c$ remain, which lead to a violation
of causality~\cite{Kamenev2011}.
For $\beta' = \beta$ instead, we obtain
\begin{multline}
  S_d[\mathcal{T}_{\beta} \Phi] = -i \epsilon
  \int_{\omega, \mathbf{q}} \left( - \sh \Phi_c^{\dagger}(\omega,\mathbf{q}) + \ch
    \Phi_q^{\dagger}(\omega,\mathbf{q}) \right)
  \begin{Bmatrix}
    h_o \oq \\ h_e \oq \sigma_z
  \end{Bmatrix}
  \\ \times \left( \sh - \ct \ch \right) \Phi_q(\omega,\mathbf{q}).
\end{multline}
\end{widetext}
By means of the identity 
\begin{equation}
  \sinh x  - \coth x  \cosh x  = - 1/\sinh x ,
\end{equation}
and after some straightforward algebraic manipulations one eventually finds that
$S_d[\mathcal{T}_{\beta} \Phi] = S_d[\Phi]$, i.e., that $S_d$ (with a certain
$\beta$) is invariant under $\mathcal{T}_{\beta}$.

\section{Invariance of dissipative vertices}
\label{sec:invar-diss-vert}

As pointed out in the main text, a first constraint that has to be imposed on
the functions $f_{1,2,3}$ appearing in the dissipative vertex in
Eq.~\eqref{eq:39} follows from the requirement of causality of the
Schwinger-Keldysh action. The latter must vanish when
$\psi_{+} = \psi_{-}$~\cite{Kamenev2011}, which implies the condition
\begin{multline}
  \label{eq:30}  
  \int_{\omega_1, \ldots ,\omega_4} \!\!\!\!\!\!\!\!\ \delta(\omega_1 - \omega_2
  + \omega_3 - \omega_4) \psi_{+}^*(\omega_1) \psi_{+}(\omega_2)
  \psi_{+}^*(\omega_3) \psi_{+}(\omega_4) \\ \times \left[
    f_1(\omega_1,\omega_2,\omega_3,\omega_4)+
    f_2(\omega_1,\omega_2,\omega_3,\omega_4)  \right. \\ \left.  +
    f_3(\omega_1,\omega_2,\omega_3,\omega_4) \right] = 0.    
\end{multline}
Now let us consider Eq.~\eqref{eq:39} with the transformed fields
$\mathcal{T}_{\beta} \Psi_{\sigma}$, i.e., $S_d[\mathcal{T}_{\beta}
\Psi]$. Requiring it to be equal to $S_d[\Psi]$, 
we find that the following conditions should be fulfilled:
\begin{equation}
  \label{eq:32}
  \begin{split}
    & f_1(\omega_1,\omega_2,\omega_3,\omega_4) -
    f_1(\omega_2,\omega_1,\omega_4,\omega_3) = 0, \\
    & f_2(\omega_1,\omega_2,\omega_3,\omega_4) -
    f_2(\omega_2,\omega_1,\omega_4,\omega_3) = 0, \\
    & f_3(\omega_1,\omega_2,\omega_3,\omega_4) - e^{\beta \left( \omega_1 -
        \omega_2 \right)} f_3(\omega_2,\omega_1,\omega_4,\omega_3) = 0,    
  \end{split}
\end{equation}
where we used the conservation of frequencies implied by the $\delta$-function
in Eq.~\eqref{eq:39} to simplify the exponent in the last line. Specifically,
the necessary conditions are that the expressions on the l.h.s.~of these
relations should vanish when integrated over frequencies after having been
multiplied by the corresponding combinations of fields in Eq.~\eqref{eq:39} and
by the $\delta$-function on frequencies.
The relations in Eq.~\eqref{eq:32} are, however, sufficient conditions
for the equality of $S_d[\mathcal{T}_{\beta} \Psi]$ and $S_d[\Psi]$.

To begin with, we investigate the possible existence of a frequency-independent
solution of Eqs.~\eqref{eq:30} and \eqref{eq:32} for $f_{1,2,3}$; these two
equations then imply
\begin{equation}
  \label{eq:37}
  f_1 = - f_2 = \mathrm{constant} \quad \mbox{and} \quad f_3 = 0.
\end{equation}
However, this solution can be seen to lack physical relevance for the following
reason: any physically sensible dissipative contribution to the
Schwinger-Keldysh action compatible with the thermal symmetry can be considered
as originating from integrating out a thermal bath which is appropriately
coupled to the system. Anticipating the discussion of
Sec.~\ref{sec:non-equil-nature}, we note that such dissipative contributions
always involve terms which are not diagonal in the contour indices (cf.\
Eq.~\eqref{eq:121}). $S_d$ with $f_{1,2,3}$ given by Eq.~\eqref{eq:37}, however,
is \emph{not} of this form. In fact, inserting Eq.~\eqref{eq:37} in
Eq.~\eqref{eq:39} yields a vertex that is equal to the two-body interaction in
Eq.~\eqref{eq:29} apart from an overall factor of $i$, i.e., such a vertex would
originate from an \emph{imaginary} two-body coupling in a Hamiltonian. Clearly,
this would violate hermiticity, rendering the Hamiltonian unphysical.

While this demonstrates that --- as anticipated in the main text --- a
frequency-independent number-conserving quartic vertex is not compatible with
equilibrium conditions, solutions of Eqs.~\eqref{eq:30} and~\eqref{eq:32}
\emph{do} exist with $f_i$ depending on frequency.
One particular solution is given by Eq.~\eqref{eq:42} of the main text.

\section{Representation of correlation functions in the Schwinger-Keldysh
  formalism}
\label{sec:repr-corr-funct}

\paragraph{Two-time correlation functions.}

In order to derive the representation of a two-time correlation function in the
Schwinger-Keldysh formalism reported in Eq.~\eqref{eq:22}, we insert the
explicit expressions~\eqref{eq:24} for the Heisenberg operators $A(t_A)$ and
$B(t_B)$ in the trace which defines the l.h.s.~of Eq.~\eqref{eq:22} according to
Eq.~\eqref{eq:16}. Then, by introducing an additional and arbitrary time $t_f$
such that $t_i < t_{A, B} < t_f$, and by using the cyclic property of the trace
one can write
\begin{equation}
  \label{eq:25}
  \langle A(t_A) B(t_B) \rangle = \tr
  \left( e^{-i H \left( t_f - t_B \right)} B e^{-i H \left( t_B - t_i \right)}
    \rho e^{i H \left( t_A - t_i \right)} A e^{i H \left( t_f - t_A \right)}
  \right).
\end{equation}
The evolution of the density matrix is adjoint to the evolution of Heisenberg
operators, i.e., $\rho(t) = e^{-i Ht}\rho e^{i Ht}$. Thus, the operator
$e^{-i H ( t - t' )}$ ($e^{i H ( t - t')}$) acting from the left (right) on the
density matrix $\rho$ corresponds to the evolution in time from $t'$ to $t$. In
the correlation function~\eqref{eq:25}, the time evolution from $t_i$ to $t_f$
on the left (right) of $\rho$ is intercepted by the operator $B$ at time $t_B$
($A$ at time $t_A$).
In order to convert the r.h.s.~of Eq.~\eqref{eq:25} into a path integral, the
standard procedure (see, e.g., Refs.~\onlinecite{Kamenev2011,Altland/Simons}) to be
followed consists in writing the exponentials of the evolution operators as
infinite products of infinitesimal and subsequent temporal evolutions (Trotter
decomposition), in-between of which one can introduce completeness relations in
terms of coherent states carrying the additional label ``$+$'' on the left of the
density matrix, and a ``$-$'' on its right. These coherent states are eventually
labeled by a temporal index on the forward ($+$) and backward ($-$) branches of
the close-time path which characterizes the resulting action. Correspondingly,
the operators on the left and on the right of the density matrix ($B$ and $A$,
respectively, in Eq.~\eqref{eq:25}) turn out to be evaluated on the fields
(i.e., coherent states) which are defined, respectively, on the forward and
backward branches of the closed time path and this yields immediately the
equality in Eq.~\eqref{eq:22}, where the ordering of the matrix elements $A_-$
and $B_+$ on its r.h.s.~is inconsequential.
For the sake of completeness, we note that the expression as a Schwinger-Keldysh
functional integral of a two-time function is not unique: in fact, it is
straightforward to check that, by rearranging operators in Eq.~\eqref{eq:25},
one can equivalently arrive at
\begin{equation}
  \langle A(t_A) B(t_B) \rangle =
  \begin{cases}
    \langle A_{+}(t_A) B_{+}(t_B) \rangle & \text{for } t_A > t_B, \\
    \langle A_{-}(t_A) B_{-}(t_B) \rangle & \text{for } t_A < t_B. \\
  \end{cases}
\end{equation}
However, as discussed below, the choice of Eq.~\eqref{eq:22} naturally lends
itself to a generalization to multi-time correlation functions.

\paragraph{Multi-time correlation functions.}

The functional integral on the r.h.s.~of Eq.~\eqref{eq:27} relation can be
constructed from a straightforward generalization of Eq.~\eqref{eq:25}: after a
reshuffling of the operators such that $A$ and $B$ appear respectively on the
left and right of the density matrix --- as explained above --- the temporal
evolution can be artificially extended from $t_i$ to $t_f$ and it is intercepted
on the l.h.s.~of the density matrix by operators $b_1, \dotsc, b_M$ at times
$t_{B,1}, \dotsc, t_{B,M}$ and on the r.h.s.~by operators $a_1, \dotsc, a_N$ at
times $t_{A,1}, \dotsc, t_{A,N}$. Again, the resulting expression for the
correlation function can be converted directly into a path integral by a Trotter
decomposition of the subsequent evolutions and by inserting completeness
relations in terms of coherent states carrying the label ``$+$'' corresponding
to the forward contour on the l.h.s.~of the density matrix and the label ``$-$''
for the backward contour on the r.h.s., which eventually leads to
Eq.~\eqref{eq:27}.

\section{Jacobian of the equilibrium transformation}
\label{sec:jacob-equil-transf}

In order to prove that $\abs{\Det \mathcal{J} } = 1$ it is convenient to
calculate the Jacobian $\mathcal{J}$ associated with Eq.~\eqref{eq:1} in
frequency and momentum space, which reads
\begin{equation}
  \label{eq:28}
  \mathcal{J}(\omega,\mathbf{q},\omega',\mathbf{q}') =
  \left( 2 \pi \right)^d \delta^{(d)}(\mathbf{q} + \mathbf{q}') J(\omega,\omega'),
\end{equation}
where
\begin{equation}
  J(\omega,\omega') = 2 \pi \delta(\omega - \omega')
  \begin{pmatrix}
    0 & e^{\beta \omega/2} & 0 & 0 \\
    e^{-\beta \omega/2} & 0 & 0 & 0 \\
    0 & 0 & 0 & e^{\beta \omega/2} \\
    0 & 0 & e^{-\beta \omega/2} & 0
  \end{pmatrix}.
\end{equation}
The eigenvectors $v_i$ and eigenvalues $\lambda_i$ of the frequency-dependent
part, i.e., the solutions of the equation
\begin{equation}
  \int \frac{\mathrm{d} \omega'}{2 \pi} J(\omega,\omega') v_i(\omega') = \lambda_i
  v_i(\omega),
\end{equation}
are
\begin{equation}
  \begin{aligned}
    v_1(\omega) & = \left( 0, 0, -e^{\beta \omega/2}, 1
    \right)^T, & v_2(\omega) & = \left( -e^{\beta \omega/2},1, 0, 0 \right)^T, \\
    v_3(\omega) & = \left( 0, 0, e^{\beta \omega/2}, 1 \right)^T, & v_4(\omega) & =
    \left( e^{- \beta \omega/2}, 1, 0, 0 \right)^T,
  \end{aligned}
\end{equation}
with $\lambda_1 = \lambda_2 = -1$, and $\lambda_3 = \lambda_4 = 1$, so that
$\Det J =\lambda_1 \lambda_2\lambda_3\lambda_4= 1$.
As for the momentum-dependent part of the Jacobian matrix
Eq.~\eqref{eq:28}, we note that its eigenvectors can be constructed with any
function $f(\mathbf{q})$ by taking the even and odd combinations
$f(\mathbf{q}) \pm f(-\mathbf{q})$:
\begin{equation}
  \int_{\mathbf{q}'} \left( 2 \pi \right)^d \delta^{(d)}(\mathbf{q} +
  \mathbf{q}') \left( f(\mathbf{q}') \pm f(-\mathbf{q}') \right) = \pm \left(
    f(\mathbf{q}) \pm f(-\mathbf{q}) \right).
\end{equation}
Thus the eigenvalues of this part are $\pm 1$, and hence the absolute value of
the Jacobian matrix is $\abs{\Det \mathcal{J}} = 1$.

\bibliography{thermal_symmetry_bibliography}

\begin{thebibliography}{107}%
\makeatletter
\providecommand \@ifxundefined [1]{%
 \@ifx{#1\undefined}
}%
\providecommand \@ifnum [1]{%
 \ifnum #1\expandafter \@firstoftwo
 \else \expandafter \@secondoftwo
 \fi
}%
\providecommand \@ifx [1]{%
 \ifx #1\expandafter \@firstoftwo
 \else \expandafter \@secondoftwo
 \fi
}%
\providecommand \natexlab [1]{#1}%
\providecommand \enquote  [1]{``#1''}%
\providecommand \bibnamefont  [1]{#1}%
\providecommand \bibfnamefont [1]{#1}%
\providecommand \citenamefont [1]{#1}%
\providecommand \href@noop [0]{\@secondoftwo}%
\providecommand \href [0]{\begingroup \@sanitize@url \@href}%
\providecommand \@href[1]{\@@startlink{#1}\@@href}%
\providecommand \@@href[1]{\endgroup#1\@@endlink}%
\providecommand \@sanitize@url [0]{\catcode `\\12\catcode `\$12\catcode
  `\&12\catcode `\#12\catcode `\^12\catcode `\_12\catcode `\%12\relax}%
\providecommand \@@startlink[1]{}%
\providecommand \@@endlink[0]{}%
\providecommand \url  [0]{\begingroup\@sanitize@url \@url }%
\providecommand \@url [1]{\endgroup\@href {#1}{\urlprefix }}%
\providecommand \urlprefix  [0]{URL }%
\providecommand \Eprint [0]{\href }%
\providecommand \doibase [0]{http://dx.doi.org/}%
\providecommand \selectlanguage [0]{\@gobble}%
\providecommand \bibinfo  [0]{\@secondoftwo}%
\providecommand \bibfield  [0]{\@secondoftwo}%
\providecommand \translation [1]{[#1]}%
\providecommand \BibitemOpen [0]{}%
\providecommand \bibitemStop [0]{}%
\providecommand \bibitemNoStop [0]{.\EOS\space}%
\providecommand \EOS [0]{\spacefactor3000\relax}%
\providecommand \BibitemShut  [1]{\csname bibitem#1\endcsname}%
\let\auto@bib@innerbib\@empty
\bibitem [{\citenamefont {Polkovnikov}\ \emph {et~al.}(2011)\citenamefont
  {Polkovnikov}, \citenamefont {Sengupta}, \citenamefont {Silva},\ and\
  \citenamefont {Vengalattore}}]{Polkovnikovrev}%
  \BibitemOpen
  \bibfield  {author} {\bibinfo {author} {\bibfnamefont {A.}~\bibnamefont
  {Polkovnikov}}, \bibinfo {author} {\bibfnamefont {K.}~\bibnamefont
  {Sengupta}}, \bibinfo {author} {\bibfnamefont {A.}~\bibnamefont {Silva}}, \
  and\ \bibinfo {author} {\bibfnamefont {M.}~\bibnamefont {Vengalattore}},\
  }\href {\doibase 10.1103/RevModPhys.83.863} {\bibfield  {journal} {\bibinfo
  {journal} {Rev. Mod. Phys.}\ }\textbf {\bibinfo {volume} {83}},\ \bibinfo
  {pages} {863} (\bibinfo {year} {2011})}\BibitemShut {NoStop}%
\bibitem [{\citenamefont {Yukalov}(2011)}]{Yukalov11}%
  \BibitemOpen
  \bibfield  {author} {\bibinfo {author} {\bibfnamefont {V.}~\bibnamefont
  {Yukalov}},\ }\href {\doibase 10.1002/lapl.201110002} {\bibfield  {journal}
  {\bibinfo  {journal} {Laser Phys. Lett.}\ }\textbf {\bibinfo {volume} {8}},\
  \bibinfo {pages} {485} (\bibinfo {year} {2011})}\BibitemShut {NoStop}%
\bibitem [{\citenamefont {Eisert}\ \emph {et~al.}(2015)\citenamefont {Eisert},
  \citenamefont {Friesdorf},\ and\ \citenamefont {Gogolin}}]{Eisert2015}%
  \BibitemOpen
  \bibfield  {author} {\bibinfo {author} {\bibfnamefont {J.}~\bibnamefont
  {Eisert}}, \bibinfo {author} {\bibfnamefont {M.}~\bibnamefont {Friesdorf}}, \
  and\ \bibinfo {author} {\bibfnamefont {C.}~\bibnamefont {Gogolin}},\ }\href
  {\doibase 10.1038/nphys3215} {\bibfield  {journal} {\bibinfo  {journal} {Nat.
  Phys.}\ }\textbf {\bibinfo {volume} {11}},\ \bibinfo {pages} {124} (\bibinfo
  {year} {2015})}\BibitemShut {NoStop}%
\bibitem [{\citenamefont {Lamacraft}\ and\ \citenamefont
  {Moore}(2012)}]{Lamacraft12}%
  \BibitemOpen
  \bibfield  {author} {\bibinfo {author} {\bibfnamefont {A.}~\bibnamefont
  {Lamacraft}}\ and\ \bibinfo {author} {\bibfnamefont {J.}~\bibnamefont
  {Moore}},\ }in\ \href@noop {} {\emph {\bibinfo {booktitle} {Ultracold Bosonic
  and Fermionic Gases}}},\ \bibinfo {editor} {edited by\ \bibinfo {editor}
  {\bibfnamefont {A.}~\bibnamefont {Fetter}}, \bibinfo {editor} {\bibfnamefont
  {K.}~\bibnamefont {Levin}}, \ and\ \bibinfo {editor} {\bibfnamefont
  {D.}~\bibnamefont {Stamper-Kurn}}}\ (\bibinfo  {publisher} {Elsevier},\
  \bibinfo {address} {Oxford},\ \bibinfo {year} {2012})\ Chap.~\bibinfo
  {chapter} {7}\BibitemShut {NoStop}%
\bibitem [{\citenamefont {Bloch}\ \emph {et~al.}(2008)\citenamefont {Bloch},
  \citenamefont {Dalibard},\ and\ \citenamefont {Zwerger}}]{Bloch2008}%
  \BibitemOpen
  \bibfield  {author} {\bibinfo {author} {\bibfnamefont {I.}~\bibnamefont
  {Bloch}}, \bibinfo {author} {\bibfnamefont {J.}~\bibnamefont {Dalibard}}, \
  and\ \bibinfo {author} {\bibfnamefont {W.}~\bibnamefont {Zwerger}},\ }\href
  {\doibase 10.1103/RevModPhys.80.885} {\bibfield  {journal} {\bibinfo
  {journal} {Rev. Mod. Phys.}\ }\textbf {\bibinfo {volume} {80}},\ \bibinfo
  {pages} {885} (\bibinfo {year} {2008})}\BibitemShut {NoStop}%
\bibitem [{\citenamefont {Jaynes}(1957)}]{Jaynes1957}%
  \BibitemOpen
  \bibfield  {author} {\bibinfo {author} {\bibfnamefont {E.~T.}\ \bibnamefont
  {Jaynes}},\ }\href {\doibase 10.1103/PhysRev.106.620} {\bibfield  {journal}
  {\bibinfo  {journal} {Phys. Rev.}\ }\textbf {\bibinfo {volume} {106}},\
  \bibinfo {pages} {620} (\bibinfo {year} {1957})}\BibitemShut {NoStop}%
\bibitem [{\citenamefont {Kinoshita}\ \emph {et~al.}(2006)\citenamefont
  {Kinoshita}, \citenamefont {Wenger},\ and\ \citenamefont
  {Weiss}}]{Kinoshita2006}%
  \BibitemOpen
  \bibfield  {author} {\bibinfo {author} {\bibfnamefont {T.}~\bibnamefont
  {Kinoshita}}, \bibinfo {author} {\bibfnamefont {T.}~\bibnamefont {Wenger}}, \
  and\ \bibinfo {author} {\bibfnamefont {D.~S.}\ \bibnamefont {Weiss}},\ }\href
  {\doibase 10.1038/nature04693} {\bibfield  {journal} {\bibinfo  {journal}
  {Nature}\ }\textbf {\bibinfo {volume} {440}},\ \bibinfo {pages} {900}
  (\bibinfo {year} {2006})}\BibitemShut {NoStop}%
\bibitem [{\citenamefont {Rigol}\ \emph {et~al.}(2007)\citenamefont {Rigol},
  \citenamefont {Dunjko}, \citenamefont {Yurovsky},\ and\ \citenamefont
  {Olshanii}}]{Rigol2007}%
  \BibitemOpen
  \bibfield  {author} {\bibinfo {author} {\bibfnamefont {M.}~\bibnamefont
  {Rigol}}, \bibinfo {author} {\bibfnamefont {V.}~\bibnamefont {Dunjko}},
  \bibinfo {author} {\bibfnamefont {V.}~\bibnamefont {Yurovsky}}, \ and\
  \bibinfo {author} {\bibfnamefont {M.}~\bibnamefont {Olshanii}},\ }\href
  {\doibase 10.1103/PhysRevLett.98.050405} {\bibfield  {journal} {\bibinfo
  {journal} {Phys. Rev. Lett.}\ }\textbf {\bibinfo {volume} {98}},\ \bibinfo
  {pages} {050405} (\bibinfo {year} {2007})}\BibitemShut {NoStop}%
\bibitem [{\citenamefont {Kollar}\ \emph {et~al.}(2011)\citenamefont {Kollar},
  \citenamefont {Wolf},\ and\ \citenamefont {Eckstein}}]{Kollar2011}%
  \BibitemOpen
  \bibfield  {author} {\bibinfo {author} {\bibfnamefont {M.}~\bibnamefont
  {Kollar}}, \bibinfo {author} {\bibfnamefont {F.}~\bibnamefont {Wolf}}, \ and\
  \bibinfo {author} {\bibfnamefont {M.}~\bibnamefont {Eckstein}},\ }\href
  {\doibase 10.1103/PhysRevB.84.054304} {\bibfield  {journal} {\bibinfo
  {journal} {Phys. Rev. B}\ }\textbf {\bibinfo {volume} {84}},\ \bibinfo
  {pages} {054304} (\bibinfo {year} {2011})}\BibitemShut {NoStop}%
\bibitem [{\citenamefont {Caux}\ and\ \citenamefont {Konik}(2012)}]{Caux2012}%
  \BibitemOpen
  \bibfield  {author} {\bibinfo {author} {\bibfnamefont {J.-S.}\ \bibnamefont
  {Caux}}\ and\ \bibinfo {author} {\bibfnamefont {R.}~\bibnamefont {Konik}},\
  }\href {\doibase 10.1103/PhysRevLett.109.175301} {\bibfield  {journal}
  {\bibinfo  {journal} {Phys. Rev. Lett.}\ }\textbf {\bibinfo {volume} {109}},\
  \bibinfo {pages} {175301} (\bibinfo {year} {2012})}\BibitemShut {NoStop}%
\bibitem [{\citenamefont {Basko}\ \emph {et~al.}(2006)\citenamefont {Basko},
  \citenamefont {Aleiner},\ and\ \citenamefont {Altshuler}}]{Basko2006}%
  \BibitemOpen
  \bibfield  {author} {\bibinfo {author} {\bibfnamefont {D.}~\bibnamefont
  {Basko}}, \bibinfo {author} {\bibfnamefont {I.}~\bibnamefont {Aleiner}}, \
  and\ \bibinfo {author} {\bibfnamefont {B.}~\bibnamefont {Altshuler}},\ }\href
  {\doibase 10.1016/j.aop.2005.11.014} {\bibfield  {journal} {\bibinfo
  {journal} {Ann. Phys. (N. Y).}\ }\textbf {\bibinfo {volume} {321}},\ \bibinfo
  {pages} {1126} (\bibinfo {year} {2006})}\BibitemShut {NoStop}%
\bibitem [{\citenamefont {Pal}\ and\ \citenamefont {Huse}(2010)}]{PalHuse10}%
  \BibitemOpen
  \bibfield  {author} {\bibinfo {author} {\bibfnamefont {A.}~\bibnamefont
  {Pal}}\ and\ \bibinfo {author} {\bibfnamefont {D.~A.}\ \bibnamefont {Huse}},\
  }\href {\doibase 10.1103/PhysRevB.82.174411} {\bibfield  {journal} {\bibinfo
  {journal} {Phys. Rev. B}\ }\textbf {\bibinfo {volume} {82}},\ \bibinfo
  {pages} {174411} (\bibinfo {year} {2010})}\BibitemShut {NoStop}%
\bibitem [{\citenamefont {Serbyn}\ \emph {et~al.}(2013)\citenamefont {Serbyn},
  \citenamefont {Papi\ifmmode~\acute{c}\else \'{c}\fi{}},\ and\ \citenamefont
  {Abanin}}]{Serbyn2013}%
  \BibitemOpen
  \bibfield  {author} {\bibinfo {author} {\bibfnamefont {M.}~\bibnamefont
  {Serbyn}}, \bibinfo {author} {\bibfnamefont {Z.}~\bibnamefont
  {Papi\ifmmode~\acute{c}\else \'{c}\fi{}}}, \ and\ \bibinfo {author}
  {\bibfnamefont {D.}~\bibnamefont {Abanin}},\ }\href {\doibase
  10.1103/PhysRevLett.111.127201} {\bibfield  {journal} {\bibinfo  {journal}
  {Phys. Rev. Lett.}\ }\textbf {\bibinfo {volume} {111}},\ \bibinfo {pages}
  {127201} (\bibinfo {year} {2013})}\BibitemShut {NoStop}%
\bibitem [{\citenamefont {Vosk}\ and\ \citenamefont
  {Altman}(2013)}]{VoskAltman13}%
  \BibitemOpen
  \bibfield  {author} {\bibinfo {author} {\bibfnamefont {R.}~\bibnamefont
  {Vosk}}\ and\ \bibinfo {author} {\bibfnamefont {E.}~\bibnamefont {Altman}},\
  }\href {\doibase 10.1103/PhysRevLett.110.067204} {\bibfield  {journal}
  {\bibinfo  {journal} {Phys. Rev. Lett.}\ }\textbf {\bibinfo {volume} {110}},\
  \bibinfo {pages} {067204} (\bibinfo {year} {2013})}\BibitemShut {NoStop}%
\bibitem [{\citenamefont {Rossini}\ \emph {et~al.}(2009)\citenamefont
  {Rossini}, \citenamefont {Silva}, \citenamefont {Mussardo},\ and\
  \citenamefont {Santoro}}]{Rossini2009}%
  \BibitemOpen
  \bibfield  {author} {\bibinfo {author} {\bibfnamefont {D.}~\bibnamefont
  {Rossini}}, \bibinfo {author} {\bibfnamefont {A.}~\bibnamefont {Silva}},
  \bibinfo {author} {\bibfnamefont {G.}~\bibnamefont {Mussardo}}, \ and\
  \bibinfo {author} {\bibfnamefont {G.~E.}\ \bibnamefont {Santoro}},\ }\href
  {\doibase 10.1103/PhysRevLett.102.127204} {\bibfield  {journal} {\bibinfo
  {journal} {Phys. Rev. Lett.}\ }\textbf {\bibinfo {volume} {102}},\ \bibinfo
  {pages} {127204} (\bibinfo {year} {2009})}\BibitemShut {NoStop}%
\bibitem [{\citenamefont {Mitra}\ and\ \citenamefont
  {Giamarchi}(2011)}]{Mitra2011}%
  \BibitemOpen
  \bibfield  {author} {\bibinfo {author} {\bibfnamefont {A.}~\bibnamefont
  {Mitra}}\ and\ \bibinfo {author} {\bibfnamefont {T.}~\bibnamefont
  {Giamarchi}},\ }\href {\doibase 10.1103/PhysRevLett.107.150602} {\bibfield
  {journal} {\bibinfo  {journal} {Phys. Rev. Lett.}\ }\textbf {\bibinfo
  {volume} {107}},\ \bibinfo {pages} {150602} (\bibinfo {year}
  {2011})}\BibitemShut {NoStop}%
\bibitem [{\citenamefont {Foini}\ \emph {et~al.}(2011)\citenamefont {Foini},
  \citenamefont {Cugliandolo},\ and\ \citenamefont
  {Gambassi}}]{PhysRevB.84.212404}%
  \BibitemOpen
  \bibfield  {author} {\bibinfo {author} {\bibfnamefont {L.}~\bibnamefont
  {Foini}}, \bibinfo {author} {\bibfnamefont {L.~F.}\ \bibnamefont
  {Cugliandolo}}, \ and\ \bibinfo {author} {\bibfnamefont {A.}~\bibnamefont
  {Gambassi}},\ }\href {\doibase 10.1103/PhysRevB.84.212404} {\bibfield
  {journal} {\bibinfo  {journal} {Phys. Rev. B}\ }\textbf {\bibinfo {volume}
  {84}},\ \bibinfo {pages} {212404} (\bibinfo {year} {2011})}\BibitemShut
  {NoStop}%
\bibitem [{\citenamefont {Foini}\ \emph {et~al.}(2012)\citenamefont {Foini},
  \citenamefont {Cugliandolo},\ and\ \citenamefont {Gambassi}}]{Foini2012}%
  \BibitemOpen
  \bibfield  {author} {\bibinfo {author} {\bibfnamefont {L.}~\bibnamefont
  {Foini}}, \bibinfo {author} {\bibfnamefont {L.~F.}\ \bibnamefont
  {Cugliandolo}}, \ and\ \bibinfo {author} {\bibfnamefont {A.}~\bibnamefont
  {Gambassi}},\ }\href {\doibase 10.1088/1742-5468/2012/09/P09011} {\bibfield
  {journal} {\bibinfo  {journal} {J. Stat. Mech. Theory Exp.}\ }\textbf
  {\bibinfo {volume} {2012}},\ \bibinfo {pages} {P09011} (\bibinfo {year}
  {2012})}\BibitemShut {NoStop}%
\bibitem [{\citenamefont {Carusotto}\ and\ \citenamefont
  {Ciuti}(2013)}]{Carusotto2013}%
  \BibitemOpen
  \bibfield  {author} {\bibinfo {author} {\bibfnamefont {I.}~\bibnamefont
  {Carusotto}}\ and\ \bibinfo {author} {\bibfnamefont {C.}~\bibnamefont
  {Ciuti}},\ }\href {\doibase 10.1103/RevModPhys.85.299} {\bibfield  {journal}
  {\bibinfo  {journal} {Rev. Mod. Phys.}\ }\textbf {\bibinfo {volume} {85}},\
  \bibinfo {pages} {299} (\bibinfo {year} {2013})}\BibitemShut {NoStop}%
\bibitem [{\citenamefont {Byrnes}\ \emph {et~al.}(2014)\citenamefont {Byrnes},
  \citenamefont {Kim},\ and\ \citenamefont {Yamamoto}}]{Byrnes2014}%
  \BibitemOpen
  \bibfield  {author} {\bibinfo {author} {\bibfnamefont {T.}~\bibnamefont
  {Byrnes}}, \bibinfo {author} {\bibfnamefont {N.~Y.}\ \bibnamefont {Kim}}, \
  and\ \bibinfo {author} {\bibfnamefont {Y.}~\bibnamefont {Yamamoto}},\ }\href
  {\doibase 10.1038/nphys3143} {\bibfield  {journal} {\bibinfo  {journal} {Nat.
  Phys.}\ }\textbf {\bibinfo {volume} {10}},\ \bibinfo {pages} {803} (\bibinfo
  {year} {2014})}\BibitemShut {NoStop}%
\bibitem [{\citenamefont {Hartmann}\ \emph {et~al.}(2008)\citenamefont
  {Hartmann}, \citenamefont {Brand\~{a}o},\ and\ \citenamefont
  {Plenio}}]{Hartmann2008}%
  \BibitemOpen
  \bibfield  {author} {\bibinfo {author} {\bibfnamefont {M.}~\bibnamefont
  {Hartmann}}, \bibinfo {author} {\bibfnamefont {F.}~\bibnamefont
  {Brand\~{a}o}}, \ and\ \bibinfo {author} {\bibfnamefont {M.}~\bibnamefont
  {Plenio}},\ }\href {\doibase 10.1002/lpor.200810046} {\bibfield  {journal}
  {\bibinfo  {journal} {Laser Photonics Rev.}\ }\textbf {\bibinfo {volume}
  {2}},\ \bibinfo {pages} {527} (\bibinfo {year} {2008})}\BibitemShut {NoStop}%
\bibitem [{\citenamefont {Houck}\ \emph {et~al.}(2012)\citenamefont {Houck},
  \citenamefont {T\"{u}reci},\ and\ \citenamefont {Koch}}]{Houck2012}%
  \BibitemOpen
  \bibfield  {author} {\bibinfo {author} {\bibfnamefont {A.~A.}\ \bibnamefont
  {Houck}}, \bibinfo {author} {\bibfnamefont {H.~E.}\ \bibnamefont
  {T\"{u}reci}}, \ and\ \bibinfo {author} {\bibfnamefont {J.}~\bibnamefont
  {Koch}},\ }\href {\doibase 10.1038/nphys2251} {\bibfield  {journal} {\bibinfo
   {journal} {Nat. Phys.}\ }\textbf {\bibinfo {volume} {8}},\ \bibinfo {pages}
  {292} (\bibinfo {year} {2012})}\BibitemShut {NoStop}%
\bibitem [{\citenamefont {Blatt}\ and\ \citenamefont {Roos}(2012)}]{Blatt2012}%
  \BibitemOpen
  \bibfield  {author} {\bibinfo {author} {\bibfnamefont {R.}~\bibnamefont
  {Blatt}}\ and\ \bibinfo {author} {\bibfnamefont {C.~F.}\ \bibnamefont
  {Roos}},\ }\href {\doibase 10.1038/nphys2252} {\bibfield  {journal} {\bibinfo
   {journal} {Nat. Phys.}\ }\textbf {\bibinfo {volume} {8}},\ \bibinfo {pages}
  {277} (\bibinfo {year} {2012})}\BibitemShut {NoStop}%
\bibitem [{\citenamefont {Marquardt}\ and\ \citenamefont
  {Girvin}(2009)}]{Physics.2.40}%
  \BibitemOpen
  \bibfield  {author} {\bibinfo {author} {\bibfnamefont {F.}~\bibnamefont
  {Marquardt}}\ and\ \bibinfo {author} {\bibfnamefont {S.~M.}\ \bibnamefont
  {Girvin}},\ }\href {\doibase 10.1103/Physics.2.40} {\bibfield  {journal}
  {\bibinfo  {journal} {Physics (College. Park. Md).}\ }\textbf {\bibinfo
  {volume} {2}},\ \bibinfo {pages} {40} (\bibinfo {year} {2009})}\BibitemShut
  {NoStop}%
\bibitem [{\citenamefont {Mitra}\ \emph {et~al.}(2006)\citenamefont {Mitra},
  \citenamefont {Takei}, \citenamefont {Kim},\ and\ \citenamefont
  {Millis}}]{Mitra2006}%
  \BibitemOpen
  \bibfield  {author} {\bibinfo {author} {\bibfnamefont {A.}~\bibnamefont
  {Mitra}}, \bibinfo {author} {\bibfnamefont {S.}~\bibnamefont {Takei}},
  \bibinfo {author} {\bibfnamefont {Y.~B.}\ \bibnamefont {Kim}}, \ and\
  \bibinfo {author} {\bibfnamefont {A.~J.}\ \bibnamefont {Millis}},\ }\href
  {\doibase 10.1103/PhysRevLett.97.236808} {\bibfield  {journal} {\bibinfo
  {journal} {Phys. Rev. Lett.}\ }\textbf {\bibinfo {volume} {97}},\ \bibinfo
  {pages} {236808} (\bibinfo {year} {2006})}\BibitemShut {NoStop}%
\bibitem [{\citenamefont {Diehl}\ \emph {et~al.}(2008)\citenamefont {Diehl},
  \citenamefont {Micheli}, \citenamefont {Kantian}, \citenamefont {Kraus},
  \citenamefont {B\"{u}chler},\ and\ \citenamefont {Zoller}}]{Diehl2008}%
  \BibitemOpen
  \bibfield  {author} {\bibinfo {author} {\bibfnamefont {S.}~\bibnamefont
  {Diehl}}, \bibinfo {author} {\bibfnamefont {A.}~\bibnamefont {Micheli}},
  \bibinfo {author} {\bibfnamefont {A.}~\bibnamefont {Kantian}}, \bibinfo
  {author} {\bibfnamefont {B.}~\bibnamefont {Kraus}}, \bibinfo {author}
  {\bibfnamefont {H.~P.}\ \bibnamefont {B\"{u}chler}}, \ and\ \bibinfo {author}
  {\bibfnamefont {P.}~\bibnamefont {Zoller}},\ }\href {\doibase
  10.1038/nphys1073} {\bibfield  {journal} {\bibinfo  {journal} {Nat. Phys.}\
  }\textbf {\bibinfo {volume} {4}},\ \bibinfo {pages} {878} (\bibinfo {year}
  {2008})}\BibitemShut {NoStop}%
\bibitem [{\citenamefont {Diehl}\ \emph {et~al.}(2010)\citenamefont {Diehl},
  \citenamefont {Tomadin}, \citenamefont {Micheli}, \citenamefont {Fazio},\
  and\ \citenamefont {Zoller}}]{Diehl2010a}%
  \BibitemOpen
  \bibfield  {author} {\bibinfo {author} {\bibfnamefont {S.}~\bibnamefont
  {Diehl}}, \bibinfo {author} {\bibfnamefont {A.}~\bibnamefont {Tomadin}},
  \bibinfo {author} {\bibfnamefont {A.}~\bibnamefont {Micheli}}, \bibinfo
  {author} {\bibfnamefont {R.}~\bibnamefont {Fazio}}, \ and\ \bibinfo {author}
  {\bibfnamefont {P.}~\bibnamefont {Zoller}},\ }\href {\doibase
  10.1103/PhysRevLett.105.015702} {\bibfield  {journal} {\bibinfo  {journal}
  {Phys. Rev. Lett.}\ }\textbf {\bibinfo {volume} {105}},\ \bibinfo {pages}
  {015702} (\bibinfo {year} {2010})}\BibitemShut {NoStop}%
\bibitem [{\citenamefont {{Dalla Torre}}\ \emph {et~al.}(2010)\citenamefont
  {{Dalla Torre}}, \citenamefont {Demler}, \citenamefont {Giamarchi},\ and\
  \citenamefont {Altman}}]{DallaTorre2010}%
  \BibitemOpen
  \bibfield  {author} {\bibinfo {author} {\bibfnamefont {E.~G.}\ \bibnamefont
  {{Dalla Torre}}}, \bibinfo {author} {\bibfnamefont {E.}~\bibnamefont
  {Demler}}, \bibinfo {author} {\bibfnamefont {T.}~\bibnamefont {Giamarchi}}, \
  and\ \bibinfo {author} {\bibfnamefont {E.}~\bibnamefont {Altman}},\ }\href
  {\doibase 10.1038/nphys1754} {\bibfield  {journal} {\bibinfo  {journal} {Nat.
  Phys.}\ }\textbf {\bibinfo {volume} {6}},\ \bibinfo {pages} {806} (\bibinfo
  {year} {2010})}\BibitemShut {NoStop}%
\bibitem [{\citenamefont {{Dalla Torre}}\ \emph {et~al.}(2012)\citenamefont
  {{Dalla Torre}}, \citenamefont {Demler}, \citenamefont {Giamarchi},\ and\
  \citenamefont {Altman}}]{DallaTorre2012}%
  \BibitemOpen
  \bibfield  {author} {\bibinfo {author} {\bibfnamefont {E.~G.}\ \bibnamefont
  {{Dalla Torre}}}, \bibinfo {author} {\bibfnamefont {E.}~\bibnamefont
  {Demler}}, \bibinfo {author} {\bibfnamefont {T.}~\bibnamefont {Giamarchi}}, \
  and\ \bibinfo {author} {\bibfnamefont {E.}~\bibnamefont {Altman}},\ }\href
  {\doibase 10.1103/PhysRevB.85.184302} {\bibfield  {journal} {\bibinfo
  {journal} {Phys. Rev. B}\ }\textbf {\bibinfo {volume} {85}},\ \bibinfo
  {pages} {184302} (\bibinfo {year} {2012})}\BibitemShut {NoStop}%
\bibitem [{\citenamefont {{Dalla Torre}}\ \emph {et~al.}(2013)\citenamefont
  {{Dalla Torre}}, \citenamefont {Diehl}, \citenamefont {Lukin}, \citenamefont
  {Sachdev},\ and\ \citenamefont {Strack}}]{DallaTorre2013}%
  \BibitemOpen
  \bibfield  {author} {\bibinfo {author} {\bibfnamefont {E.~G.~D.}\
  \bibnamefont {{Dalla Torre}}}, \bibinfo {author} {\bibfnamefont
  {S.}~\bibnamefont {Diehl}}, \bibinfo {author} {\bibfnamefont {M.~D.}\
  \bibnamefont {Lukin}}, \bibinfo {author} {\bibfnamefont {S.}~\bibnamefont
  {Sachdev}}, \ and\ \bibinfo {author} {\bibfnamefont {P.}~\bibnamefont
  {Strack}},\ }\href {\doibase 10.1103/PhysRevA.87.023831} {\bibfield
  {journal} {\bibinfo  {journal} {Phys. Rev. A}\ }\textbf {\bibinfo {volume}
  {87}},\ \bibinfo {pages} {023831} (\bibinfo {year} {2013})}\BibitemShut
  {NoStop}%
\bibitem [{\citenamefont {Chiocchetta}\ and\ \citenamefont
  {Carusotto}(2013)}]{Chiocchetta2013}%
  \BibitemOpen
  \bibfield  {author} {\bibinfo {author} {\bibfnamefont {A.}~\bibnamefont
  {Chiocchetta}}\ and\ \bibinfo {author} {\bibfnamefont {I.}~\bibnamefont
  {Carusotto}},\ }\href {\doibase 10.1209/0295-5075/102/67007} {\bibfield
  {journal} {\bibinfo  {journal} {Europhys. Lett.}\ }\textbf {\bibinfo {volume}
  {102}},\ \bibinfo {pages} {67007} (\bibinfo {year} {2013})}\BibitemShut
  {NoStop}%
\bibitem [{\citenamefont {Sieberer}\ \emph {et~al.}(2013)\citenamefont
  {Sieberer}, \citenamefont {Huber}, \citenamefont {Altman},\ and\
  \citenamefont {Diehl}}]{Sieberer2013}%
  \BibitemOpen
  \bibfield  {author} {\bibinfo {author} {\bibfnamefont {L.~M.}\ \bibnamefont
  {Sieberer}}, \bibinfo {author} {\bibfnamefont {S.~D.}\ \bibnamefont {Huber}},
  \bibinfo {author} {\bibfnamefont {E.}~\bibnamefont {Altman}}, \ and\ \bibinfo
  {author} {\bibfnamefont {S.}~\bibnamefont {Diehl}},\ }\href {\doibase
  10.1103/PhysRevLett.110.195301} {\bibfield  {journal} {\bibinfo  {journal}
  {Phys. Rev. Lett.}\ }\textbf {\bibinfo {volume} {110}},\ \bibinfo {pages}
  {195301} (\bibinfo {year} {2013})}\BibitemShut {NoStop}%
\bibitem [{\citenamefont {T\"auber}\ and\ \citenamefont
  {Diehl}(2014)}]{Tauber14}%
  \BibitemOpen
  \bibfield  {author} {\bibinfo {author} {\bibfnamefont {U.~C.}\ \bibnamefont
  {T\"auber}}\ and\ \bibinfo {author} {\bibfnamefont {S.}~\bibnamefont
  {Diehl}},\ }\href {\doibase 10.1103/PhysRevX.4.021010} {\bibfield  {journal}
  {\bibinfo  {journal} {Phys. Rev. X}\ }\textbf {\bibinfo {volume} {4}},\
  \bibinfo {pages} {021010} (\bibinfo {year} {2014})}\BibitemShut {NoStop}%
\bibitem [{\citenamefont {Wouters}\ and\ \citenamefont
  {Carusotto}(2006)}]{Wouters2006}%
  \BibitemOpen
  \bibfield  {author} {\bibinfo {author} {\bibfnamefont {M.}~\bibnamefont
  {Wouters}}\ and\ \bibinfo {author} {\bibfnamefont {I.}~\bibnamefont
  {Carusotto}},\ }\href {\doibase 10.1103/PhysRevB.74.245316} {\bibfield
  {journal} {\bibinfo  {journal} {Phys. Rev. B}\ }\textbf {\bibinfo {volume}
  {74}},\ \bibinfo {pages} {245316} (\bibinfo {year} {2006})}\BibitemShut
  {NoStop}%
\bibitem [{\citenamefont {Mitra}\ and\ \citenamefont
  {Giamarchi}(2012)}]{Mitra2012}%
  \BibitemOpen
  \bibfield  {author} {\bibinfo {author} {\bibfnamefont {A.}~\bibnamefont
  {Mitra}}\ and\ \bibinfo {author} {\bibfnamefont {T.}~\bibnamefont
  {Giamarchi}},\ }\href {\doibase 10.1103/PhysRevB.85.075117} {\bibfield
  {journal} {\bibinfo  {journal} {Phys. Rev. B}\ }\textbf {\bibinfo {volume}
  {85}},\ \bibinfo {pages} {075117} (\bibinfo {year} {2012})}\BibitemShut
  {NoStop}%
\bibitem [{\citenamefont {\"{O}ztop}\ \emph {et~al.}(2012)\citenamefont
  {\"{O}ztop}, \citenamefont {Bordyuh}, \citenamefont {M\"{u}stecaplıoğlu},\
  and\ \citenamefont {T\"{u}reci}}]{Oztop2012}%
  \BibitemOpen
  \bibfield  {author} {\bibinfo {author} {\bibfnamefont {B.}~\bibnamefont
  {\"{O}ztop}}, \bibinfo {author} {\bibfnamefont {M.}~\bibnamefont {Bordyuh}},
  \bibinfo {author} {\bibfnamefont {O.~E.}\ \bibnamefont
  {M\"{u}stecaplıoğlu}}, \ and\ \bibinfo {author} {\bibfnamefont {H.~E.}\
  \bibnamefont {T\"{u}reci}},\ }\href {\doibase 10.1088/1367-2630/14/8/085011}
  {\bibfield  {journal} {\bibinfo  {journal} {New J. Phys.}\ }\textbf {\bibinfo
  {volume} {14}},\ \bibinfo {pages} {085011} (\bibinfo {year}
  {2012})}\BibitemShut {NoStop}%
\bibitem [{\citenamefont {Maghrebi}\ and\ \citenamefont
  {Gorshkov}()}]{Maghrebi2015}%
  \BibitemOpen
  \bibfield  {author} {\bibinfo {author} {\bibfnamefont {M.~F.}\ \bibnamefont
  {Maghrebi}}\ and\ \bibinfo {author} {\bibfnamefont {A.~V.}\ \bibnamefont
  {Gorshkov}},\ }\href {http://arxiv.org/abs/1507.01939} {\bibinfo  {journal}
  {arXiv:1507.01939}\ }\BibitemShut {NoStop}%
\bibitem [{\citenamefont {Chou}\ \emph {et~al.}(1985)\citenamefont {Chou},
  \citenamefont {Su}, \citenamefont {Hao},\ and\ \citenamefont
  {Yu}}]{Chou1985}%
  \BibitemOpen
\bibfield  {journal} {  }\bibfield  {author} {\bibinfo {author} {\bibfnamefont
  {K.-c.}\ \bibnamefont {Chou}}, \bibinfo {author} {\bibfnamefont {Z.-b.}\
  \bibnamefont {Su}}, \bibinfo {author} {\bibfnamefont {B.-l.}\ \bibnamefont
  {Hao}}, \ and\ \bibinfo {author} {\bibfnamefont {L.}~\bibnamefont {Yu}},\
  }\href {\doibase 10.1016/0370-1573(85)90136-X} {\bibfield  {journal}
  {\bibinfo  {journal} {Phys. Rep.}\ }\textbf {\bibinfo {volume} {118}},\
  \bibinfo {pages} {1} (\bibinfo {year} {1985})}\BibitemShut {NoStop}%
\bibitem [{\citenamefont {Wang}\ and\ \citenamefont {Heinz}(2002)}]{Wang2002}%
  \BibitemOpen
  \bibfield  {author} {\bibinfo {author} {\bibfnamefont {E.}~\bibnamefont
  {Wang}}\ and\ \bibinfo {author} {\bibfnamefont {U.}~\bibnamefont {Heinz}},\
  }\href {\doibase 10.1103/PhysRevD.66.025008} {\bibfield  {journal} {\bibinfo
  {journal} {Phys. Rev. D}\ }\textbf {\bibinfo {volume} {66}},\ \bibinfo
  {pages} {025008} (\bibinfo {year} {2002})}\BibitemShut {NoStop}%
\bibitem [{\citenamefont {Jakobs}\ \emph {et~al.}(2010)\citenamefont {Jakobs},
  \citenamefont {Pletyukhov},\ and\ \citenamefont {Schoeller}}]{Jakobs2010}%
  \BibitemOpen
  \bibfield  {author} {\bibinfo {author} {\bibfnamefont {S.~G.}\ \bibnamefont
  {Jakobs}}, \bibinfo {author} {\bibfnamefont {M.}~\bibnamefont {Pletyukhov}},
  \ and\ \bibinfo {author} {\bibfnamefont {H.}~\bibnamefont {Schoeller}},\
  }\href {\doibase 10.1088/1751-8113/43/10/103001} {\bibfield  {journal}
  {\bibinfo  {journal} {J. Phys. A Math. Theor.}\ }\textbf {\bibinfo {volume}
  {43}},\ \bibinfo {pages} {103001} (\bibinfo {year} {2010})}\BibitemShut
  {NoStop}%
\bibitem [{\citenamefont {Messiah}(1965)}]{Messiah:II}%
  \BibitemOpen
  \bibfield  {author} {\bibinfo {author} {\bibfnamefont {A.}~\bibnamefont
  {Messiah}},\ }\href@noop {} {\emph {\bibinfo {title} {Quantum Mechanics
  II}}},\ \bibinfo {edition} {3rd}\ ed.\ (\bibinfo  {publisher} {North-Holland
  Pubsishing Company},\ \bibinfo {address} {Amsterdam},\ \bibinfo {year}
  {1965})\BibitemShut {NoStop}%
\bibitem [{\citenamefont {Kubo}(1957)}]{Kubo1957}%
  \BibitemOpen
  \bibfield  {author} {\bibinfo {author} {\bibfnamefont {R.}~\bibnamefont
  {Kubo}},\ }\href {\doibase 10.1143/JPSJ.12.570} {\bibfield  {journal}
  {\bibinfo  {journal} {J. Phys. Soc. Japan}\ }\textbf {\bibinfo {volume}
  {12}},\ \bibinfo {pages} {570} (\bibinfo {year} {1957})}\BibitemShut
  {NoStop}%
\bibitem [{\citenamefont {Martin}\ and\ \citenamefont
  {Schwinger}(1959)}]{Martin1959}%
  \BibitemOpen
  \bibfield  {author} {\bibinfo {author} {\bibfnamefont {P.}~\bibnamefont
  {Martin}}\ and\ \bibinfo {author} {\bibfnamefont {J.}~\bibnamefont
  {Schwinger}},\ }\href {\doibase 10.1103/PhysRev.115.1342} {\bibfield
  {journal} {\bibinfo  {journal} {Phys. Rev.}\ }\textbf {\bibinfo {volume}
  {115}},\ \bibinfo {pages} {1342} (\bibinfo {year} {1959})}\BibitemShut
  {NoStop}%
\bibitem [{\citenamefont {Janssen}(1976)}]{Janssen1976}%
  \BibitemOpen
  \bibfield  {author} {\bibinfo {author} {\bibfnamefont {H.}~\bibnamefont
  {Janssen}},\ }\href {\doibase 10.1007/BF01316547} {\bibfield  {journal}
  {\bibinfo  {journal} {Z. Phys. B Condens. Matter}\ }\textbf {\bibinfo
  {volume} {23}},\ \bibinfo {pages} {377} (\bibinfo {year} {1976})}\BibitemShut
  {NoStop}%
\bibitem [{\citenamefont {Bausch}\ \emph {et~al.}(1976)\citenamefont {Bausch},
  \citenamefont {Janssen},\ and\ \citenamefont {Wagner}}]{Bausch1976}%
  \BibitemOpen
  \bibfield  {author} {\bibinfo {author} {\bibfnamefont {R.}~\bibnamefont
  {Bausch}}, \bibinfo {author} {\bibfnamefont {H.~K.}\ \bibnamefont {Janssen}},
  \ and\ \bibinfo {author} {\bibfnamefont {H.}~\bibnamefont {Wagner}},\ }\href
  {\doibase 10.1007/BF01312880} {\bibfield  {journal} {\bibinfo  {journal} {Z.
  Phys. B Condens. Matter}\ }\textbf {\bibinfo {volume} {24}},\ \bibinfo
  {pages} {113} (\bibinfo {year} {1976})}\BibitemShut {NoStop}%
\bibitem [{\citenamefont {Janssen}(1979)}]{Janssen1979}%
  \BibitemOpen
  \bibfield  {author} {\bibinfo {author} {\bibfnamefont {H.}~\bibnamefont
  {Janssen}},\ }in\ \href@noop {} {\emph {\bibinfo {booktitle} {Dynamical
  Critical Phenomena and Related Topics}}},\ \bibinfo {series} {Lecture Notes
  in Physics}, Vol.\ \bibinfo {volume} {104},\ \bibinfo {editor} {edited by\
  \bibinfo {editor} {\bibfnamefont {C.}~\bibnamefont {Enz}}}\ (\bibinfo
  {publisher} {Springer},\ \bibinfo {address} {Berlin, Heidelberg},\ \bibinfo
  {year} {1979})\BibitemShut {NoStop}%
\bibitem [{\citenamefont {Janssen}(1992)}]{Janssen1992}%
  \BibitemOpen
  \bibfield  {author} {\bibinfo {author} {\bibfnamefont {H.}~\bibnamefont
  {Janssen}},\ }in\ \href@noop {} {\emph {\bibinfo {booktitle} {From Phase
  Transitions to Chaos, Topics in Modern Statistical Physics}}},\ \bibinfo
  {editor} {edited by\ \bibinfo {editor} {\bibfnamefont {G.}~\bibnamefont
  {Gy\"{o}rgyi}}, \bibinfo {editor} {\bibfnamefont {I.}~\bibnamefont {Kondor}},
  \bibinfo {editor} {\bibfnamefont {L.}~\bibnamefont {Sasv\'{a}ri}}, \ and\
  \bibinfo {editor} {\bibfnamefont {T.}~\bibnamefont {T\'{e}l}}}\ (\bibinfo
  {publisher} {World Scientific},\ \bibinfo {address} {Singapore},\ \bibinfo
  {year} {1992})\BibitemShut {NoStop}%
\bibitem [{\citenamefont {Aron}\ \emph {et~al.}(2010)\citenamefont {Aron},
  \citenamefont {Biroli},\ and\ \citenamefont
  {Cugliandolo}}]{aron10:_symmet_langev}%
  \BibitemOpen
  \bibfield  {author} {\bibinfo {author} {\bibfnamefont {C.}~\bibnamefont
  {Aron}}, \bibinfo {author} {\bibfnamefont {G.}~\bibnamefont {Biroli}}, \ and\
  \bibinfo {author} {\bibfnamefont {L.~F.}\ \bibnamefont {Cugliandolo}},\
  }\href {http://stacks.iop.org/1742-5468/2010/i=11/a=P11018} {\bibfield
  {journal} {\bibinfo  {journal} {J. Stat. Mech.}\ }\textbf {\bibinfo {volume}
  {2010}},\ \bibinfo {pages} {P11018} (\bibinfo {year} {2010})}\BibitemShut
  {NoStop}%
\bibitem [{\citenamefont {Aron}\ \emph {et~al.}()\citenamefont {Aron},
  \citenamefont {Barci}, \citenamefont {Cugliandolo}, \citenamefont {Arenas},\
  and\ \citenamefont {Lozano}}]{Aron2014}%
  \BibitemOpen
  \bibfield  {author} {\bibinfo {author} {\bibfnamefont {C.}~\bibnamefont
  {Aron}}, \bibinfo {author} {\bibfnamefont {D.~G.}\ \bibnamefont {Barci}},
  \bibinfo {author} {\bibfnamefont {L.~F.}\ \bibnamefont {Cugliandolo}},
  \bibinfo {author} {\bibfnamefont {Z.~G.}\ \bibnamefont {Arenas}}, \ and\
  \bibinfo {author} {\bibfnamefont {G.~S.}\ \bibnamefont {Lozano}},\ }\href
  {http://arxiv.org/abs/1412.7564} {\bibinfo  {journal} {arXiv:1412.7564}\
  }\BibitemShut {NoStop}%
\bibitem [{\citenamefont {Schwinger}(1961)}]{Schwinger1961}%
  \BibitemOpen
\bibfield  {journal} {  }\bibfield  {author} {\bibinfo {author} {\bibfnamefont
  {J.}~\bibnamefont {Schwinger}},\ }\href {\doibase 10.1063/1.1703727}
  {\bibfield  {journal} {\bibinfo  {journal} {J. Math. Phys.}\ }\textbf
  {\bibinfo {volume} {2}},\ \bibinfo {pages} {407} (\bibinfo {year}
  {1961})}\BibitemShut {NoStop}%
\bibitem [{\citenamefont {Bakshi}\ and\ \citenamefont
  {Mahanthappa}(1963{\natexlab{a}})}]{Bakshi1963}%
  \BibitemOpen
  \bibfield  {author} {\bibinfo {author} {\bibfnamefont {P.~M.}\ \bibnamefont
  {Bakshi}}\ and\ \bibinfo {author} {\bibfnamefont {K.~T.}\ \bibnamefont
  {Mahanthappa}},\ }\href {\doibase 10.1063/1.1703883} {\bibfield  {journal}
  {\bibinfo  {journal} {J. Math. Phys.}\ }\textbf {\bibinfo {volume} {4}},\
  \bibinfo {pages} {1} (\bibinfo {year} {1963}{\natexlab{a}})}\BibitemShut
  {NoStop}%
\bibitem [{\citenamefont {Bakshi}\ and\ \citenamefont
  {Mahanthappa}(1963{\natexlab{b}})}]{Bakshi1963a}%
  \BibitemOpen
  \bibfield  {author} {\bibinfo {author} {\bibfnamefont {P.~M.}\ \bibnamefont
  {Bakshi}}\ and\ \bibinfo {author} {\bibfnamefont {K.~T.}\ \bibnamefont
  {Mahanthappa}},\ }\href {\doibase 10.1063/1.1703879} {\bibfield  {journal}
  {\bibinfo  {journal} {J. Math. Phys.}\ }\textbf {\bibinfo {volume} {4}},\
  \bibinfo {pages} {12} (\bibinfo {year} {1963}{\natexlab{b}})}\BibitemShut
  {NoStop}%
\bibitem [{\citenamefont {Mahanthappa}(1962)}]{Mahanthappa1962}%
  \BibitemOpen
  \bibfield  {author} {\bibinfo {author} {\bibfnamefont {K.}~\bibnamefont
  {Mahanthappa}},\ }\href {\doibase 10.1103/PhysRev.126.329} {\bibfield
  {journal} {\bibinfo  {journal} {Phys. Rev.}\ }\textbf {\bibinfo {volume}
  {126}},\ \bibinfo {pages} {329} (\bibinfo {year} {1962})}\BibitemShut
  {NoStop}%
\bibitem [{\citenamefont {Keldysh}(1965)}]{Keldysh1965}%
  \BibitemOpen
  \bibfield  {author} {\bibinfo {author} {\bibfnamefont {L.~V.}\ \bibnamefont
  {Keldysh}},\ }\href@noop {} {\bibfield  {journal} {\bibinfo  {journal} {Sov.
  Phys. JETP}\ }\textbf {\bibinfo {volume} {20}},\ \bibinfo {pages} {1018}
  (\bibinfo {year} {1965})}\BibitemShut {NoStop}%
\bibitem [{\citenamefont {Kamenev}(2011)}]{Kamenev2011}%
  \BibitemOpen
  \bibfield  {author} {\bibinfo {author} {\bibfnamefont {A.}~\bibnamefont
  {Kamenev}},\ }\href@noop {} {\emph {\bibinfo {title} {{Field Theory of
  Non-Equilibrium Systems}}}}\ (\bibinfo  {publisher} {Cambridge University
  Press},\ \bibinfo {address} {Cambridge},\ \bibinfo {year} {2011})\BibitemShut
  {NoStop}%
\bibitem [{\citenamefont {Altland}\ and\ \citenamefont
  {Simons}(2010)}]{Altland/Simons}%
  \BibitemOpen
  \bibfield  {author} {\bibinfo {author} {\bibfnamefont {A.}~\bibnamefont
  {Altland}}\ and\ \bibinfo {author} {\bibfnamefont {B.}~\bibnamefont
  {Simons}},\ }\href@noop {} {\emph {\bibinfo {title} {Condensed Matter Field
  Theory}}},\ \bibinfo {edition} {2nd}\ ed.\ (\bibinfo  {publisher} {Cambridge
  University Press},\ \bibinfo {address} {Cambridge},\ \bibinfo {year}
  {2010})\BibitemShut {NoStop}%
\bibitem [{\citenamefont {Stoof}(1999)}]{Stoof1999}%
  \BibitemOpen
  \bibfield  {author} {\bibinfo {author} {\bibfnamefont {H.~T.~C.}\
  \bibnamefont {Stoof}},\ }\href {\doibase 10.1023/A:1021897703053} {\bibfield
  {journal} {\bibinfo  {journal} {J. Low Temp. Phys.}\ }\textbf {\bibinfo
  {volume} {114}},\ \bibinfo {pages} {11} (\bibinfo {year} {1999})}\BibitemShut
  {NoStop}%
\bibitem [{\citenamefont {Altland}\ \emph {et~al.}(2010)\citenamefont
  {Altland}, \citenamefont {{De Martino}}, \citenamefont {Egger},\ and\
  \citenamefont {Narozhny}}]{Altland2010}%
  \BibitemOpen
  \bibfield  {author} {\bibinfo {author} {\bibfnamefont {A.}~\bibnamefont
  {Altland}}, \bibinfo {author} {\bibfnamefont {A.}~\bibnamefont {{De
  Martino}}}, \bibinfo {author} {\bibfnamefont {R.}~\bibnamefont {Egger}}, \
  and\ \bibinfo {author} {\bibfnamefont {B.}~\bibnamefont {Narozhny}},\ }\href
  {\doibase 10.1103/PhysRevB.82.115323} {\bibfield  {journal} {\bibinfo
  {journal} {Phys. Rev. B}\ }\textbf {\bibinfo {volume} {82}},\ \bibinfo
  {pages} {115323} (\bibinfo {year} {2010})}\BibitemShut {NoStop}%
\bibitem [{\citenamefont {Kossakowski}(1972)}]{Kossakowski1972}%
  \BibitemOpen
  \bibfield  {author} {\bibinfo {author} {\bibfnamefont {A.}~\bibnamefont
  {Kossakowski}},\ }\href {\doibase 10.1016/0034-4877(72)90010-9} {\bibfield
  {journal} {\bibinfo  {journal} {Reports Math. Phys.}\ }\textbf {\bibinfo
  {volume} {3}},\ \bibinfo {pages} {247} (\bibinfo {year} {1972})}\BibitemShut
  {NoStop}%
\bibitem [{\citenamefont {Lindblad}(1976)}]{Lindblad1976}%
  \BibitemOpen
  \bibfield  {author} {\bibinfo {author} {\bibfnamefont {G.}~\bibnamefont
  {Lindblad}},\ }\href {\doibase 10.1007/BF01608499} {\bibfield  {journal}
  {\bibinfo  {journal} {Commun. Math. Phys.}\ }\textbf {\bibinfo {volume}
  {48}},\ \bibinfo {pages} {119} (\bibinfo {year} {1976})}\BibitemShut
  {NoStop}%
\bibitem [{\citenamefont {Lifshitz}\ and\ \citenamefont
  {Pitaevskii}(1980)}]{Lifshitz1980}%
  \BibitemOpen
  \bibfield  {author} {\bibinfo {author} {\bibfnamefont {E.~M.}\ \bibnamefont
  {Lifshitz}}\ and\ \bibinfo {author} {\bibfnamefont {L.~P.}\ \bibnamefont
  {Pitaevskii}},\ }\href@noop {} {\emph {\bibinfo {title} {{Statistical
  Physics, Part 2: Theory of the Condensed State}}}},\ \bibinfo {edition}
  {2nd}\ ed.\ (\bibinfo  {publisher} {Pergamon Press},\ \bibinfo {address} {New
  York},\ \bibinfo {year} {1980})\BibitemShut {NoStop}%
\bibitem [{Note1()}]{Note1}%
  \BibitemOpen
  \bibinfo {note} {In Ref.~\protect \rev@citealp {Altland2010}, the symmetry is
  stated in terms of the real phase variables of complex fields. Then, the
  complex conjugation in Eq.~\protect \textup {\hbox {\mathsurround \z@
  \protect \normalfont (\ignorespaces \ref {eq:0}\unskip \@@italiccorr )}}
  should be replaced by a change of sign.}\BibitemShut {Stop}%
\bibitem [{\citenamefont {Hohenberg}\ and\ \citenamefont
  {Halperin}(1977)}]{Hohenberg1977}%
  \BibitemOpen
  \bibfield  {author} {\bibinfo {author} {\bibfnamefont {P.~C.}\ \bibnamefont
  {Hohenberg}}\ and\ \bibinfo {author} {\bibfnamefont {B.~I.}\ \bibnamefont
  {Halperin}},\ }\href {\doibase 10.1103/RevModPhys.49.435} {\bibfield
  {journal} {\bibinfo  {journal} {Rev. Mod. Phys.}\ }\textbf {\bibinfo {volume}
  {49}},\ \bibinfo {pages} {435} (\bibinfo {year} {1977})}\BibitemShut
  {NoStop}%
\bibitem [{\citenamefont {Folk}\ and\ \citenamefont {Moser}(2006)}]{Folk2006}%
  \BibitemOpen
  \bibfield  {author} {\bibinfo {author} {\bibfnamefont {R.}~\bibnamefont
  {Folk}}\ and\ \bibinfo {author} {\bibfnamefont {G.}~\bibnamefont {Moser}},\
  }\href {\doibase 10.1088/0305-4470/39/24/R01} {\bibfield  {journal} {\bibinfo
   {journal} {J. Phys. A. Math. Gen.}\ }\textbf {\bibinfo {volume} {39}},\
  \bibinfo {pages} {R207} (\bibinfo {year} {2006})}\BibitemShut {NoStop}%
\bibitem [{\citenamefont {Iucci}\ and\ \citenamefont
  {Cazalilla}(2009)}]{Iucci2009}%
  \BibitemOpen
  \bibfield  {author} {\bibinfo {author} {\bibfnamefont {A.}~\bibnamefont
  {Iucci}}\ and\ \bibinfo {author} {\bibfnamefont {M.~A.}\ \bibnamefont
  {Cazalilla}},\ }\href {\doibase 10.1103/PhysRevA.80.063619} {\bibfield
  {journal} {\bibinfo  {journal} {Phys. Rev. A}\ }\textbf {\bibinfo {volume}
  {80}},\ \bibinfo {pages} {063619} (\bibinfo {year} {2009})}\BibitemShut
  {NoStop}%
\bibitem [{\citenamefont {Barthel}\ and\ \citenamefont
  {Schollw\"ock}(2008)}]{Barthel2008}%
  \BibitemOpen
  \bibfield  {author} {\bibinfo {author} {\bibfnamefont {T.}~\bibnamefont
  {Barthel}}\ and\ \bibinfo {author} {\bibfnamefont {U.}~\bibnamefont
  {Schollw\"ock}},\ }\href {\doibase 10.1103/PhysRevLett.100.100601} {\bibfield
   {journal} {\bibinfo  {journal} {Phys. Rev. Lett.}\ }\textbf {\bibinfo
  {volume} {100}},\ \bibinfo {pages} {100601} (\bibinfo {year}
  {2008})}\BibitemShut {NoStop}%
\bibitem [{\citenamefont {Goldstein}\ and\ \citenamefont
  {Andrei}()}]{Goldstein2014}%
  \BibitemOpen
  \bibfield  {author} {\bibinfo {author} {\bibfnamefont {G.}~\bibnamefont
  {Goldstein}}\ and\ \bibinfo {author} {\bibfnamefont {N.}~\bibnamefont
  {Andrei}},\ }\href {http://arxiv.org/abs/1405.4224} {\bibinfo  {journal}
  {arXiv:1405.4224}\ }\BibitemShut {NoStop}%
\bibitem [{\citenamefont {Pozsgay}\ \emph {et~al.}(2014)\citenamefont
  {Pozsgay}, \citenamefont {Mesty\'{a}n}, \citenamefont {Werner}, \citenamefont
  {Kormos}, \citenamefont {Zar\'{a}nd},\ and\ \citenamefont
  {Tak\'{a}cs}}]{Pozsgay2014}%
  \BibitemOpen
\bibfield  {journal} {  }\bibfield  {author} {\bibinfo {author} {\bibfnamefont
  {B.}~\bibnamefont {Pozsgay}}, \bibinfo {author} {\bibfnamefont
  {M.}~\bibnamefont {Mesty\'{a}n}}, \bibinfo {author} {\bibfnamefont
  {M.}~\bibnamefont {Werner}}, \bibinfo {author} {\bibfnamefont
  {M.}~\bibnamefont {Kormos}}, \bibinfo {author} {\bibfnamefont
  {G.}~\bibnamefont {Zar\'{a}nd}}, \ and\ \bibinfo {author} {\bibfnamefont
  {G.}~\bibnamefont {Tak\'{a}cs}},\ }\href {\doibase
  10.1103/PhysRevLett.113.117203} {\bibfield  {journal} {\bibinfo  {journal}
  {Phys. Rev. Lett.}\ }\textbf {\bibinfo {volume} {113}},\ \bibinfo {pages}
  {117203} (\bibinfo {year} {2014})}\BibitemShut {NoStop}%
\bibitem [{\citenamefont {Mierzejewski}\ \emph {et~al.}(2014)\citenamefont
  {Mierzejewski}, \citenamefont {Prelov\v{s}ek},\ and\ \citenamefont
  {Prosen}}]{Mierzejewski2014}%
  \BibitemOpen
  \bibfield  {author} {\bibinfo {author} {\bibfnamefont {M.}~\bibnamefont
  {Mierzejewski}}, \bibinfo {author} {\bibfnamefont {P.}~\bibnamefont
  {Prelov\v{s}ek}}, \ and\ \bibinfo {author} {\bibfnamefont {T.}~\bibnamefont
  {Prosen}},\ }\href {\doibase 10.1103/PhysRevLett.113.020602} {\bibfield
  {journal} {\bibinfo  {journal} {Phys. Rev. Lett.}\ }\textbf {\bibinfo
  {volume} {113}},\ \bibinfo {pages} {020602} (\bibinfo {year}
  {2014})}\BibitemShut {NoStop}%
\bibitem [{\citenamefont {Wouters}\ \emph {et~al.}(2014)\citenamefont
  {Wouters}, \citenamefont {De~Nardis}, \citenamefont {Brockmann},
  \citenamefont {Fioretto}, \citenamefont {Rigol},\ and\ \citenamefont
  {Caux}}]{Wouters2014}%
  \BibitemOpen
  \bibfield  {author} {\bibinfo {author} {\bibfnamefont {B.}~\bibnamefont
  {Wouters}}, \bibinfo {author} {\bibfnamefont {J.}~\bibnamefont {De~Nardis}},
  \bibinfo {author} {\bibfnamefont {M.}~\bibnamefont {Brockmann}}, \bibinfo
  {author} {\bibfnamefont {D.}~\bibnamefont {Fioretto}}, \bibinfo {author}
  {\bibfnamefont {M.}~\bibnamefont {Rigol}}, \ and\ \bibinfo {author}
  {\bibfnamefont {J.-S.}\ \bibnamefont {Caux}},\ }\href {\doibase
  10.1103/PhysRevLett.113.117202} {\bibfield  {journal} {\bibinfo  {journal}
  {Phys. Rev. Lett.}\ }\textbf {\bibinfo {volume} {113}},\ \bibinfo {pages}
  {117202} (\bibinfo {year} {2014})}\BibitemShut {NoStop}%
\bibitem [{\citenamefont {Essler}\ \emph {et~al.}()\citenamefont {Essler},
  \citenamefont {Mussardo},\ and\ \citenamefont {Panfil}}]{Essler2014}%
  \BibitemOpen
  \bibfield  {author} {\bibinfo {author} {\bibfnamefont {F.~H.~L.}\
  \bibnamefont {Essler}}, \bibinfo {author} {\bibfnamefont {G.}~\bibnamefont
  {Mussardo}}, \ and\ \bibinfo {author} {\bibfnamefont {M.}~\bibnamefont
  {Panfil}},\ }\href {http://arxiv.org/abs/1411.5352} {\bibinfo  {journal}
  {arXiv:1411.5352}\ }\BibitemShut {NoStop}%
\bibitem [{\citenamefont {Fisher}\ \emph {et~al.}(1989)\citenamefont {Fisher},
  \citenamefont {Grinstein},\ and\ \citenamefont {Fisher}}]{Fisher1989}%
  \BibitemOpen
\bibfield  {journal} {  }\bibfield  {author} {\bibinfo {author} {\bibfnamefont
  {M.~P.~A.}\ \bibnamefont {Fisher}}, \bibinfo {author} {\bibfnamefont
  {G.}~\bibnamefont {Grinstein}}, \ and\ \bibinfo {author} {\bibfnamefont
  {D.~S.}\ \bibnamefont {Fisher}},\ }\href {\doibase 10.1103/PhysRevB.40.546}
  {\bibfield  {journal} {\bibinfo  {journal} {Phys. Rev. B}\ }\textbf {\bibinfo
  {volume} {40}},\ \bibinfo {pages} {546} (\bibinfo {year} {1989})}\BibitemShut
  {NoStop}%
\bibitem [{\citenamefont {Martin}\ \emph {et~al.}(1973)\citenamefont {Martin},
  \citenamefont {Siggia},\ and\ \citenamefont {Rose}}]{MSR1973}%
  \BibitemOpen
  \bibfield  {author} {\bibinfo {author} {\bibfnamefont {P.~C.}\ \bibnamefont
  {Martin}}, \bibinfo {author} {\bibfnamefont {E.~D.}\ \bibnamefont {Siggia}},
  \ and\ \bibinfo {author} {\bibfnamefont {H.~A.}\ \bibnamefont {Rose}},\
  }\href {\doibase 10.1103/PhysRevA.8.423} {\bibfield  {journal} {\bibinfo
  {journal} {Phys. Rev. A}\ }\textbf {\bibinfo {volume} {8}},\ \bibinfo {pages}
  {423} (\bibinfo {year} {1973})}\BibitemShut {NoStop}%
\bibitem [{\citenamefont {{De Dominicis, C.}}(1976)}]{DeDominicis1976}%
  \BibitemOpen
  \bibfield  {author} {\bibinfo {author} {\bibnamefont {{De Dominicis, C.}}},\
  }\href {\doibase 10.1051/jphyscol:1976138} {\bibfield  {journal} {\bibinfo
  {journal} {J. Phys. Colloques}\ }\textbf {\bibinfo {volume} {37}},\ \bibinfo
  {pages} {C1} (\bibinfo {year} {1976})}\BibitemShut {NoStop}%
\bibitem [{\citenamefont {De~Dominicis}(1978)}]{DeDominicis1978}%
  \BibitemOpen
  \bibfield  {author} {\bibinfo {author} {\bibfnamefont {C.}~\bibnamefont
  {De~Dominicis}},\ }\href {\doibase 10.1103/PhysRevB.18.4913} {\bibfield
  {journal} {\bibinfo  {journal} {Phys. Rev. B}\ }\textbf {\bibinfo {volume}
  {18}},\ \bibinfo {pages} {4913} (\bibinfo {year} {1978})}\BibitemShut
  {NoStop}%
\bibitem [{\citenamefont {T{\"a}uber}(2014)}]{Tauber2014}%
  \BibitemOpen
  \bibfield  {author} {\bibinfo {author} {\bibfnamefont {U.~C.}\ \bibnamefont
  {T{\"a}uber}},\ }\href@noop {} {\emph {\bibinfo {title} {{Critical Dynamics:
  A Field Theory Approach to Equilibrium and Non-Equilibrium Scaling
  Behavior}}}}\ (\bibinfo  {publisher} {Cambridge University Press},\ \bibinfo
  {address} {Cambridge},\ \bibinfo {year} {2014})\BibitemShut {NoStop}%
\bibitem [{\citenamefont {Sieberer}\ \emph {et~al.}(2014)\citenamefont
  {Sieberer}, \citenamefont {Huber}, \citenamefont {Altman},\ and\
  \citenamefont {Diehl}}]{Sieberer2014}%
  \BibitemOpen
  \bibfield  {author} {\bibinfo {author} {\bibfnamefont {L.~M.}\ \bibnamefont
  {Sieberer}}, \bibinfo {author} {\bibfnamefont {S.~D.}\ \bibnamefont {Huber}},
  \bibinfo {author} {\bibfnamefont {E.}~\bibnamefont {Altman}}, \ and\ \bibinfo
  {author} {\bibfnamefont {S.}~\bibnamefont {Diehl}},\ }\href {\doibase
  10.1103/PhysRevB.89.134310} {\bibfield  {journal} {\bibinfo  {journal} {Phys.
  Rev. B}\ }\textbf {\bibinfo {volume} {89}},\ \bibinfo {pages} {134310}
  (\bibinfo {year} {2014})}\BibitemShut {NoStop}%
\bibitem [{\citenamefont {Gardiner}(2004)}]{GardinerBook}%
  \BibitemOpen
  \bibfield  {author} {\bibinfo {author} {\bibfnamefont {C.}~\bibnamefont
  {Gardiner}},\ }\href@noop {} {\emph {\bibinfo {title} {Handbook of stochastic
  methods for physics, chemistry and the natural sciences}}},\ \bibinfo
  {edition} {3rd}\ ed.,\ Springer Series in Synergetics\ (\bibinfo  {publisher}
  {Springer},\ \bibinfo {address} {Berlin, Heidelberg},\ \bibinfo {year}
  {2004})\BibitemShut {NoStop}%
\bibitem [{\citenamefont {Zinn-Justin}(2002)}]{Zinn-Justin}%
  \BibitemOpen
  \bibfield  {author} {\bibinfo {author} {\bibfnamefont {J.}~\bibnamefont
  {Zinn-Justin}},\ }\href@noop {} {\emph {\bibinfo {title} {{Quantum Field
  Theory and Critical Phenomena}}}},\ \bibinfo {edition} {4th}\ ed.,\ \bibinfo
  {series} {International Series of Monographs on Physics}\ No.\ \bibinfo
  {number} {113}\ (\bibinfo  {publisher} {Oxford University Press},\ \bibinfo
  {address} {Oxford},\ \bibinfo {year} {2002})\BibitemShut {NoStop}%
\bibitem [{\citenamefont {Agarwal}(1973)}]{Agarwal1973}%
  \BibitemOpen
  \bibfield  {author} {\bibinfo {author} {\bibfnamefont {G.~S.}\ \bibnamefont
  {Agarwal}},\ }\href {\doibase 10.1007/BF01391504} {\bibfield  {journal}
  {\bibinfo  {journal} {Z. Phys.}\ }\textbf {\bibinfo {volume} {258}},\
  \bibinfo {pages} {409} (\bibinfo {year} {1973})}\BibitemShut {NoStop}%
\bibitem [{\citenamefont {Alicki}(1976)}]{Alicki1976}%
  \BibitemOpen
  \bibfield  {author} {\bibinfo {author} {\bibfnamefont {R.}~\bibnamefont
  {Alicki}},\ }\href {\doibase 10.1016/0034-4877(76)90046-X} {\bibfield
  {journal} {\bibinfo  {journal} {Reports Math. Phys.}\ }\textbf {\bibinfo
  {volume} {10}},\ \bibinfo {pages} {249} (\bibinfo {year} {1976})}\BibitemShut
  {NoStop}%
\bibitem [{\citenamefont {Kossakowski}\ \emph {et~al.}(1977)\citenamefont
  {Kossakowski}, \citenamefont {Frigerio}, \citenamefont {Gorini},\ and\
  \citenamefont {Verri}}]{Kossakowski1977}%
  \BibitemOpen
  \bibfield  {author} {\bibinfo {author} {\bibfnamefont {A.}~\bibnamefont
  {Kossakowski}}, \bibinfo {author} {\bibfnamefont {A.}~\bibnamefont
  {Frigerio}}, \bibinfo {author} {\bibfnamefont {V.}~\bibnamefont {Gorini}}, \
  and\ \bibinfo {author} {\bibfnamefont {M.}~\bibnamefont {Verri}},\ }\href
  {\doibase 10.1007/BF01625769} {\bibfield  {journal} {\bibinfo  {journal}
  {Commun. Math. Phys.}\ }\textbf {\bibinfo {volume} {57}},\ \bibinfo {pages}
  {97} (\bibinfo {year} {1977})}\BibitemShut {NoStop}%
\bibitem [{\citenamefont {Frigerio}\ and\ \citenamefont
  {Gorini}(1984)}]{Frigerio1984}%
  \BibitemOpen
  \bibfield  {author} {\bibinfo {author} {\bibfnamefont {A.}~\bibnamefont
  {Frigerio}}\ and\ \bibinfo {author} {\bibfnamefont {V.}~\bibnamefont
  {Gorini}},\ }\href {\doibase 10.1007/BF01212293} {\bibfield  {journal}
  {\bibinfo  {journal} {Commun. Math. Phys.}\ }\textbf {\bibinfo {volume}
  {93}},\ \bibinfo {pages} {517} (\bibinfo {year} {1984})}\BibitemShut
  {NoStop}%
\bibitem [{\citenamefont {Majewski}(1984)}]{Majewski1984}%
  \BibitemOpen
  \bibfield  {author} {\bibinfo {author} {\bibfnamefont {W.~A.}\ \bibnamefont
  {Majewski}},\ }\href {\doibase 10.1063/1.526164} {\bibfield  {journal}
  {\bibinfo  {journal} {J. Math. Phys.}\ }\textbf {\bibinfo {volume} {25}},\
  \bibinfo {pages} {614} (\bibinfo {year} {1984})}\BibitemShut {NoStop}%
\bibitem [{\citenamefont {Alicki}\ and\ \citenamefont
  {Lendi}(2007)}]{Alicki2007}%
  \BibitemOpen
  \bibfield  {author} {\bibinfo {author} {\bibfnamefont {R.}~\bibnamefont
  {Alicki}}\ and\ \bibinfo {author} {\bibfnamefont {K.}~\bibnamefont {Lendi}},\
  }\href@noop {} {\emph {\bibinfo {title} {Quantum Dynamical Semigroups and
  Applications}}},\ \bibinfo {series} {Lecture Notes in Physics}, Vol.\
  \bibinfo {volume} {717}\ (\bibinfo  {publisher} {Springer},\ \bibinfo
  {address} {Berlin, Heidelberg},\ \bibinfo {year} {2007})\BibitemShut
  {NoStop}%
\bibitem [{\citenamefont {Carmichael}\ and\ \citenamefont
  {Walls}(1976)}]{Carmichael1976}%
  \BibitemOpen
  \bibfield  {author} {\bibinfo {author} {\bibfnamefont {H.~J.}\ \bibnamefont
  {Carmichael}}\ and\ \bibinfo {author} {\bibfnamefont {D.~F.}\ \bibnamefont
  {Walls}},\ }\href {\doibase 10.1007/BF01318974} {\bibfield  {journal}
  {\bibinfo  {journal} {Z. Phys. B Condens. Matter}\ }\textbf {\bibinfo
  {volume} {23}},\ \bibinfo {pages} {299} (\bibinfo {year} {1976})}\BibitemShut
  {NoStop}%
\bibitem [{\citenamefont {Deutsch}(1991)}]{Deutsch1991}%
  \BibitemOpen
  \bibfield  {author} {\bibinfo {author} {\bibfnamefont {J.~M.}\ \bibnamefont
  {Deutsch}},\ }\href {\doibase 10.1103/PhysRevA.43.2046} {\bibfield  {journal}
  {\bibinfo  {journal} {Phys. Rev. A}\ }\textbf {\bibinfo {volume} {43}},\
  \bibinfo {pages} {2046} (\bibinfo {year} {1991})}\BibitemShut {NoStop}%
\bibitem [{\citenamefont {Srednicki}(1994)}]{Srednicki1994}%
  \BibitemOpen
  \bibfield  {author} {\bibinfo {author} {\bibfnamefont {M.}~\bibnamefont
  {Srednicki}},\ }\href {\doibase 10.1103/PhysRevE.50.888} {\bibfield
  {journal} {\bibinfo  {journal} {Phys. Rev. E}\ }\textbf {\bibinfo {volume}
  {50}},\ \bibinfo {pages} {888} (\bibinfo {year} {1994})}\BibitemShut
  {NoStop}%
\bibitem [{\citenamefont {Srednicki}(1999)}]{Srednicki1999}%
  \BibitemOpen
  \bibfield  {author} {\bibinfo {author} {\bibfnamefont {M.}~\bibnamefont
  {Srednicki}},\ }\href {http://stacks.iop.org/0305-4470/32/i=7/a=007}
  {\bibfield  {journal} {\bibinfo  {journal} {J. Phys. A: Math. Gen.}\ }\textbf
  {\bibinfo {volume} {32}},\ \bibinfo {pages} {1163} (\bibinfo {year}
  {1999})}\BibitemShut {NoStop}%
\bibitem [{\citenamefont {Khatami}\ \emph {et~al.}(2013)\citenamefont
  {Khatami}, \citenamefont {Pupillo}, \citenamefont {Srednicki},\ and\
  \citenamefont {Rigol}}]{Khatami2013}%
  \BibitemOpen
  \bibfield  {author} {\bibinfo {author} {\bibfnamefont {E.}~\bibnamefont
  {Khatami}}, \bibinfo {author} {\bibfnamefont {G.}~\bibnamefont {Pupillo}},
  \bibinfo {author} {\bibfnamefont {M.}~\bibnamefont {Srednicki}}, \ and\
  \bibinfo {author} {\bibfnamefont {M.}~\bibnamefont {Rigol}},\ }\href
  {\doibase 10.1103/PhysRevLett.111.050403} {\bibfield  {journal} {\bibinfo
  {journal} {Phys. Rev. Lett.}\ }\textbf {\bibinfo {volume} {111}},\ \bibinfo
  {pages} {050403} (\bibinfo {year} {2013})}\BibitemShut {NoStop}%
\bibitem [{Note2()}]{Note2}%
  \BibitemOpen
  \bibinfo {note} {Here we used the notion of ``Ward-Takahashi identity'' in
  the slightly generalized sense which encompasses the case of identities
  between correlation functions resulting from discrete symmetries (such as
  Eq.~\protect \textup {\hbox {\mathsurround \z@ \protect \normalfont
  (\ignorespaces \ref {eq:1}\unskip \@@italiccorr )}} in Sec.~\ref
  {sec:from-kms-symmetry}, which leads to, c.f., Eqs.~\protect \textup {\hbox
  {\mathsurround \z@ \protect \normalfont (\ignorespaces \ref {eq:52}\unskip
  \@@italiccorr )}} and \protect \textup {\hbox {\mathsurround \z@ \protect
  \normalfont (\ignorespaces \ref {eq:53}\unskip \@@italiccorr )}}) beyond the
  usual case of continuous symmetries~\cite {Zinn-Justin}.}\BibitemShut {Stop}%
\bibitem [{Note3()}]{Note3}%
  \BibitemOpen
  \bibinfo {note} {We note that this condition is usually expressed in the form
  {$$\delimiter "426830A A(t_A) B(t_B) \delimiter "526930B = \delimiter
  "426830A B(t_B) A(t_A + i \beta ) \delimiter "526930B .$$} However, an
  equilibrium state is also stationary and therefore both time arguments on the
  r.h.s.~can be translated by $-i \beta /2$ which leads immediately to
  Eq.~\protect \textup {\hbox {\mathsurround \z@ \protect \normalfont
  (\ignorespaces \ref {eq:46}\unskip \@@italiccorr )}}. Here we are assuming
  that the analytic continuation of real-time correlation functions into the
  complex plane is possible and unambiguous.}\BibitemShut {Stop}%
\bibitem [{Note4()}]{Note4}%
  \BibitemOpen
  \bibinfo {note} {It is straightforward to verify that the transformation
  $\protect \mathsf {T}$ is not modified in the presence of complex time
  arguments.}\BibitemShut {Stop}%
\bibitem [{\citenamefont {Gardiner}\ and\ \citenamefont
  {Zoller}(2000)}]{Gardiner/Zoller}%
  \BibitemOpen
  \bibfield  {author} {\bibinfo {author} {\bibfnamefont {C.~W.}\ \bibnamefont
  {Gardiner}}\ and\ \bibinfo {author} {\bibfnamefont {P.}~\bibnamefont
  {Zoller}},\ }\href@noop {} {\emph {\bibinfo {title} {Quantum Noise}}},\
  \bibinfo {edition} {2nd}\ ed.,\ \bibinfo {series} {Springer series in
  synergetics}, Vol.~\bibinfo {volume} {56}\ (\bibinfo  {publisher}
  {Springer},\ \bibinfo {address} {Berlin, Heidelberg},\ \bibinfo {year}
  {2000})\BibitemShut {NoStop}%
\bibitem [{\citenamefont {Talkner}(1986)}]{Talkner1986}%
  \BibitemOpen
  \bibfield  {author} {\bibinfo {author} {\bibfnamefont {P.}~\bibnamefont
  {Talkner}},\ }\href {\doibase 10.1016/0003-4916(86)90207-1} {\bibfield
  {journal} {\bibinfo  {journal} {Ann. Phys. (N. Y).}\ }\textbf {\bibinfo
  {volume} {167}},\ \bibinfo {pages} {390} (\bibinfo {year}
  {1986})}\BibitemShut {NoStop}%
\bibitem [{\citenamefont {Ford}\ and\ \citenamefont
  {O'Connell}(1996)}]{Ford1996}%
  \BibitemOpen
  \bibfield  {author} {\bibinfo {author} {\bibfnamefont {G.~W.}\ \bibnamefont
  {Ford}}\ and\ \bibinfo {author} {\bibfnamefont {R.~F.}\ \bibnamefont
  {O'Connell}},\ }\href {\doibase 10.1103/PhysRevLett.77.798} {\bibfield
  {journal} {\bibinfo  {journal} {Phys. Rev. Lett.}\ }\textbf {\bibinfo
  {volume} {77}},\ \bibinfo {pages} {798} (\bibinfo {year} {1996})}\BibitemShut
  {NoStop}%
\bibitem [{\citenamefont {Andreanov}\ \emph {et~al.}(2006)\citenamefont
  {Andreanov}, \citenamefont {Biroli},\ and\ \citenamefont
  {Lefèvre}}]{Andreanov2006}%
  \BibitemOpen
  \bibfield  {author} {\bibinfo {author} {\bibfnamefont {A.}~\bibnamefont
  {Andreanov}}, \bibinfo {author} {\bibfnamefont {G.}~\bibnamefont {Biroli}}, \
  and\ \bibinfo {author} {\bibfnamefont {A.}~\bibnamefont {Lefèvre}},\ }\href
  {http://stacks.iop.org/1742-5468/2006/i=07/a=P07008} {\bibfield  {journal}
  {\bibinfo  {journal} {J. Stat. Mech. Theory Exp.}\ }\textbf {\bibinfo
  {volume} {2006}},\ \bibinfo {pages} {P07008} (\bibinfo {year}
  {2006})}\BibitemShut {NoStop}%
\bibitem [{\citenamefont {Berges}\ \emph {et~al.}(2002)\citenamefont {Berges},
  \citenamefont {Tetradis},\ and\ \citenamefont
  {Wetterich}}]{berges02:_nonper}%
  \BibitemOpen
  \bibfield  {author} {\bibinfo {author} {\bibfnamefont {J.}~\bibnamefont
  {Berges}}, \bibinfo {author} {\bibfnamefont {N.}~\bibnamefont {Tetradis}}, \
  and\ \bibinfo {author} {\bibfnamefont {C.}~\bibnamefont {Wetterich}},\ }\href
  {\doibase 10.1016/S0370-1573(01)00098-9} {\bibfield  {journal} {\bibinfo
  {journal} {Phys. Rept.}\ }\textbf {\bibinfo {volume} {363}},\ \bibinfo
  {pages} {223} (\bibinfo {year} {2002})}\BibitemShut {NoStop}%
\bibitem [{\citenamefont {Pawlowski}(2007)}]{pawlowski07:_aspec}%
  \BibitemOpen
  \bibfield  {author} {\bibinfo {author} {\bibfnamefont {J.~M.}\ \bibnamefont
  {Pawlowski}},\ }\href {\doibase 10.1016/j.aop.2007.01.007} {\bibfield
  {journal} {\bibinfo  {journal} {Ann. Phys. (N. Y).}\ }\textbf {\bibinfo
  {volume} {322}},\ \bibinfo {pages} {2831} (\bibinfo {year}
  {2007})}\BibitemShut {NoStop}%
\bibitem [{\citenamefont
  {Delamotte}(2012)}]{delamotte08:_introd_nonper_renor_group}%
  \BibitemOpen
  \bibfield  {author} {\bibinfo {author} {\bibfnamefont {B.}~\bibnamefont
  {Delamotte}},\ }in\ \href {\doibase 10.1007/978-3-642-27320-9\_2} {\emph
  {\bibinfo {booktitle} {Renormalization Group and Effective Field Theory
  Approaches to Many-Body Systems SE - 2}}},\ \bibinfo {series} {Lecture Notes
  in Physics}, Vol.\ \bibinfo {volume} {852},\ \bibinfo {editor} {edited by\
  \bibinfo {editor} {\bibfnamefont {A.}~\bibnamefont {Schwenk}}\ and\ \bibinfo
  {editor} {\bibfnamefont {J.}~\bibnamefont {Polonyi}}}\ (\bibinfo  {publisher}
  {Springer Berlin Heidelberg},\ \bibinfo {year} {2012})\ pp.\ \bibinfo {pages}
  {49--132}\BibitemShut {NoStop}%
\bibitem [{\citenamefont {Rosten}(2012)}]{rosten12:_fundam}%
  \BibitemOpen
  \bibfield  {author} {\bibinfo {author} {\bibfnamefont {O.~J.}\ \bibnamefont
  {Rosten}},\ }\href {\doibase 10.1016/j.physrep.2011.12.003} {\bibfield
  {journal} {\bibinfo  {journal} {Phys. Rep.}\ }\textbf {\bibinfo {volume}
  {511}},\ \bibinfo {pages} {177} (\bibinfo {year} {2012})}\BibitemShut
  {NoStop}%
\bibitem [{\citenamefont {Boettcher}\ \emph {et~al.}(2012)\citenamefont
  {Boettcher}, \citenamefont {Pawlowski},\ and\ \citenamefont
  {Diehl}}]{boettcher12:_ultrac_funct_renor_group}%
  \BibitemOpen
  \bibfield  {author} {\bibinfo {author} {\bibfnamefont {I.}~\bibnamefont
  {Boettcher}}, \bibinfo {author} {\bibfnamefont {J.~M.}\ \bibnamefont
  {Pawlowski}}, \ and\ \bibinfo {author} {\bibfnamefont {S.}~\bibnamefont
  {Diehl}},\ }\href {\doibase 10.1016/j.nuclphysbps.2012.06.004} {\bibfield
  {journal} {\bibinfo  {journal} {Nucl. Phys. B - Proc. Suppl.}\ }\textbf
  {\bibinfo {volume} {228}},\ \bibinfo {pages} {63} (\bibinfo {year}
  {2012})}\BibitemShut {NoStop}%
\bibitem [{\citenamefont {Canet}\ and\ \citenamefont
  {Chat\'{e}}(2007)}]{Canet2007}%
  \BibitemOpen
  \bibfield  {author} {\bibinfo {author} {\bibfnamefont {L.}~\bibnamefont
  {Canet}}\ and\ \bibinfo {author} {\bibfnamefont {H.}~\bibnamefont
  {Chat\'{e}}},\ }\href {\doibase 10.1088/1751-8113/40/9/002} {\bibfield
  {journal} {\bibinfo  {journal} {J. Phys. A: Math. Theor.}\ }\textbf {\bibinfo
  {volume} {40}},\ \bibinfo {pages} {1937} (\bibinfo {year}
  {2007})}\BibitemShut {NoStop}%
\bibitem [{\citenamefont {Mesterh\'{a}zy}\ \emph {et~al.}(2013)\citenamefont
  {Mesterh\'{a}zy}, \citenamefont {Stockemer}, \citenamefont {Palhares},\ and\
  \citenamefont {Berges}}]{Mesterhazy2013}%
  \BibitemOpen
  \bibfield  {author} {\bibinfo {author} {\bibfnamefont {D.}~\bibnamefont
  {Mesterh\'{a}zy}}, \bibinfo {author} {\bibfnamefont {J.~H.}\ \bibnamefont
  {Stockemer}}, \bibinfo {author} {\bibfnamefont {L.~F.}\ \bibnamefont
  {Palhares}}, \ and\ \bibinfo {author} {\bibfnamefont {J.}~\bibnamefont
  {Berges}},\ }\href {\doibase 10.1103/PhysRevB.88.174301} {\bibfield
  {journal} {\bibinfo  {journal} {Phys. Rev. B}\ }\textbf {\bibinfo {volume}
  {88}},\ \bibinfo {pages} {174301} (\bibinfo {year} {2013})}\BibitemShut
  {NoStop}%
\bibitem [{\citenamefont {Mesterh\'{a}zy}\ \emph {et~al.}()\citenamefont
  {Mesterh\'{a}zy}, \citenamefont {Stockemer},\ and\ \citenamefont
  {Tanizaki}}]{Mesterhazy2015}%
  \BibitemOpen
  \bibfield  {author} {\bibinfo {author} {\bibfnamefont {D.}~\bibnamefont
  {Mesterh\'{a}zy}}, \bibinfo {author} {\bibfnamefont {J.~H.}\ \bibnamefont
  {Stockemer}}, \ and\ \bibinfo {author} {\bibfnamefont {Y.}~\bibnamefont
  {Tanizaki}},\ }\href {http://arxiv.org/abs/1504.07268} {\bibinfo  {journal}
  {arXiv:1504.07268}\ }\BibitemShut {NoStop}%
\bibitem [{\citenamefont {Campisi}\ \emph {et~al.}(2011)\citenamefont
  {Campisi}, \citenamefont {H\"anggi},\ and\ \citenamefont
  {Talkner}}]{Campisi2011}%
  \BibitemOpen
\bibfield  {journal} {  }\bibfield  {author} {\bibinfo {author} {\bibfnamefont
  {M.}~\bibnamefont {Campisi}}, \bibinfo {author} {\bibfnamefont
  {P.}~\bibnamefont {H\"anggi}}, \ and\ \bibinfo {author} {\bibfnamefont
  {P.}~\bibnamefont {Talkner}},\ }\href {\doibase 10.1103/RevModPhys.83.771}
  {\bibfield  {journal} {\bibinfo  {journal} {Rev. Mod. Phys.}\ }\textbf
  {\bibinfo {volume} {83}},\ \bibinfo {pages} {771} (\bibinfo {year}
  {2011})}\BibitemShut {NoStop}%
\bibitem [{\citenamefont {Esposito}\ \emph {et~al.}(2009)\citenamefont
  {Esposito}, \citenamefont {Harbola},\ and\ \citenamefont
  {Mukamel}}]{Esposito2014}%
  \BibitemOpen
  \bibfield  {author} {\bibinfo {author} {\bibfnamefont {M.}~\bibnamefont
  {Esposito}}, \bibinfo {author} {\bibfnamefont {U.}~\bibnamefont {Harbola}}, \
  and\ \bibinfo {author} {\bibfnamefont {S.}~\bibnamefont {Mukamel}},\ }\href
  {\doibase 10.1103/RevModPhys.81.1665} {\bibfield  {journal} {\bibinfo
  {journal} {Rev. Mod. Phys.}\ }\textbf {\bibinfo {volume} {81}},\ \bibinfo
  {pages} {1665} (\bibinfo {year} {2009})}\BibitemShut {NoStop}%
\end{thebibliography}%

\end{document}